\documentclass[usenatbib]{aa}

\usepackage{graphicx}
\usepackage{txfonts}
\usepackage{hyperref}
\usepackage{xcolor}
\hypersetup{
colorlinks,
linkcolor={red!80!black},
citecolor={blue!80!black},
urlcolor={blue!80!black}
}

\usepackage{physics}
\usepackage{subcaption}
\usepackage{soul}

\newenvironment{myfont}{\fontfamily{ppl}\selectfont}{\par}
\DeclareTextFontCommand{\textmyfont}{\myfont}

\newcommand{\code}[1]{\texttt{#1}}

\def\nifs{\iso{56}Ni}

\def\cofs{\iso{56}Co}

\def\cm3{cm$^{-3}$}
\def\kms{\mbox{km~s$^{-1}$}}

\def\msun{$M_{\odot}$}

\def\one{\ts {\,\sc i}}
\def\two{\ts {\,\sc ii}}
\def\three{\ts {\,\sc iii}}
\def\four{\ts {\,\sc iv}}

\def\beq{\begin{equation}}
\def\eeq{\end{equation}}

\def\lesssim{\mathrel{\hbox{\rlap{\hbox{\lower4pt\hbox{$\sim$}}}\hbox{$<$}}}}
\def\gtrsim{\mathrel{\hbox{\rlap{\hbox{\lower4pt\hbox{$\sim$}}}\hbox{$>$}}}}

\def\one{{\,\sc i}}
\def\two{{\,\sc ii}}
\def\three{{\,\sc iii}}
\def\four{{\,\sc iv}}

\def\v1d{{\code{V1D}}}
\def\sumo{{\code{SUMO}}}

\def\cmfgen{{\code{CMFGEN}}}

\def\mic{\,$\mu$m}

\def\niidoub{[N\two]\,0.657\,$\mu$m}
\def\oidoub{[O\one]\,0.632\,$\mu$m}

\def\neiifs{[Ne\two]\,12.810\,$\mu$m}
\def\neiiifs{[Ne\three]\,15.550\,$\mu$m}

\def\ariimir{[Ar\two]\,6.983\,$\mu$m}

\def\caiidoub{[Ca\two]\,0.731\,$\mu$m}

\def\nkiiopt{[Ni\two]\,0.738\,$\mu$m}

\def\nkiimir{[Ni\two]\,6.634\,$\mu$m}

\def\nkiiimir{[Ni\three]\,7.347\,$\mu$m}

\newcommand{\iso}[2]{\ensuremath{^{#1}\rm{#2}}}

\begin{document}

   \title{Infrared diagnostics of late-time core-collapse supernova spectra}

   \titlerunning{Infrared diagnostics of nebular-phase CCSN spectra}

\author{Luc Dessart
  }

\institute{
Institut d'Astrophysique de Paris, CNRS-Sorbonne Universit\'e, 98 bis boulevard Arago, F-75014 Paris, France.\label{inst1}
  }

   \date{}

  \abstract{
We present nonlocal thermodynamic equilibrium radiative transfer calculations of red supergiant and He-star explosions, extending previous work to focus on the infrared emission from atoms and ions in the ejecta during the nebular-phase (i.e., $\sim$\,200 to $\sim$\,500\,d) -- molecules and dust are ignored. We cover non-rotating solar-metallicity progenitors spanning an initial mass between 10 and about 40\,\msun\ and exploding as Type II or Ibc supernovae (SNe). Both photometrically and spectroscopically, the SN II models evolve distinctly from the SN Ibc models primarily because of the greater ejecta kinetic-energy-to-mass ratio in the latter, which leads to a greater $\gamma$-ray escape together with a lower density and a higher ionization in our H-deficient ejecta. Type II SN models remain optically luminous at all times, whereas Type Ibc models progressively brighten in the infrared (which holds 80\,\% of their luminosity at 500\,d), causing strong infrared lines such as \neiifs\ and \nkiimir\ to evolve essentially at constant luminosity. The strength of \neiifs\ exhibits a complicated dependence with either He-core or preSN mass because of the additional impact of ejecta ionization -- this line radiates alone up to 20\,\% of the SN luminosity after $\sim$\,300\,d in our Type Ibc SN models. The numerous infrared Ni lines are found to be good tracers of the material that underwent explosive nucleosynthesis and can thus be used directly to constrain the level of \nifs\ mixing in core-collapse SNe. The evolution of the integrated flux in infrared Fe and Co lines shows a great amount of diversity, which compromises their use as a diagnostic of the \nifs-decay power source in our models. Future spectroscopic observations of core-collapse SNe by JWST will provide unprecedented information on the emission from atoms and ions in their ejecta, delivering critical constraints on the inner workings of massive star explosions.
}
   \keywords{
  radiative transfer --
  supernovae: general --
  Infrared: general --
  line: formation
               }

   \maketitle

\section{Introduction}

Up until recently, infrared spectroscopic observations of core-collapse supernovae (SNe) have largely been limited to the near infrared, and primarily used for assessing the presence of molecular emission from the CO first overtone or the presence of He\one\ lines at 1.083 and 2.058\mic\ (see, e.g., \citealt{spyromilio_co_87A_88},
\citealt{gerardy_co_00ew_02}, \citealt{hunter_etal_07gr_07bi}, \citealt{rho_co_17eaw_18}, \citealt{hsiao_nir_19}, \citealt{rho_20oi_21}, or \citealt{shahbandeh_nir_sesn_22}). The longer wavelength range was explored for SN\,1987A by \citet{wooden_87A_ir_93}, and later opened by Spitzer for a sample of nearby objects, including Type II  SN\,2004dj \citep{kotak_04dj_05,meikle_04dj_11}, SN\,2005af \citep{kotak_05af_06}, SN\,2004et \citep{kotak_04et_09}, showing evidence of CO and SiO emission at nebular times, as well as a wealth of unique emission lines, typically without optical counterparts and arising from (stable) Ni, Ar, or Ne. Ongoing observations with the James Webb Space Telescope (JWST) are now bringing a revolution in the field by providing high-quality observations from the near- to the mid-infrared with an exquisite signal even for not so nearby transients. So far though, such data for core-collapse SNe have been mostly photometric and employed for constraining their dust properties (see, e.g., \citealt{shahbandeh_jwst_23}, \citealt{zsiros_80K_24}, \citealt{szalai_93j_25}, \citealt{sarangi_05af_25}).

There is, however, much interest in using infrared spectra to infer properties of the ejecta, the explosion, and the pre-SN star, thus employing spectral lines as probes of the emitting gas, as done recently for Type Ia SNe 2021aefx and 2022pul (see, e.g., \citealt{derkacy_21aefx_23}, \citealt{kwok_21aefx_23}, \citealt{ashall_21aefx_24}, \citealt{kwok_22pul_24}). Doing this early in the nebular phase of a Type II SN, that is prior to dust formation and the partial obscuration of the metal-rich inner ejecta (thus before 500\,d if we rely on what occurred in SN\,1987A; \citealt{lucy_dust_89}) and before any late-time power contribution from ejecta-CSM interaction appears (see, e.g., \citealt{weil_17eaw_20}), would allow for the inference of key diagnostics not easily accessible with optical data. Similar information for stripped-envelope SNe is yet to come.

Previous radiative-transfer modeling of the infrared properties of core-collapse SNe seems to go back to SN\,1987A. From the modeling of the evolution of the iron, cobalt, and nickel lines during the first two years, \citet{li_87A_93} inferred that Fe clumps occupy 30\,\% of the volume in the inner ejecta as a result of the \nifs-bubble effect \citep{woosley_87A_late_88,basko_56ni_94}. \citet{KF98a,KF98b} modeled the evolution of the ejecta temperature, ionization, and the associated optical and infrared line emission using physical models of the explosion for SN\,1987A, covering epochs from 200\,d out to several years, and thereby constrained some of the ejecta composition. Other works focused on CO and SiO molecule formation and cooling \citep{liu_dalgarno_92,liu_dalgarno_94,liu_dalgarno_95}. More recently, \citet{jerkstrand_04et_12} performed radiative-transfer modeling with the Monte Carlo code \sumo\ \citep{jerkstrand_87a_11} of nebular-phase optical and infrared observations of SN\,2004et, with a simplified treatment of molecules and dust. \citet{liljegren_co_20} and \citet{liljegren_ibc_mol_23} implemented molecular networks within the radiative-transfer codes \sumo, and more recently \citet{mcleod_mol_24} operated a similar upgrade in the grid-based code \cmfgen\ \citep{HD12}, in order to solve for molecular formation and cooling together with the standard processes that control the physics of the gas and the radiation. These latter studies confirm earlier work that molecules such as CO and SiO are efficient coolants and can thus control the temperature of the material where they form (i.e., from the O/C or O/Si rich regions of the ejecta) and affect the local emission arising from atoms and ions, potentially shifting some of the optical flux to the infrared (e.g., through emission in the CO fundamental). The modeling of molecular formation and its impact on the radiative transfer is, however, challenging and computationally expensive.

In the present study, molecules and dust were therefore ignored in order to focus exclusively on the cooling of the gas by atoms and ions. Such simulations are prerequisites for more detailed calculations that would include molecules and dust, a topic that is left to future work. For simplicity, and also because our earlier work showed a relative insensitivity of our model results to clumping in nebular-phase spectra, we adopted a smooth ejecta throughout this work -- including clumping can be done in the future starting from the  present calculations. We thus merely extended the calculations of red supergiant star explosions presented in \citet{dessart_sn2p_21} and those of He-star explosions presented in \citet{dessart_snibc_21,dessart_snibc_23}, to document their infrared properties over a broad time span from about 200 to about 500\,d. Unlike \citet{liljegren_ibc_mol_23}, who focused on molecular diagnostics for stripped-envelope SNe, we explored the rich information contained in the infrared spectra from 1 to 30\mic\ for a wide range of H-rich and H-poor ejecta spanning a broad range of progenitor masses and nucleosynthetic yields, studying in detail the emission from atoms and ions in the gas. We searched for specific emission line diagnostics associated with O, Na, Mg, Ne, Si, S, Ar, Ca, Fe, Co, and Ni, and investigated what information they can provide about the progenitor and its explosion. We also confronted the information revealed by the infrared and optical ranges, for example, between \neiifs\ and \oidoub, in order to build a panchromatic view of these ejecta models. The present simulations are confronted to the nebular phase optical-to-infrared spectral observations of SN\,2024ggi in \citet{dessart_24ggi_25}.

In the next section, we present the numerical setup of the simulations performed for this work. These are largely inherited from \citet{dessart_sn2p_21} and \citet{dessart_snibc_21,dessart_snibc_23}, but some specific adjustments were made for this new focus on the infrared. We then describe in detail two ejecta models, one for a Type II SN (model s15p2; Section~\ref{sect_s15p2}) and another for a Type Ib SN (model he6p0; Section~\ref{sect_he6p0}). In Section~\ref{sect_neon}, we discuss the possibility of using \neiifs\ as an alternative to \oidoub, which has been extensively used to constrain the preSN or progenitor mass (see, e.g., \citealt{fransson_chevalier_89}; \citealt{maguire_2p_12}; \citealt{jerkstrand_04et_12}; \citealt{dessart_sn2p_21}). In Section~\ref{sect_ir_lines}, we describe the evolution of a number of key infrared lines such as \nkiimir\ or \ariimir. In Section~\ref{sect_fe_co}, we study the  evolution of Fe and Co line fluxes to find evidence of the \nifs-decay chain as the power source in our ejecta models. We present our conclusions in Section~\ref{sect_conc}. {In the appendix, we provide additional illustrations and results (Appendix~\ref{sect_add}) as well as a comparison to previous work (Appendix~\ref{sect_comp}).

\begin{table*}
    \caption{Model properties for the red supergiant star and He-star explosion models used in this study.
\label{tab_init}
}
\begin{center}
\begin{tabular}{l@{\hspace{2mm}}|c@{\hspace{2mm}}c@{\hspace{2mm}}c@{\hspace{2mm}}c@{\hspace{2mm}}c@{\hspace{2mm}}c@{\hspace{2mm}}c@{\hspace{2mm}}c@{\hspace{2mm}}c@{\hspace{2mm}}c@{\hspace{2mm}}c@{\hspace{2mm}}c@{\hspace{2mm}}c@{\hspace{2mm}}}
\hline
Model  & $M_{\rm ej}$ & $E_{\rm kin}$& $V_{\rm m}$ & \iso{1}H  &  \iso{4}He  & \iso{16}O & Ne  &   \iso{24}Mg &  \iso{36}Ar &  \iso{40}Ca &  \iso{56}Ni$_{t=0}$ & \iso{57}Ni$_{t=0}$ & \iso{58}Ni \\
       & [\msun]   &        [foe]    & [\kms]    &     [\msun] & [\msun] & [\msun] & [\msun] & [\msun] & [\msun] & [\msun] & [\msun] & [\msun] & [\msun]  \\
\hline
   s10p0$^\dagger$ &        8.19 &       0.61 &       2727 &   4.94 &   2.97 &   9.65(-2) &   1.64(-2) &   5.39(-3) &   1.73(-3) &   1.37(-3) &   2.48(-2) &   1.22(-3) &   1.15(-3) \\
   s12p0$^\dagger$ &        9.32 &       0.67 &       2683 &   5.27 &   3.41 &   3.29(-1) &   6.10(-2) &   1.72(-2) &   2.66(-3) &   2.11(-3) &   3.17(-2) &   1.50(-3) &   1.78(-3) \\
   s15p2 &       10.95 &       0.84 &       2771 &   5.24 &   3.93 &   9.97(-1) &   2.20(-1) &   4.41(-2) &   1.09(-2) &   7.77(-3) &   6.33(-2) &   2.79(-3) &   3.68(-3) \\
   s18p5$^\dagger$ &       13.26 &       0.63 &       2181 &   5.73 &   4.65 &   1.89     &   3.46(-1) &   1.42(-1) &   3.97(-3) &   3.09(-3) &   6.26(-2) &   2.68(-3) &   3.51(-3) \\
   s21p5$^\dagger$ &       14.30 &       0.71 &       2228 &   5.38 &   4.94 &   2.49     &   6.16(-1) &   7.98(-2) &   1.61(-2) &   1.09(-2) &   7.22(-2) &   2.97(-3) &   3.61(-3) \\
\hline
Model  & $M_{\rm ej}$ & $E_{\rm kin}$& $V_{\rm m}$ & \iso{1}H  &  \iso{4}He  & \iso{16}O & Ne  &   \iso{24}Mg &  \iso{36}Ar &  \iso{40}Ca &  \iso{56}Ni$_{t=0}$ & \iso{57}Ni$_{t=0}$ & \iso{58}Ni \\
       & [\msun]   &        [foe]    & [\kms]    &     [\msun] & [\msun] & [\msun] & [\msun] & [\msun] & [\msun] & [\msun] & [\msun] & [\msun] & [\msun]  \\
\hline
  he6p0 &       2.82 &       1.10 &       6269 &   $\dots$ &   9.50(-1) &   9.74(-1) &   3.18(-1) &   1.01(-1) &   2.35(-3) &   2.12(-3) &   7.04(-2) &   3.48(-3) &   5.00(-3) \\
  he8p0 &       3.95 &       0.71 &       4251 &   $\dots$ &   8.37(-1) &   1.71     &   5.86(-1) &   1.10(-1) &   2.16(-3) &   2.00(-3) &   5.46(-2) &   2.65(-3) &   3.46(-3) \\
 he12p0 &       5.32 &       0.81 &       3911 &   $\dots$ &   2.34(-1) &   3.03     &   6.74(-1) &   8.73(-2) &   3.91(-3) &   3.42(-3) &   7.90(-2) &   2.69(-3) &   2.47(-3) \\
\hline
\end{tabular}
\end{center}
    {\bf Notes:} The columns list from left to right the ejecta mass, kinetic energy and mean expansion rate, as well as the yields associated with dominant species or important coolants in the infrared. Abundances of unstable isotopes (i.e., \iso{56}Ni and \iso{57}Ni) are given at time $t=$\,0 (other tabulated isotopes are stable and their abundances are constant in time). For Ne, we sum the contribution from \iso{20}Ne and \iso{22}Ne, which are the main Ne isotopes in the ejecta. $^\dagger$: These models were run with the same setup as in \citet{dessart_sn2p_21}, accounted only for the most abundant, stable isotope of each species, with the exception of \nifs\ and its decay products. Numbers in parenthesis correspond to powers of ten. (See Section~\ref{sect_setup} for discussion.)
\end{table*}

\section{Numerical setup}
\label{sect_setup}

  This is a continuation of previous work on nebular-phase radiative-transfer modeling carried out with \cmfgen\ \citep{HD12}. Starting from explosions of solar-metallicity red supergiant stars computed by \citet{sukhbold_ccsn_16}, \citet{dessart_sn2p_21} presented the properties of the Type II SN models but only at a single epoch of 350\,d after explosion. \citet{dessart_late_23} evolved a standard Type II SN model (i.e., 15.2\,\msun\ progenitor reaching core collapse as a red supergiant star and exploding with an explosion energy of about 10$^{51}$\,erg) from 350 to 1000\,d with or without the treatment of shock power from ejecta interaction with circumstellar material. Such models with CSM interaction radiate this additional shock power preferentially in the ultraviolet (or X-rays) at late times and thus with little or negligible power emerging in the infrared (broad, boxy profiles are nonetheless present in the infrared as in the optical, for example, with Pa$\alpha$ resembling the H$\alpha$ profile morphology). In all of these simulations, only the two-step decay chain associated with \nifs\ was accounted for.

  Similarly, starting from explosions of solar-metallicity He-stars computed by \citet{woosley_he_19} and \citet{ertl_ibc_20}, \citet{dessart_snibc_21,D22_lsst,dessart_snibc_23} presented radiative transfer models of Type Ib and Ic SNe covering from about 100 up until about 500\,d but with a focus on optical properties only. \citet{dessart_pm_24} studied the ultraviolet to infrared properties of He-star explosion models until about 10\,yr after explosion but under the influence of power injection from the compact remnant.

    Thus, in this work, we wish to revisit these simulations and shift our focus to the infrared properties to address forthcoming data from the JWST. Our models cover the nebular phase from about 200 out to about 500\,d. We used the same progenitor and explosion models as in these works, and the radiative transfer modeling was performed essentially with the same procedure. Thus, to avoid duplication, we provide here only a succinct summary with information exclusively on the relevant adjustments.

    The nomenclature in this study is the same as in previous works, with model s15p2 corresponding to the explosion of a solar-metallicity red supergiant star that started originally on the zero-age main-sequence (ZAMS) with a mass of 15.2\,\msun. Similarly, model he6p0 corresponds to the explosion of a star that had originally a mass of 6.0\,\msun\ on the helium ZAMS (and a mass of 23.33\,\msun\ on the hydrogen ZAMS; \citealt{woosley_he_19}). To limit the computational burden, we limited our sample to models s10p0, s12p0, s15p2, s18p5, and s21p5 for the H-rich ejecta  and to models he6p0, he8p0, and he12p0 for the H-deficient ejecta.

  In \citet{dessart_sn2p_21} and \citet{dessart_snibc_21}, the ejecta composition was setup slightly differently. For the Type II SN models, we selected the most abundant stable isotope for each species between H and Ni (for Ne, we used both \iso{20}Ne and \iso{22}Ne), together with the undecayed \nifs\ -- most species are usually most abundant in the form of one stable isotope (other forms being underabundant or unstable). We then scaled the mass fractions of all selected isotopes except \nifs\ to ensure a normalization at each depth to unity. This implies that Ar is \iso{36}Ar exclusively and that Ni is either \nifs\ (unstable) or \iso{58}Ni (stable). For the He-star explosion models (which were computed at a later date), we used a more accurate procedure by accounting for all isotopes of all important species between He and Ni (e.g., we accounted for all Ni isotopes between \iso{51}Ni and \iso{73}Ni when considering the Ni abundance) so that no renormalization was needed. In practice, this produces ejecta models with essentially the same yields as with the previous, alternate and more simplistic treatment (e.g., O or Ne), but there are subtle exceptions for some metals.

  For example, Ni is present after explosive nucleosynthesis primarily in three forms, namely \nifs, \iso{57}Ni, and \iso{58}Ni. In the ejecta models that we took from \citet{sukhbold_ccsn_16} and \citet{ertl_ibc_20}, the \iso{58}Ni yield is comparable to that of \iso{57}Ni (here, it goes from being comparable at low mass to being 40\,\% greater at a high progenitor mass -- see Table~\ref{tab_init}), and both combined typically represent 10\,\% of the original \nifs\ mass. However, in \citet{dessart_sn2p_21} and \citet{dessart_snibc_21}, we treated only the radioactive decay of \nifs\ (i.e., \iso{57}Ni was either ignored or considered stable). This implies that in the Type II SN models, the Ni abundance was essentially accurate (i.e., we ignored \iso{57}Ni initially together with its decay), whereas in the He-star explosion models, we typically overestimated the mass of stable Ni by O(10\,\%) (i.e., we accounted for \iso{57}Ni initially but ignored its decay) and correspondingly underestimated the mass of Co (i.e., what should have been \iso{57}Co at those epochs).\footnote{The half-life of \iso{57}Ni is 1.48\,d and is thus found primarily as \iso{57}Co early in the nebular phase, whereas the half-life of \iso{57}Co is 271.7\,d and and thus progressively turns into \iso{57}Fe in the time span 200 to 500\,d covered here. Relative to \nifs, the \iso{57}Ni decay chain contributes negligible power over that time span.} This slight inaccuracy in the abundance of stable Ni is not a major concern and is relevant only for the relative strength of Ni and Co lines in the infrared.

  To resolve this issue, we recomputed models s15p2, he6p0, he8p0, and he12p0 with a full, detailed account of the isotopic composition of the initial inputs of \citet{sukhbold_ccsn_16} and \citet{ertl_ibc_20}, and modeled the radiative transfer in such ejecta out to late times with a more complete set of two-step decay chains whose parent isotopes are \iso{44}Ti, \iso{48}Cr, \iso{52}Fe, \nifs, and \iso{57}Ni. Most of the models discussed in this work use this latter set of models -- we make a special note whenever we use models s10p0, s12p0, s18p5, and s21p5, which were computed with the older treatment (the difference in the results are small and limited to an O(10\,\%) offset in the flux of Ni and Co lines in the infrared). Given the uncertainties surrounding the exact masses of \iso{57}Ni and \iso{58}Ni in stellar explosions (see, for example, the discussion in \citealt{sukhbold_ccsn_16}), this is not a major concern. Tests on how composition variations impact the infrared spectra of core-collapse SNe would help when modeling forthcoming JWST observations -- this is left to future work.

  One important property of these ejecta models is that they were mixed macroscopically by shuffling in mass space the distribution of the main shells in the 1D, unmixed explosion model (for details, see \citealt{dessart_shuffle_20}). The advantage of this technique is that it introduces macroscopic mixing without any microscopic mixing. Composition profiles for such shuffled-shell ejecta are shown for completeness in the appendix for models s15p2 (Fig.~\ref{fig_s15p2_init_comp}) and he6p0 (Fig.~\ref{fig_he6p0_init_comp}). We present a summary of the ejecta properties for our model sample in Table~\ref{tab_init}.

The new calculations presented here were also performed with an up-to-date model atom for all species, and specifically included the updates that were made in \citet{blondin_21aefx_23} for the radiative-transfer modeling of optical and infrared spectra of the Type Ia SN\,2021aefx. As discussed in that work, these updates were important for the modeling of weaker lines in the infrared. Essentially, the same core model atom was used for metals in both H-rich and H-deficient models, including the treatment of Ni\one\ and a large model atom for Co and Ni. Ne\three\ was included in all calculations because of the need to allow for the potential formation of \neiiifs\ -- our simulations revealed that this line is in all cases much weaker than \neiifs. Thus, in all models, we included (in order of increasing atomic weight and then ionization) He\one\ (40,51), He\two\ (13,30), C\one\ (14,26), C\two\ (14,26), N\one\ (44,104), N\two\ (23,41), O\one\ (21,51), O\two\ (54,123), Ne\one\ (78,155), Ne\two\ (22,91), Ne\three\ (32,80), Na\one\ (22,71), Mg\one\ (39,122), Mg\two\ (31,80), Si\one\ (100,187), Si\two\ (31,59), S\one\ (106,322), S\two\ (56,324), Ar\one\ (56,110), Ar\two\ (134,415), K\one\ (25,44), Ca\one\ (76,98), Ca\two\ (21,77), Sc\one\ (26,72), Sc\two\ (38,85), Sc\three\ (25,45), Ti\two\ (37,152), Ti\three\ (33,206), Cr\two\ (28,196), Cr\three\ (30,145), Fe\one\ (413,1142), Fe\two\ (228,2698), Fe\three\ (96,1001), Co\two\ (112,1005), Co\three\ (88,1075),  Ni\one\ (56,301), Ni\two\ (59,1000), Ni\three\ (47,1000), Ba\one\ (17,31), and Ba\two\ (28,61) -- the numbers in parenthesis correspond to the number of super-levels and full-levels (for additional information on super levels, see \citealt{hm98}). In H-rich ejecta models, we included H\one\ (26,36). Because of the higher ionization in He-star explosion models, we added C\three\ (62,112), N\three\ (25,53), O\three\ (44,86), Mg\three\ (31,99), Si\three\ (33,61), S\three\ (48,98), Ar\three\ (32,346), and Fe\four\ (48,272).

\begin{figure}
\centering
\includegraphics[width=0.9\hsize]{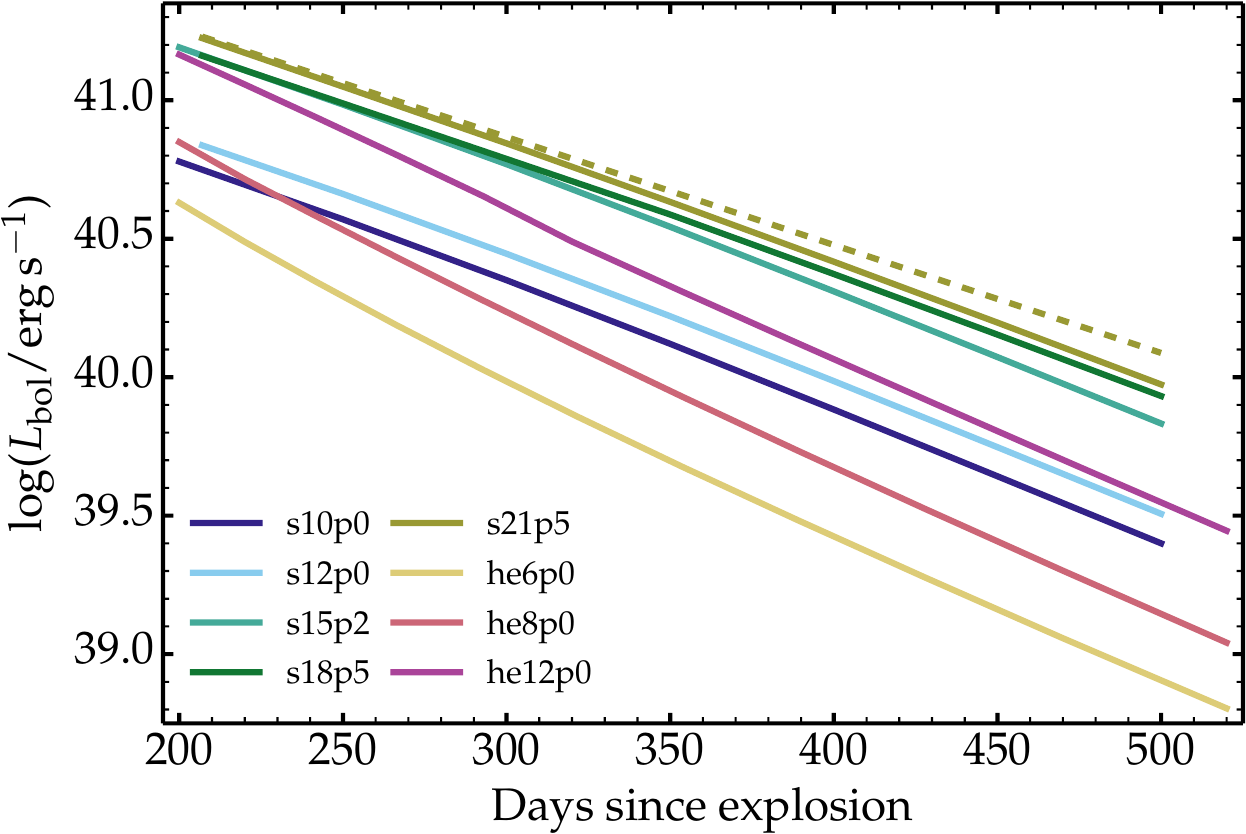}
\caption{Bolometric light curves for our model sample. The dashed line corresponds to the decay-power emitted in model s21p5.
\label{fig_lbol}
}
\end{figure}

Figure~\ref{fig_lbol} shows the \cmfgen\ results for the bolometric light curves of our model sample (i.e., by bolometric we mean all low-energy photons, specifically those falling between 0.1 and 30\mic). Apart from models s10p0 and s12p0, all models have essentially the same initial \nifs\ mass, and thus the differences in brightness and fading rate reflect the greater $\gamma$-ray escape in He-star explosion models owing to their greater expansion rate (see Table.~\ref{tab_init} as well as Fig.~\ref{fig_frac_edep}). These bolometric light curves are thus mostly a reflection of the radiative transfer of $\gamma$-rays because at such late times the ejecta are optically thin to optical and infrared photons. That is, the decay-power absorbed (rather than emitted -- see dashed line in Fig.~\ref{fig_lbol}) equals the bolometric luminosity at any time.

\begin{figure*}
\centering
\includegraphics[width=0.85\hsize]{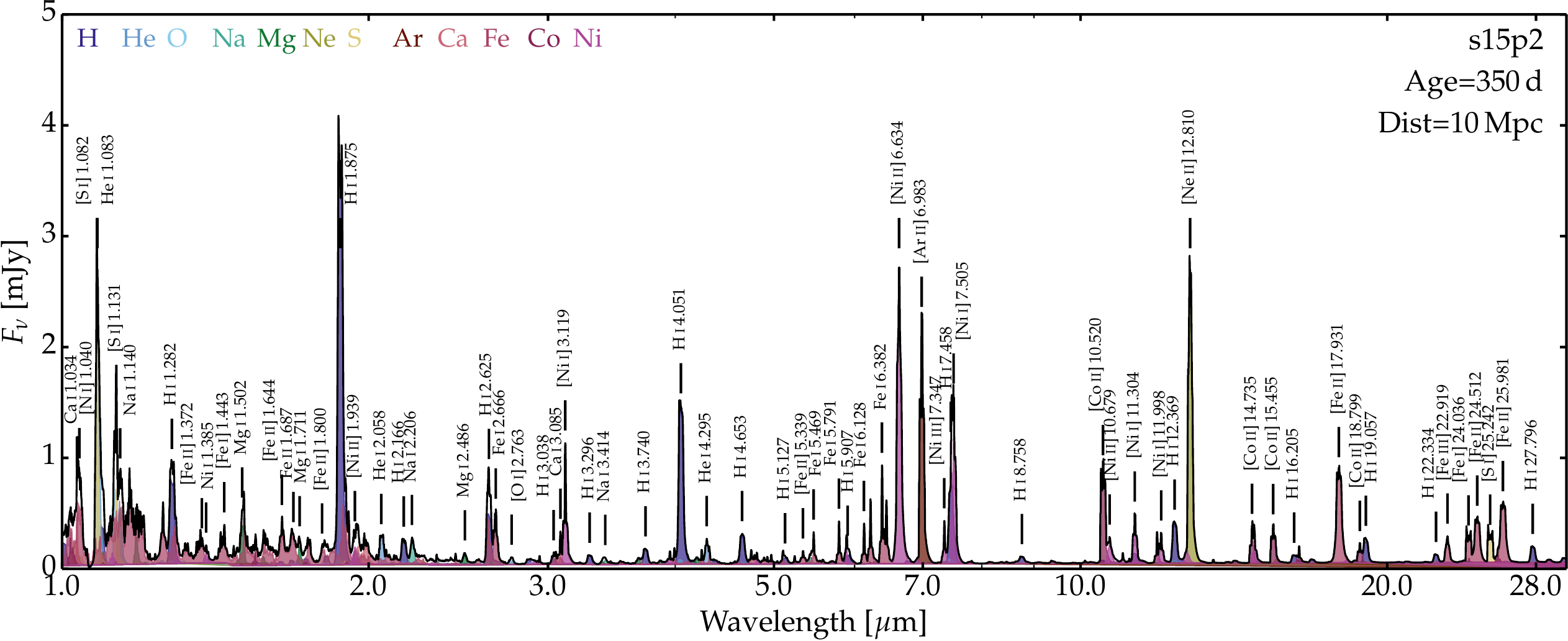}
\caption{Infrared spectrum of Type II SN model s15p2 at 350\,d for an assumed distance of 10\,Mpc. A logarithmic scale is used for the $x$-axis. Labels indicate the main contributor to each emission line (i.e., primarily H, Ne, Ar, and iron-group elements). The color coding indicates the associated species. (See Section~\ref{sect_s15p2} for discussion.)
\label{fig_spec_s15p2_350d}
}
\end{figure*}

\begin{figure*}
\centering
\includegraphics[width=0.4\hsize]{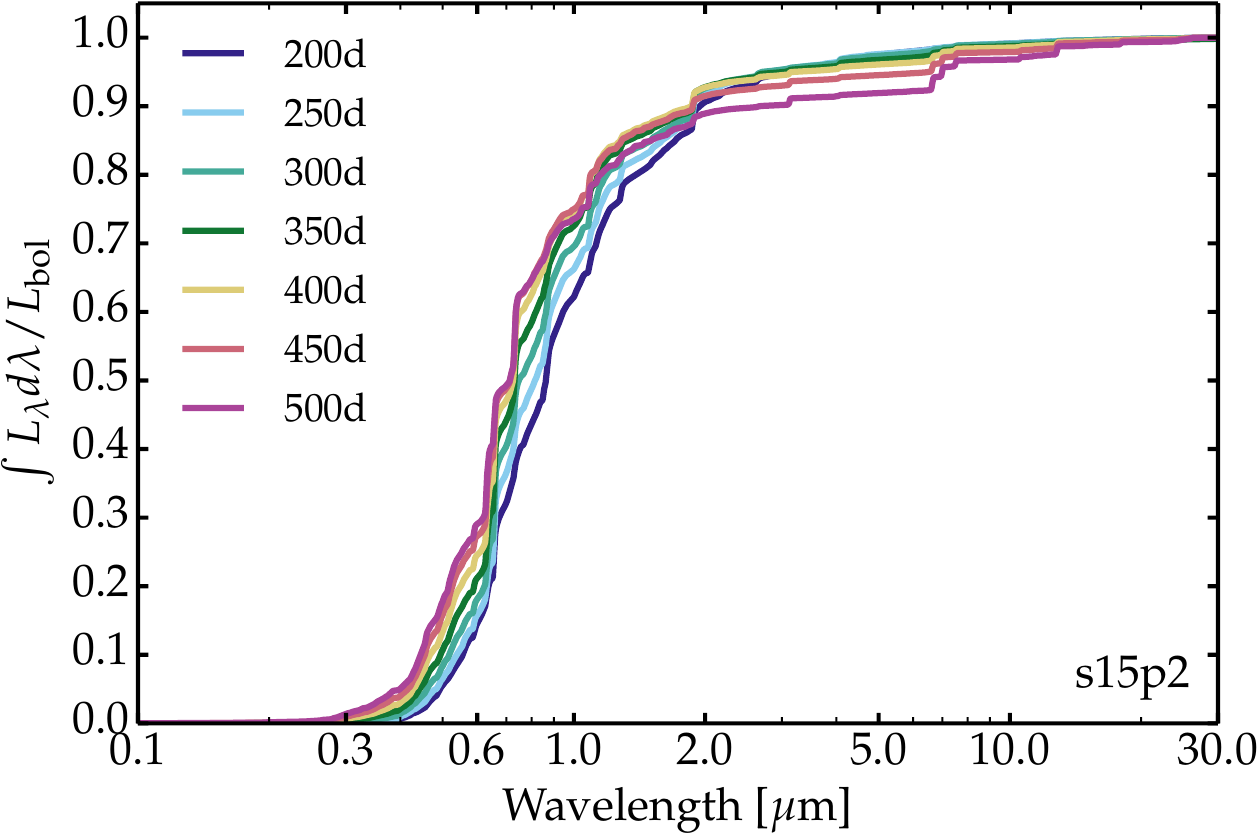}
\includegraphics[width=0.4\hsize]{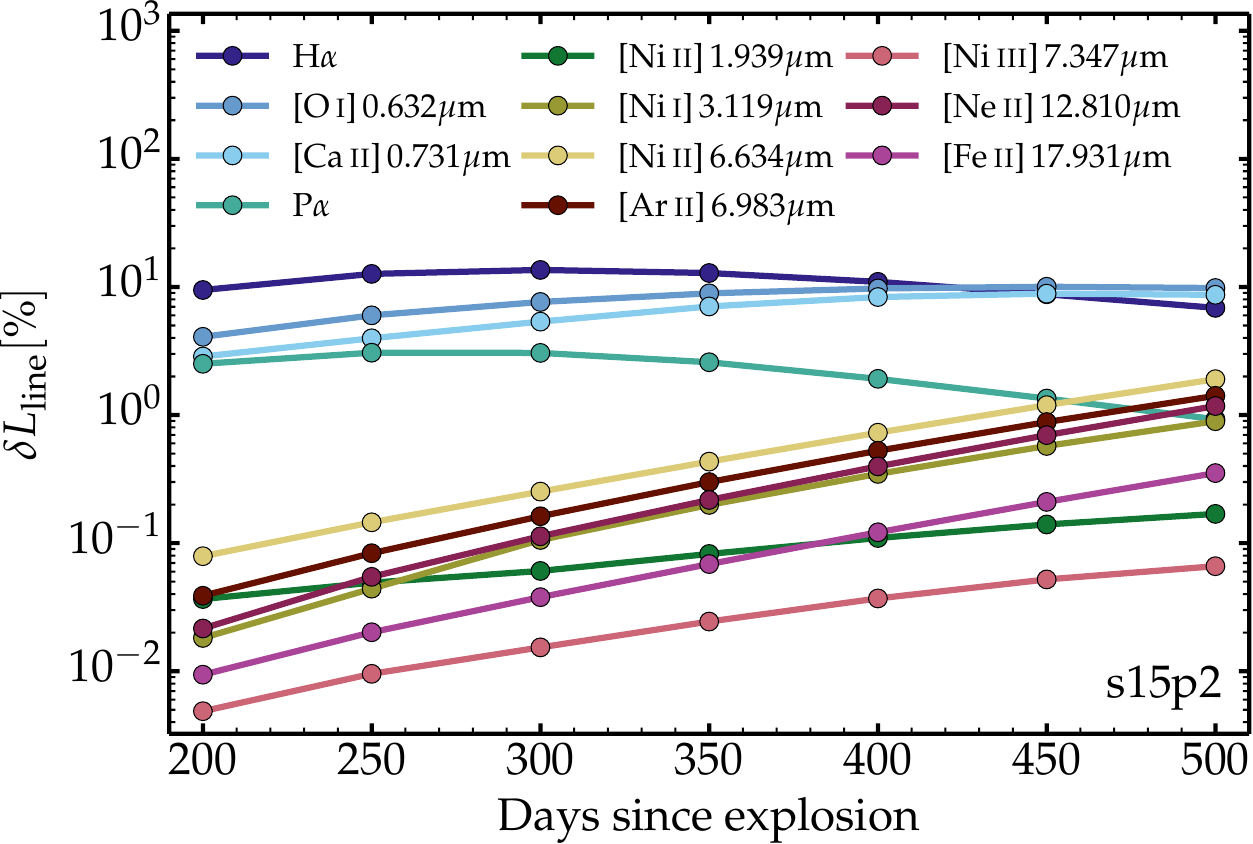}
\caption{Evolution of some spectral properties for model s15p2 over the time span from 200 to 500\,d after explosion. We show the evolution of the cumulative flux integrated from 0.1 to 30.0\mic\ (left) and the evolution of the percentage, fractional luminosity relative to the bolometric luminosity of a few strong lines in the infrared (right). (See Section~\ref{sect_s15p2} for discussion.)
\label{fig_cum_lum_s15p2}
}
\end{figure*}

\section{Results for the Type II SN model s15p2}
\label{sect_s15p2}

In this section, we discuss the results for the Type II SN model s15p2 (see ejecta properties in Table~\ref{tab_init} and composition profiles in Fig.~\ref{fig_s15p2_init_comp}). Figure~\ref{fig_spec_s15p2_350d} shows the s15p2 model spectrum from 1 to 30\mic\ and at an age of 350\,d (the corresponding optical spectrum is shown in \citealt{dessart_sn2p_21}). For a more direct comparison to observations, we show the flux $F_\nu$ versus wavelength for an adopted distance of 10\,Mpc. At this time, the continuum flux is essentially negligible relative to a myriad of emission lines of varying strengths. Labels indicate the main contributor to each emission feature in Fig.~\ref{fig_spec_s15p2_350d} but there are often multiple lines overlapping or the transitions are multiplets (in this case we report the strongest component). The color coding indicates the flux associated with a given species (for example, the color associated with ``Ni'' includes contributions from Ni\one, Ni\two, and Ni\three).

Hydrogen lines are predicted up to and beyond the Humphreys series, with the stronger emission stemming from the Paschen series. We predict Pa$\alpha$
(1.875\mic), Pa$\beta$ (1.282\mic), Br$\alpha$ (4.051\mic), Br$\beta$ (2.625\mic), Br$\gamma$ (2.166\mic), Pf$\alpha$ (7.458\mic), Pf$\beta$
(4.653\mic), Pf$\gamma$ (3.740\mic), Hu$\alpha$ (12.369\mic), Hu$\beta$ (7.500\mic), Hu$\gamma$ (5.907\mic), and transitions from the seventh series
at 19.057\mic\ or the isolated 8.758\mic. A few He\one\ lines are present, with the strongest one at 1.083\mic\ and weaker ones at 2.058\mic\ and 4.295\mic. The model predicts no sizable emission from C in the infrared, a weak emission from N with [N\one]\,1.04\mic, and a few weak lines of O with O\one\ at 1.129\mic\ and 2.763\mic\ (this is in stark contrast with the strong \oidoub\ in the optical range; \citealt{dessart_sn2p_21}) -- these lines overlap with extended emission from Fe. A few weak Na\one\ lines are present at 1.138 and 1.140\mic, 2.206 and 2.208\mic, as well as 5.427 and 5.434\mic. Most of the preceding transitions were permitted. With Ne, there is one strong forbidden-line emission at 12.810\mic, whereas \neiiifs\ has a negligible strength (about a hundredth that of \neiifs). There are a number of Mg\one\ lines but all weak (similar in strength to those of Na\one), with transitions at 1.183\mic, 1.502\mic, 1.711\mic, 2.486\mic\ (many are multiplets). Silicon is mostly Si\one\ and contributes negligible emission. In contrast, sulfur contributes one strong emission at 1.082\mic\ (which overlaps with He\one\,1.083\mic) and another at 25.242\mic\ (both are forbidden transitions). Argon contributes one strong forbidden line with [Ar\two]\,6.983\mic.

All other emission lines arise from Fe, Co, and Ni. There is a forest of Fe lines, mostly from Fe\two\ up to two microns, beyond which the model predicts a few isolated, stronger, forbidden lines for Fe\one\ at 24.036\mic, Fe\two\ at 17.931\mic, 24.512\mic, and 25.981\mic, as well as Fe\three\ at 22.919\mic. There are a number of Co lines, with Co\two\ at 10.520\mic, 14.735\mic, 15.455\mic, and 18.799\mic. Co\three\ lines are essentially absent because Co is mostly Co$^+$ in our model. Finally, there are a few strong lines of Ni, which arise at 350\,d from the sizable abundance of stable Ni isotopes. The strongest Ni line is due to Ni\two\ at 6.634\mic, with weaker transitions (often overlapping with other lines) at 1.939\mic, 10.679\mic, and 12.725\mic. There is also a Ni\three\ transition at 7.347\mic\ although weaker than [Ni\two]\,6.634\mic, as well as numerous Ni\one\ transitions at 3.119\mic, 7.505\mic, 11.304\mic\ and 11.998\mic.

The use of $F_\nu$ in Fig.~\ref{fig_spec_s15p2_350d} is convenient to show the weak transitions present in the infrared but it can be misleading when evaluating the relative strength of these transitions when widely separated or with respect to optical lines such as \oidoub, H$\alpha$, or \caiidoub. So, to recover a proper perspective on how the flux is distributed across the electromagnetic spectrum, the left panel of Fig.~\ref{fig_cum_lum_s15p2} illustrates the normalized cumulative flux distribution, integrated from 0.1\mic\ in the ultraviolet to 30\mic\ in the infrared, and shown at multiple epochs spanning 200\,d to 500\,d. What is first apparent is that the general shape of that distribution evolves little in time, with essentially zero flux falling in the ultraviolet, a fraction rising from 65\,\% to 75\,\% in the optical (below 1\mic), and the rest (35\,\% to 25\,\%) falling in the infrared. The progression of that distribution with wavelength is smooth with a few jumps becoming accentuated as time passes. These jumps correspond to the strongest lines in the spectrum and we can identify \oidoub, H$\alpha$, and \caiidoub, which combined represent over 30\,\% of the total flux emitted at 500\,d. In the infrared, the main jumps are due to Pa$\alpha$, [Ni\one]\,3.119\mic, [Ni\two]\,6.634\mic, [Ar\two]\,6.983\mic, the 7.5\mic\ blend of H\one, Ni\one\ and Ni\three\ lines, and finally \neiifs.

The right panel of Fig.~\ref{fig_cum_lum_s15p2} shows the fractional flux emitted in a number of strong lines in model s15p2 at epochs between 200 and 500\,d. The transitions located at shorter wavelength carry most of the total decay power absorbed in the ejecta (i.e., \oidoub, H$\alpha$, and \caiidoub), whereas the strongest transitions in the infrared carry individually at most 1\% (that maximum is reached at 500\,d) of the total flux (i.e., [Ni\two]\,6.634\mic, [Ar\two]\,6.983\mic, and \neiifs). So, in terms of power, the infrared plays a secondary role in model s15p2, although this would be slightly modified in the presence of molecules (see, e.g., \citealt{liljegren_ibc_mol_23} and Section~\ref{sect_conc}). Such diagnostic lines of intermediate-mass and Fe-group elements have, however, high scientific content for characterizing the progenitor and explosion properties of Type II SNe. These properties reflect how the decay power emitted is absorbed by the ejecta, in what proportion it is absorbed by the different ejecta regions (i.e., the shells of distinct composition in the preSN model are now the shuffled shells in our ejecta models), and finally what coolants dominate in each of the relevant shells (e.g., the O/Ne/Mg shell versus the H/He shell -- see Fig.~\ref{fig_frac_edep}). In model s15p2, the fractional decay power absorbed by the ejecta drops from 99.5\,\% at 200\,d to 66.3\,\% at 500\,d. Over the same time span, the fraction of that power absorbed in the H/He shell rises from 38.1\,\% to 55.4\,\%, stays roughly constant at $\sim$\,11\,\% in the He-rich shell, drops slightly in the O-rich shell from 21.6\,\% to 18.0\,\% and drops from 29.8\,\% to 15.5\,\% in the Fe/Si-rich shells. These trends reflect the increasing escape of $\gamma$ rays from the Fe/Si-rich shell which goes mostly to the benefit of the massive H-rich material enshrouding, or macroscopically mixed within the inner, metal-rich ejecta.

\begin{figure}
\centering
\includegraphics[width=0.9\hsize]{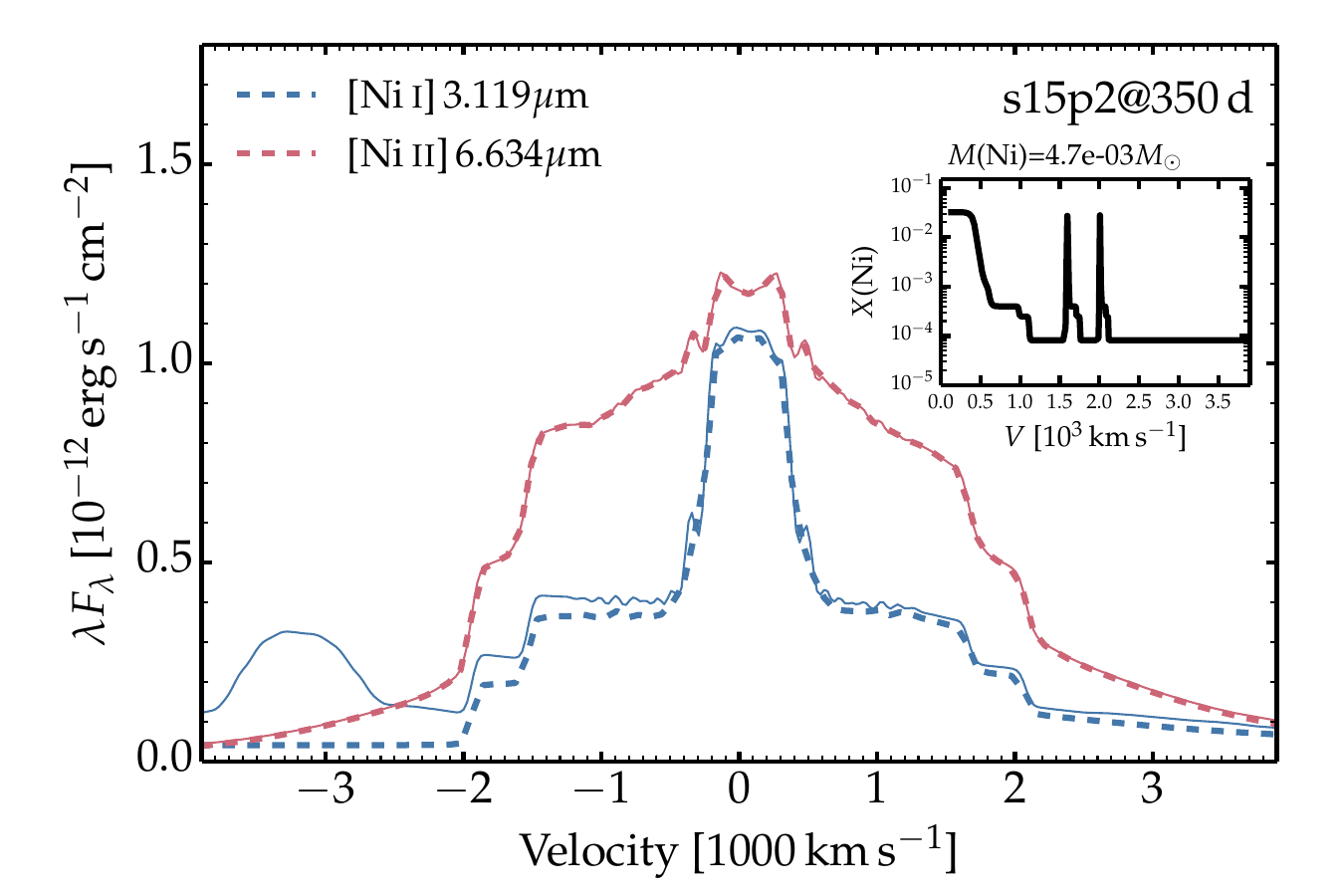}
\caption{Infrared Ni lines in the models s15p2 at 350\,d. We show the quantity $\lambda F_\lambda$ using the total flux (solid) and that due to Ni only (dashed) centered at the rest wavelength of two strong lines of [Ni\one] and [Ni\two] (see label). A distance of 10\,Mpc is assumed. The inset shows the Ni mass fraction versus velocity at 350\,d, which thus accounts for all stable Ni isotopes (those that after explosion were neither \iso{56}Ni nor \iso{57}Ni -- these arise here at an 80\,\%\ level from \iso{58}Ni). (See Section~\ref{sect_s15p2} for discussion.)
\label{fig_nick_s15p2}
}
\end{figure}

At 200\,d, the main coolants and processes in model s15p2 are nonthermal excitation of H\one\ (20.3\,\% of the total cooling), He\one\ (9.0\,\%), and O\one\ (3.7\,\%), and collisional excitation of Ca\two\ (17.3\,\%), Fe\two\ (15.7\,\%), Mg\two\ (10.3\,\%), O\one\ (5.6\,\%), Co\two\ (3.1\,\%), C\two\ (2.6\,\%), and Si\two\ (2.0\,\%). Other coolants are typically below 1\,\%. At 350\,d, the main coolants and processes in model s15p2 are nonthermal excitation of H\one\ (28.2\,\%), He\one\ (10.5\,\%), and O\one\ (3.5\,\%), and collisional excitation of Mg\two\ (12.7\,\%), Ca\two\ (10.7\,\%), Fe\two\ (10.6\,\%), O\one\ (9.3\,\%), and C\two\ (2.0\,\%). Collisional excitation of metals leading to infrared lines is growing with S\one\ at 1.7\,\%, Ni\two\ at 1.4\,\%, Fe\one\ at 1.7\,\%, Fe\three\  at 0.8\,\%, Co\two\ at 0.9\,\%, but Ar\two\ is still low with only 0.2\,\%. At 500\,d, the main coolants and processes in model s15p2 are nonthermal excitation of H\one\ (34.4\,\%), He\one\ (11.6\,\%), and O\one\ (3.8\,\%), and collisional excitation of Fe\two\ (11.8\,\%), O\one\ (9.7\,\%), Ca\two\ (8.4\,\%), Ni\two\ (3.7\,\%), Mg\two\ (3.1\,\%), S\one\ (2.6\,\%), Ni\one\ (1.5\,\%), Ar\two\ (1.2\,\%), and Ne\two\ (1\,\%). These values reflect the fractional line fluxes shown in the right panel of Fig.~\ref{fig_cum_lum_s15p2}.

Before wrapping up this section, we show in Fig.~\ref{fig_nick_s15p2} the morphology in velocity space of two Ni\one\ and Ni\two\ lines predicted in the model s15p2 at 350\,d after explosion. To give a more direct connection to the actual flux in each of these lines, we show the quantity $\lambda F_\lambda$. For each transition, we show the total flux in each spectral region and draw as a dashed line the flux associated with Ni only. In this figure, we omit \nkiiopt\ because it is strongly contaminated by \caiidoub\ and numerous Fe lines in our ejecta models. Similarly, line overlap affects [Ni\two]\,1.939\mic\ and [Ni\one]\,7.505\mic, whereas [Ni\one]\,3.119\mic\ and [Ni\two]\,6.634\mic\ are strong and the main contributor to the associated feature. The Ni composition profile shown in the inset confirms that these Ni lines primarily arise from the regions that underwent explosive nucleosynthesis and rich initially in \nifs\ (and in \iso{58}Ni) rather than from the H-rich, He-rich, or O-rich material with a primordial abundance of \iso{58}Ni. There is a small contribution to the Ni\two\ emission from the H-rich material, which is visible in the extended wings of [Ni\two]\,6.634\mic\ at $\gtrsim$\,2000\,\kms\ from line center (Fe\two\ also contributes to cooling the H-rich regions with a primordial composition). The velocity location where the flux jumps in the profile wings corresponds to the ejecta velocity at which the Ni-rich shells reside -- these jumps are inherent to the shuffled-shell method \citep{dessart_shuffle_20}. A less ordered, or even a randomized spatial distribution of the emitting material from the Fe/Si-rich shell would produce a smoother emission profile (i.e., from the spread in projected velocity of that emission). It would, however, have little impact on the actual flux in the lines because $\gamma$-rays from \nifs\ decay permeate throughout the inner ejecta (provided the explosion is not too asymmetric).

Hence, the profile morphology of Ni lines represents an excellent tracer of the Fe-group elements produced during the explosion and may serve to constrain its distribution in radial or velocity space. Cobalt lines may also be used for that purpose (i.e., as long as the \cofs\ abundance is sizable; by 300\,d, Fe is ten times more abundant in the Fe/Si-rich material than Co, and the contrast grows with time), whereas Fe lines may suffer greater contamination from the emission by the abundant, primordial Fe present in H-rich and He-rich material (wherein Fe\two\ acts as an efficient coolant).

\begin{figure*}
\centering
\includegraphics[width=0.85\hsize]{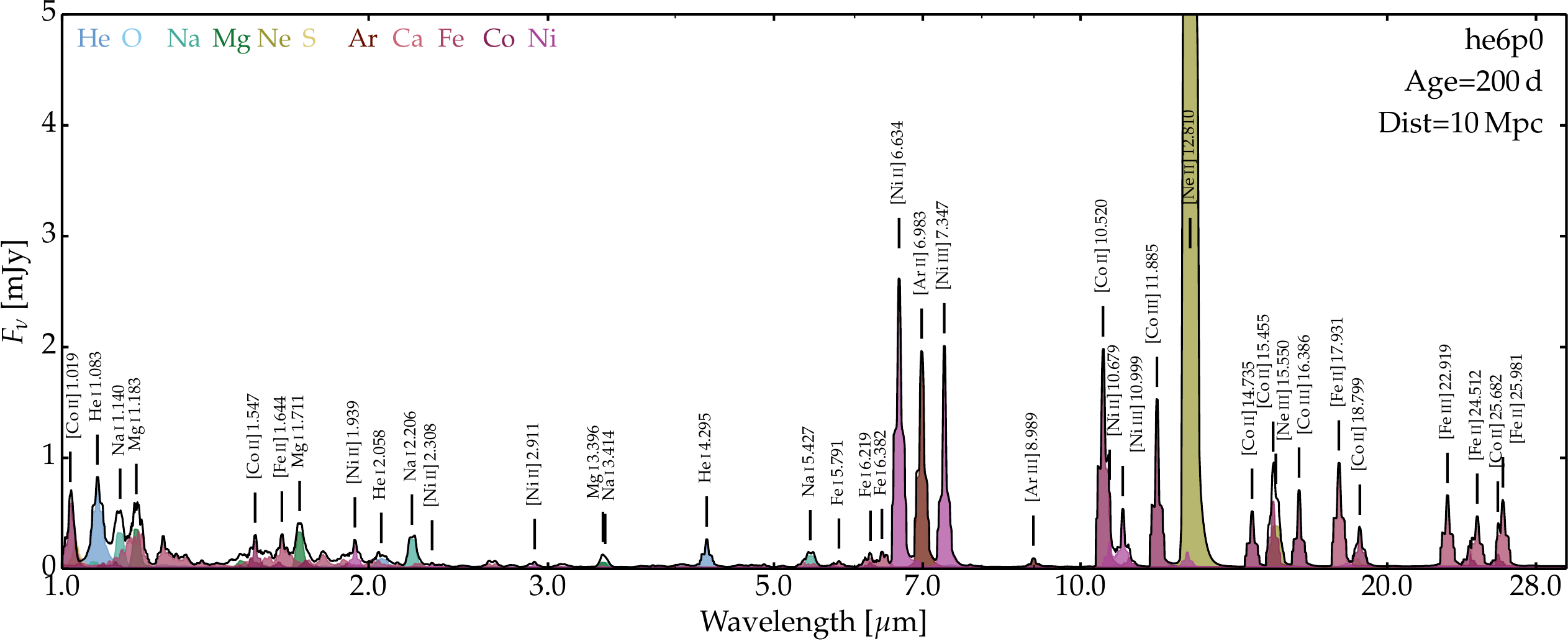}
\caption{Same as Fig.~\ref{fig_spec_s15p2_350d} bur for the Type Ib SN model he6p0 at 200\,d and a distance of 10\,Mpc.
\label{fig_spec_he6_200d}
}
\end{figure*}

\begin{figure*}
\centering
\includegraphics[width=0.4\hsize]{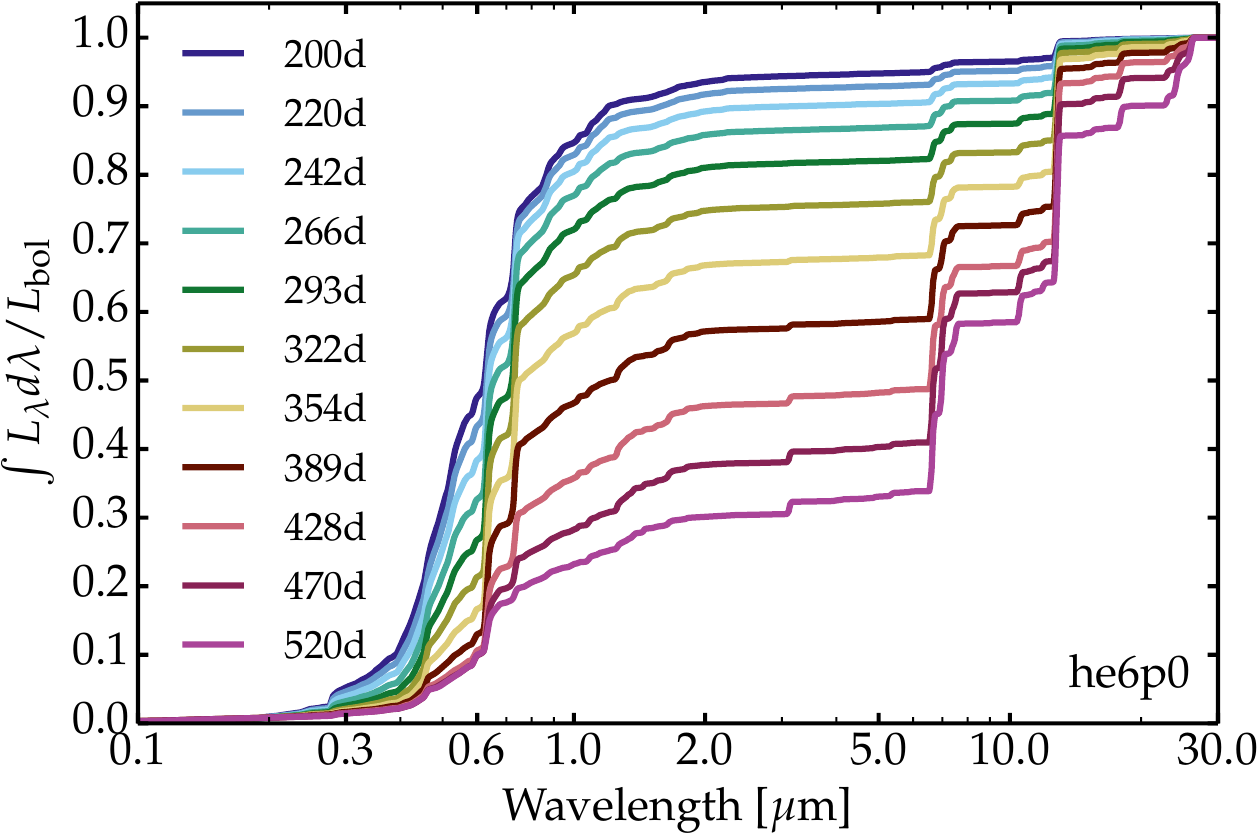}
\includegraphics[width=0.4\hsize]{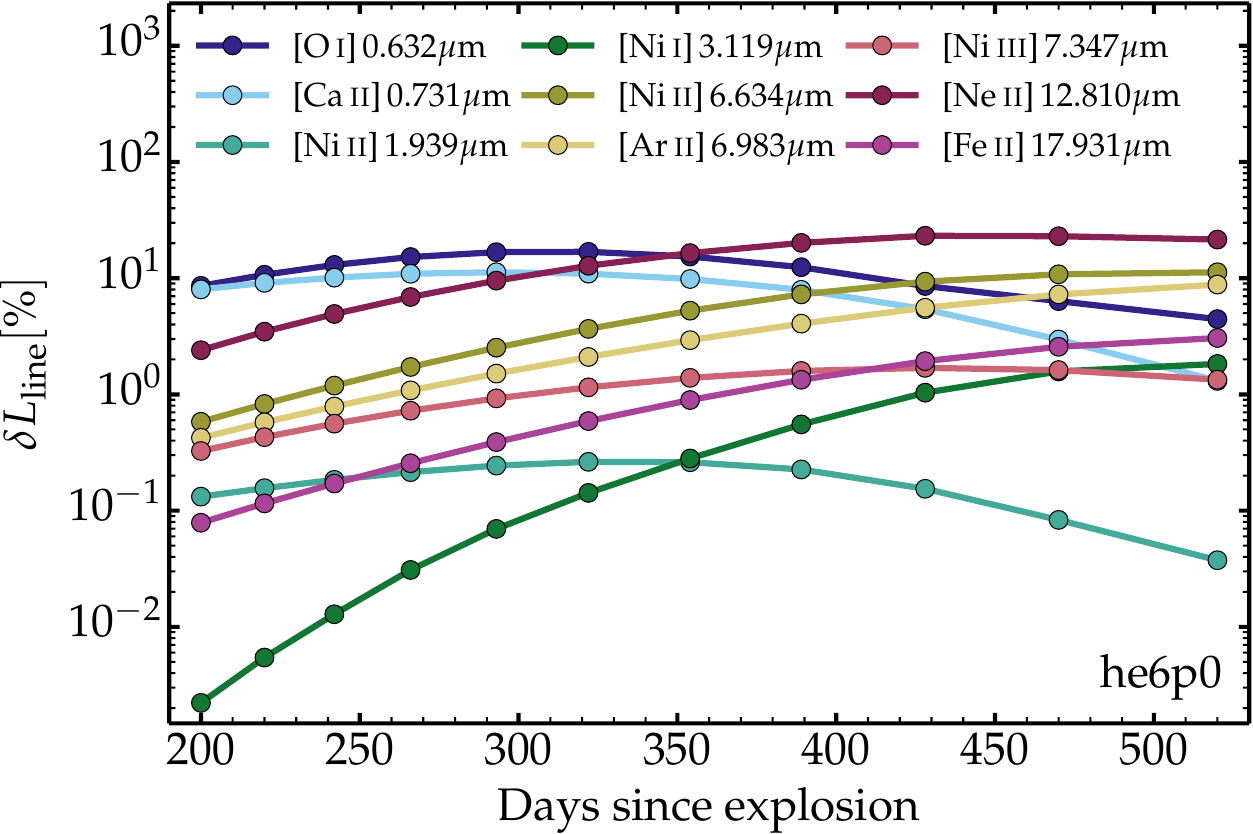}
\caption{Same as Fig.~\ref{fig_cum_lum_s15p2} but for model he6p0 over the time span from 200 to 520\,d. (See Section~\ref{sect_he6p0} for discussion.)
\label{fig_cum_lum_he6p0}
}
\end{figure*}

\section{Results for Type Ib SN model he6p0}
\label{sect_he6p0}

The results obtained for the He-star explosion model he6p0 are analogous to those discussed above for model s15p2 (some he6p0 model properties are summarized in Table~\ref{tab_init} and illustrated in Fig.~\ref{fig_he6p0_init_comp}). Obvious differences arise from the lack of H\one\ lines from this H-deficient ejecta. Another difference arises from the larger ejecta expansion rate ($V_{\rm m}$ is 6269\,\kms\ compared to 2771\,\kms\ in model s15p2, thus about 2.3 times greater) -- this arises primarily from the much smaller ejecta mass of 2.82\,\msun\ compared to 10.95\,\msun, whereas the kinetic energy is comparable (1.1 compared to 0.84\,$\times$\,10$^{51}$\,erg). This leads to a weaker continuum flux, stronger and broader emission lines, and a higher ionization of the gas. This higher ionization translates into weaker lines of neutral species (i.e., atoms), and stronger ones from once- and twice-ionized species.

Figure~\ref{fig_spec_he6_200d} shows the 1--30\mic\ infrared spectrum of model he6p0 at 200\,d after explosion and for an assumed distance of 10\,Mpc. Compared to model s15p2, the model predicts (see also labels in figure) stronger and broader He\one\ lines at 1.083\mic, 2.058\mic,  and 4.295\mic. No sizable line from C, N, and O is predicted in the infrared, though \oidoub\ and a weaker \niidoub\ are present in the optical (see \citealt{dessart_snibc_21,dessart_snibc_23}). As in model s15p2 at 350\,d, a few weak Na\one\ lines are present at 1.138 and 1.140\mic, 2.206 and 2.208\mic, as well as 5.427 and 5.434\mic. Weak Mg\one\ lines are also present at 1.183\mic, 1.502\mic, 2.486\mic, and a stronger and more isolated line is predicted at 1.711\mic. The strongest infrared line (as seen in $F_\nu$) is \neiifs, with a much weaker \neiiifs. Because of the higher ionization overall in this model, there are no Si\one\ nor S\one\ lines predicted. Argon contributes one strong emission with [Ar\two]\,6.983\mic\ and a weaker one with [Ar\three]\,8.989\mic. All other emission lines arise from Fe, Co, and Ni. As in model s15p2, there is a forest of overlapping Fe\two\ lines up to two microns but much weaker, likely because of the lower ejecta density (in spite of the earlier date). There is, however, a myriad of forbidden transitions, primarily with [Fe\two] at 17.931\mic, 24.512\mic, and 25.981\mic, as well as [Fe\three] at 22.919\mic. Cobalt lines are predicted with [Co\two] at 1.019\mic, 10.520\mic, 14.735\mic, 15.455\mic, 18.799\mic, and 25.682\mic. There are also lines of [Co\three] at 11.885\mic\ and 16.386\mic. There are a few strong lines of Ni, but with a higher ionization than in model s15p2 (i.e., Ni\one\ lines have a negligible strength). There is the strong [Ni\two]\,6.634\mic, with weaker transitions of [Ni\two] at 1.939\mic, 2.308\mic, 2.911\mic, 10.679\mic, 12.725\mic. Higher in ionization is the strong [Ni\three]\,7.347\mic, with a weaker component at 10.999\mic.

Figure~\ref{fig_cum_lum_he6p0} is a counterpart of Fig.~\ref{fig_cum_lum_s15p2} but for the model he6p0. The left panel illustrates the normalized cumulative flux distribution integrated from the ultraviolet to the infrared and shown over a time span from 200\,d to 520\,d. Compared to model s15p2, the distribution is analogous at 200\,d, with about 80\,\% of the flux falling in the optical range, but it strongly deviates as time advances. Indeed, the flux progressively shifts to longer wavelength, so that 80\,\% of the flux falls within the infrared at 520\,d -- the evolution was instead negligible in model s15p2. The vertical jumps in the distribution correspond to lines also present in the s15p2 model but that now have a much greater strength. A major jump occurs across \neiifs, and to a lesser extent across \nkiimir, \ariimir, and \nkiiimir.

The right panel of Figure~\ref{fig_cum_lum_he6p0} shows the evolution of the fractional luminosity relative to the bolometric luminosity of specific lines spanning the optical and the infrared and covering various species, atoms, and ions (see Fig.~\ref{fig_cum_lum_s15p2} for the equivalent for model s15p2). As time progresses, the strength of most of these lines, which are forbidden, increases, probably as a result of the decreasing density. But at about 350\,d, a shift occurs with a weakening of \oidoub\ and \caiidoub, and a continuous strengthening of coolants located in the infrared. By 520\,d, \nkiimir\ and \neiifs\ are responsible for respectively 10 and 20\,\% of the total ejecta cooling.

This contrast in spectral properties with model s15p2 arises in part from the different distribution of the decay power absorbed within the ejecta. In model he6p0, the fractional decay power absorbed drops from 25.6\,\% at 200\,d to only 6.9\,\% at 520\,d, time by which the fraction arising from the local absorption of positrons represents 44.5\,\%. This implies that 44.5\,\% of the decay power at 520\,d benefits only the regions originally rich in \nifs, that is the regions where explosive burning took place. The decay power absorbed in the Fe/Si-rich regions indeed grows from 25.1\,\% at 200\,d to 48.7\,\% at 520\,d. Over the same time span, the decay power absorbed in the He-rich material drops from 16.2\,\% to 12.2\,\%, and in the O-rich shell from 58.8\,\% to 39.0\,\%.

\begin{figure}
\centering
\includegraphics[width=0.9\hsize]{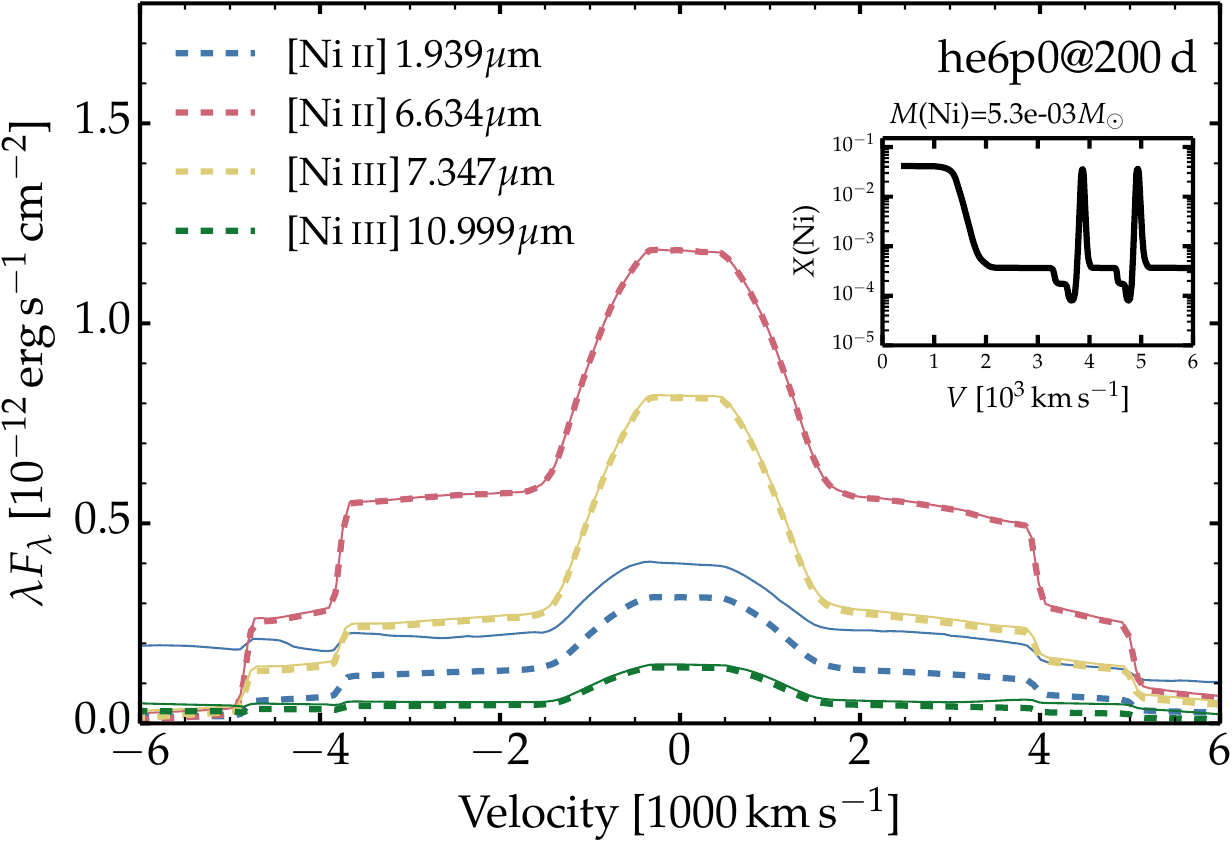}
\caption{Same as Fig.~\ref{fig_nick_s15p2} but for model he6p0 at 200\,d and for [Ni\two] and [Ni\three] lines ([Ni\one] lines have a negligible strength in this model at that time). (See Section~\ref{sect_he6p0} for discussion.)
\label{fig_nick_he6p0}
}
\end{figure}

At 200\,d, the main coolants and processes in model he6p0 are collisional excitation of Mg\two\ (32.0\,\%), Ca\two\ (9.7\,\%), C\two\ (9.2\,\%), O\one\ (8.4\,\%), Co\two\ (5.6\,\%), Fe\two\ (5.4\,\%), Fe\three\ (2.8\,\%), Ne\two\ (2.2\,\%), and Ni\two\ (2.1\,\%), and nonthermal excitation of O\two\ (5.1\,\%) and He\one\ (4.9\,\%). Other coolants are typically at or below 1\,\%. At 350\,d, the main coolants and processes in model he6p0 are collisional excitation of O\one\ (15.3\,\%), Ne\two\ (15.2\,\%), Fe\two\ (12.2\,\%), Ca\two\ (9.7\,\%), Ni\two\ (7.9\,\%), Co\two\ (4.4\,\%), Ar\two\ (2.9\,\%), Mg\two\ (2.2\,\%), N\two\ (1.9\,\%), and  C\two\ (1.7\,\%) and nonthermal excitation of He\one\ (6.8\,\%), O\one\ (3.8\,\%), and O\two\ (3.2\,\%). At 520\,d, the main coolants and processes in model he6p0 are collisional excitation of Ne\two\ (19.6\,\%), Fe\two\ (11.7\,\%), Ni\two\ (10.9\,\%), Ar\two\ (8.3\,\%), O\one\ (5.8\,\%), Ni\one\ (4.9\,\%), Co\two\ (4.8\,\%), Fe\one\ (3.5,\%), Si\two\ (1.9\,\%), Fe\three\ (1.8\,\%), and Ni\three\ (1.4\,\%), and nonthermal excitation of He\one\ (9.1\,\%) and O\one\ (5.5\,\%).

Figure~\ref{fig_nick_he6p0} is a counterpart of Fig.~\ref{fig_nick_s15p2} and illustrates the emission profiles in velocity space of a few [Ni\two] and [Ni\three] lines in the infrared for model he6p0 at 200\,d. Because of the faster ejecta expansion rate, these lines are much broader than in model s15p2, with a maximum extent reaching out to about 5000\,\kms, which represents the velocity of the outermost shell originally rich in \nifs. Because of the large $\gamma$ ray escape, little Ni emission is predicted outside of these Ni-rich regions, making such line diagnostics excellent tracers of the ashes from explosive burning in the ejecta. The width and strength (i.e., peak flux) of these lines stays essentially constant from 200 to 520\,d.

\begin{figure}
\centering
\includegraphics[width=0.85\hsize]{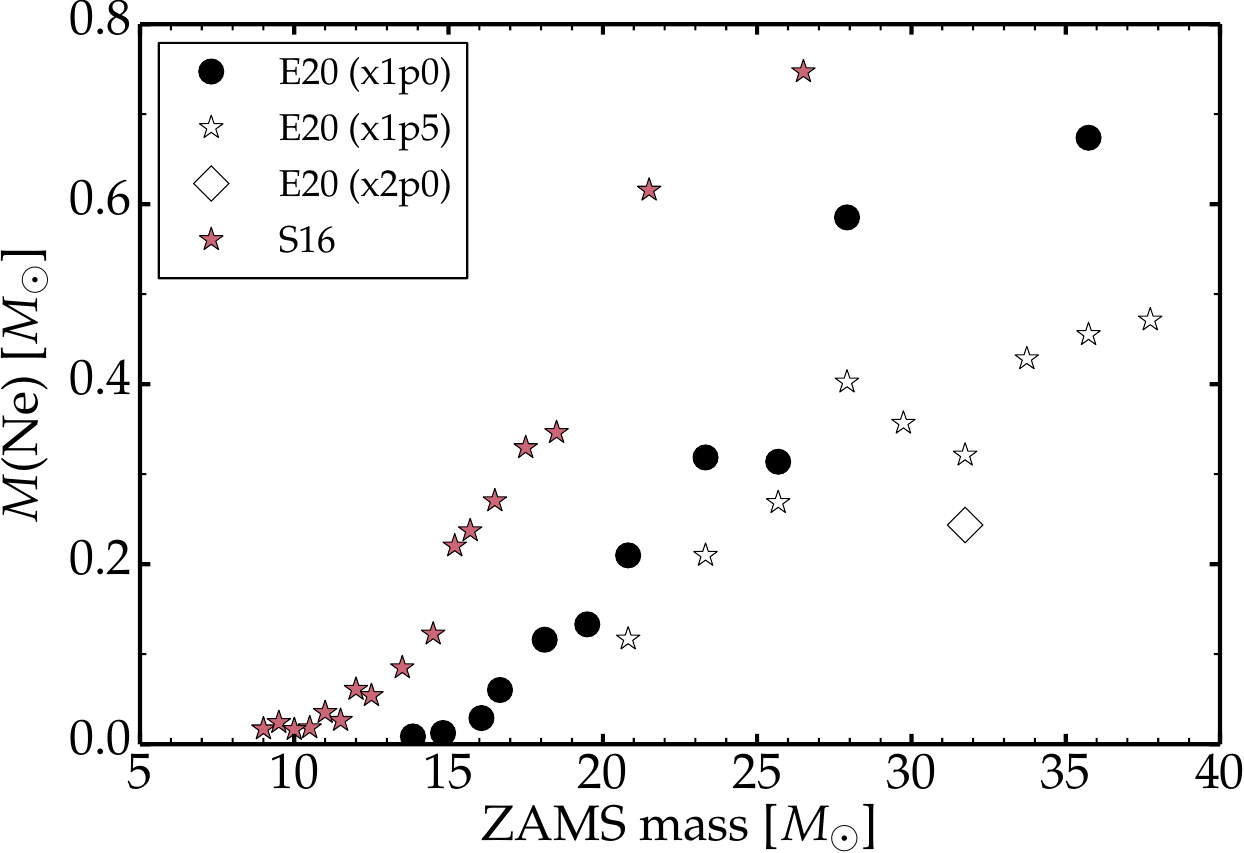}
\includegraphics[width=0.85\hsize]{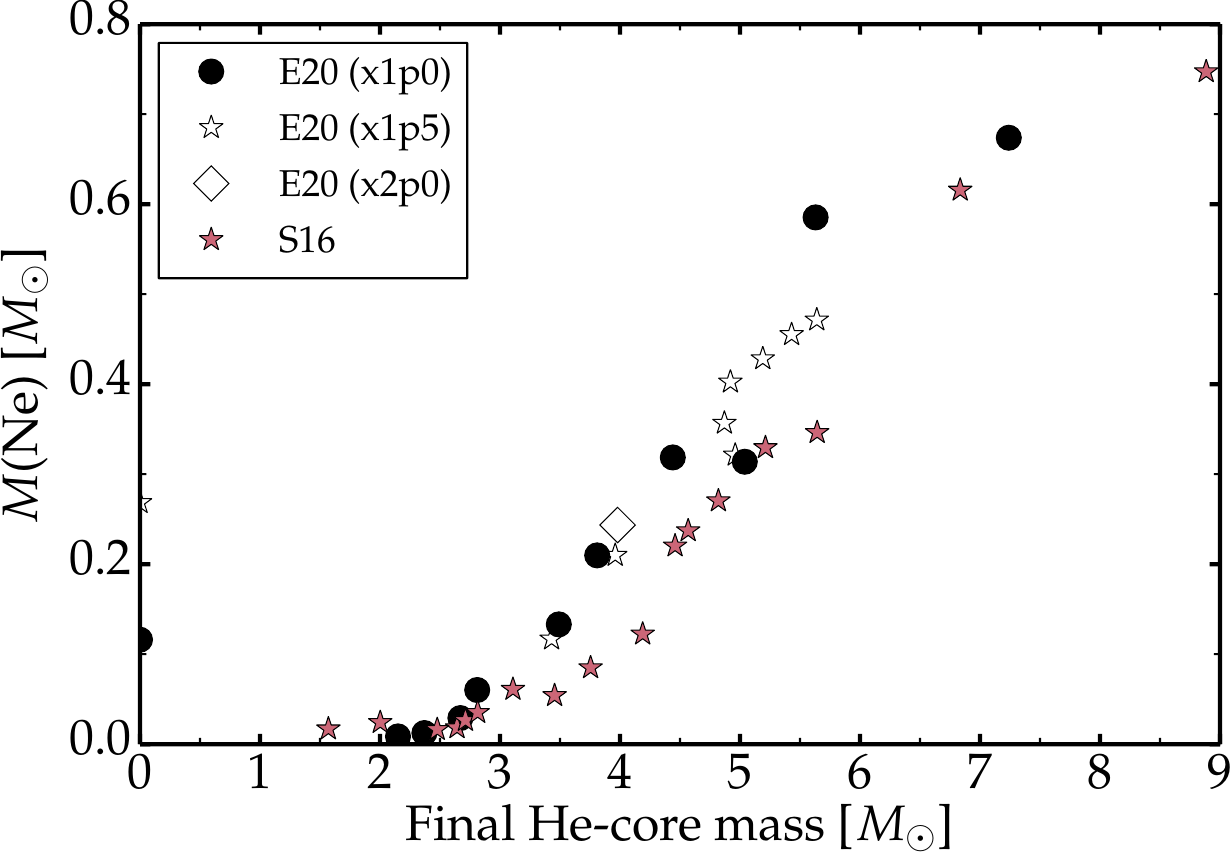}
\caption{Top: Total Ne yield versus ZAMS mass for the ejecta arising from He-star progenitor models (black symbols; sample from \citealt{dessart_snibc_21}, originally from \citealt{ertl_ibc_20}; x1p0, x1p5, and x2p0 correspond to different scalings for the adopted wind mass loss rates; see \citealt{woosley_he_19} for details) and for red supergiant star models (filled red stars; sample from \citealt{dessart_sn2p_21}, originally from \citealt{sukhbold_ccsn_16}). Bottom: Same as top, but now shown versus the preSN mass for He-star models and versus the final He-core mass for the red supergiant star models. (See Section~\ref{sect_neon} for discussion.)
\label{fig_neon_ej}
}
\end{figure}

\begin{figure}
\centering
\includegraphics[width=0.9\hsize]{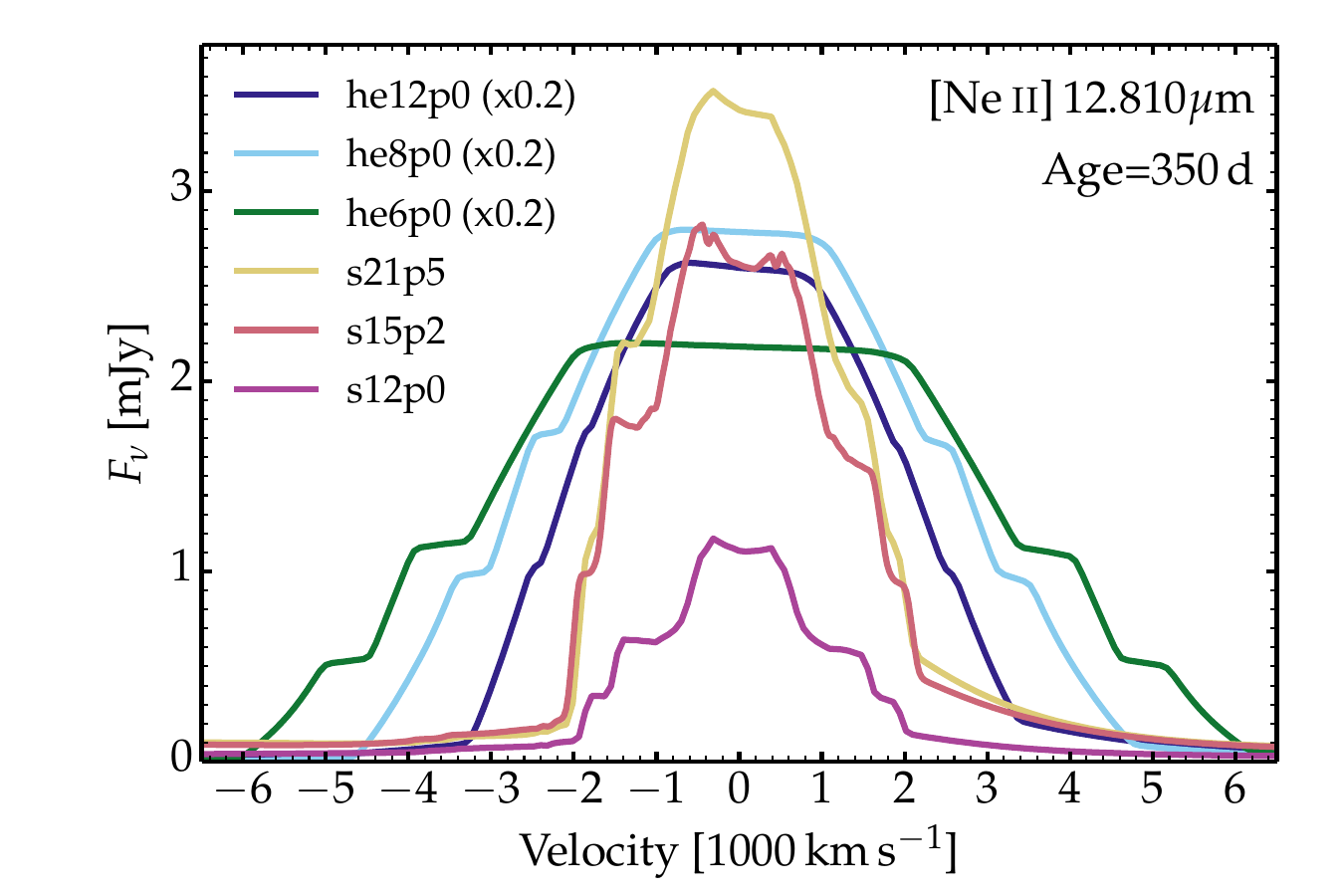}
\includegraphics[width=0.9\hsize]{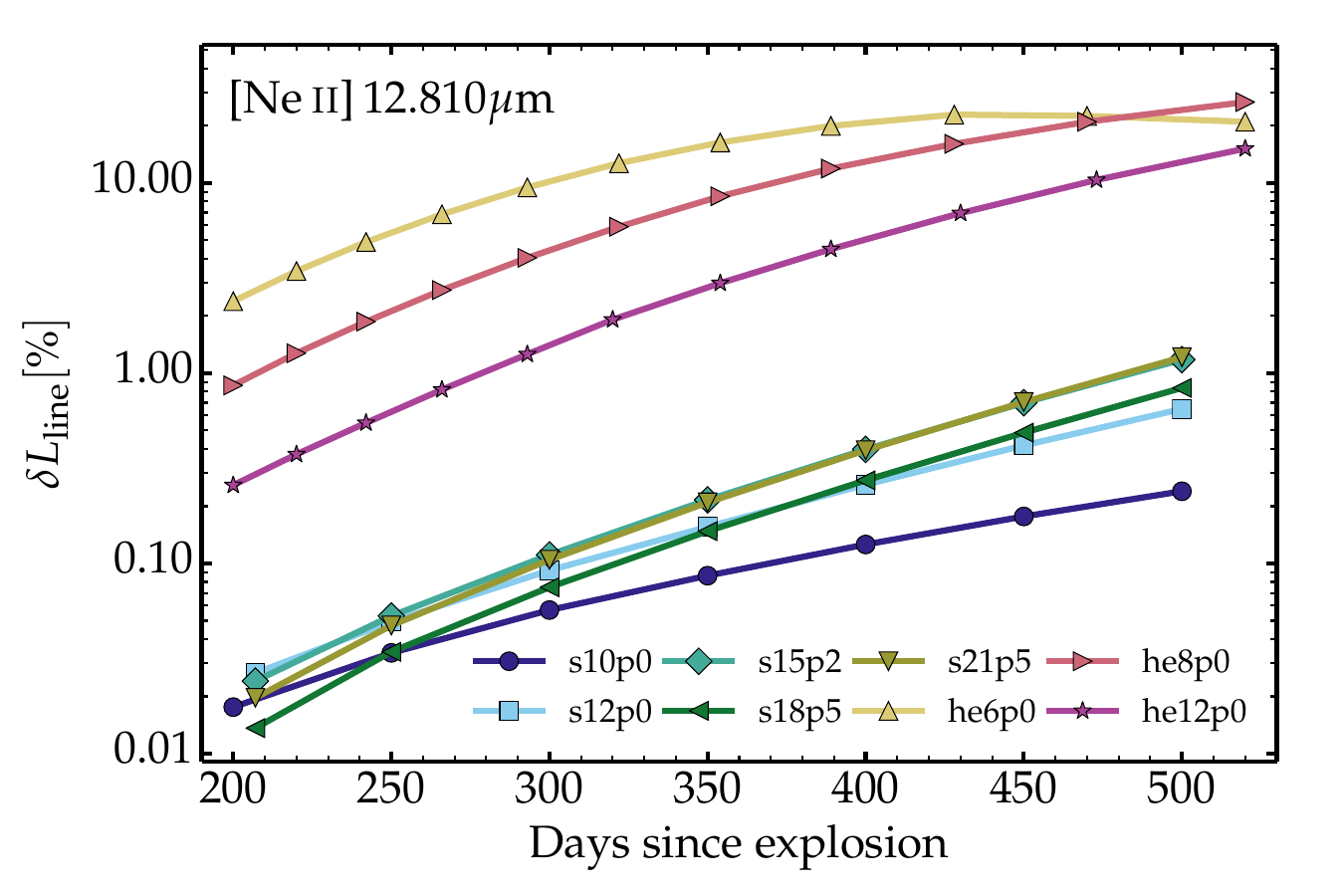}
\caption{Properties of the \neiifs\ emission line for a sample of core-collapse SN models. Top: We show the profile morphology versus wavelength at 350\,d after explosion for models s12p0, s15p2, s21p5, he6p0, he8p0, and he12p0. The spectra are scaled to a distance of 10\,Mpc and an additional scaling is applied to the He-star explosion models for better visibility. Bottom: We show the evolution of the percentage fraction of the bolometric flux that is radiated in \neiifs\ for our full model sample. The color coding differs between the two panels. A similar plot for \oidoub\ is shown in the left panel of Fig.~\ref{fig_opt_lines}. (See Section~\ref{sect_neon} for discussion.)
}
\label{fig_ne2_evol}
\end{figure}

\section{Neon as a diagnostic of progenitor mass}
\label{sect_neon}

In our core-collapse SN ejecta models, the main isotopes of Ne are \iso{20}Ne and \iso{22}Ne (other Ne isotopes with atomic mass between 16 and 25 and included in the nuclear network have a negligible yield; for discussion see \citealt{sukhbold_ccsn_16}, \citealt{woosley_he_19}). Neon is essentially unaffected by explosive burning so the Ne composition is set at the time of core collapse. \iso{22}Ne comes from He burning which turns N into Ne, whereas \iso{20}Ne comes from carbon burning and thus later in the evolution of the progenitor (see, e.g., \citealt{whw02}). The abundance of \iso{22}Ne is essentially limited by the primordial abundance of C, N, and O and thus the metallicity, whereas the abundance of \iso{20}Ne is much larger and the more so the larger the initial mass or the preSN mass. \iso{20}Ne is found in the O/Ne/Mg shell, whereas \iso{22}Ne is primarily located in the He/C/O shell and thus at larger velocities in an unmixed 1D explosion model of the preSN star. In the red supergiant star explosion models of our sample, the ratio $M$(\iso{20}Ne)/$M$(\iso{22}Ne) is a few (models s10p0 to s12p0, in which the He/C/O shell is much more massive than the O/Ne/Mg shell) up to a few tens (higher mass models s15p2 to s21p5, where the O/Ne/Mg shell is much more massive than the He/C/O shell). In the models he6p0, he8p0, and he12p0, this mass ratio is $\sim$\,30.

Figure~\ref{fig_neon_ej} shows the Ne yield for the whole set of red supergiant star and He-star explosion models presented in \citet{dessart_sn2p_21} and \citet{dessart_snibc_21}. The Ne yield for a given ZAMS mass is much lower (as for O) for the He-star progenitors of SNe Ibc than for the red supergiant progenitors of SNe II, and the more so for higher adopted Wolf-Rayet wind mass loss rates (see results for the set ``x1p0'' that uses a nominal mass loss rate and the set ``x2p0'' that uses a twice higher mass loss rate; for discussion, see \citealt{woosley_he_19}). When the Ne yield is shown versus He-core mass or preSN mass, the models then follow almost a single line, which indicates that the scatter in Ne yield obtained versus ZAMS mass results mainly from preSN mass loss. In any case, the main point shown by Fig.~\ref{fig_neon_ej} is that the Ne yield increases considerably with He-core or preSN mass, suggesting that Ne could be an alternative to O as a diagnostic of preSN mass, or progenitor mass in H-rich progenitors. For that purpose one would use \neiifs\ in the infrared as an alternative to \oidoub\ in the optical. One expects \neiifs\ to form under slightly higher temperature or ionization conditions than \oidoub, and in light of the discussion in Sections~\ref{sect_s15p2}--\ref{sect_he6p0}, we expect the \neiifs\ emission to thrive in our He-star explosion models.

The top panel of Fig.~\ref{fig_ne2_evol} illustrates the \neiifs\ emission profile in velocity space at 350\,d for a subset of our H-rich and H-deficient ejecta models. A distance of 10\,Mpc is assumed. Although the different models have a different \nifs\ mass, which can account in part for the differences shown, the major driver behind the diversity amongst these models are the decay power absorbed by the material from the O/Ne/Mg shell and its ionization. The former sets the limits of how much power may be radiated by lines forming in the O/Ne/Mg shell. The latter tunes how much the cooling of the O/Ne/Mg-rich material may operate through \oidoub\ (i.e., in case of a lower ionization) or through \neiifs\ (i.e., in case of a higher ionization).

In all models, the simulated \neiifs\ emission line extends in velocity space out to the outermost layers of the O/Ne/Mg shell, corresponding to $\gtrsim$\,2000\,\kms\ in our model s15p2 and $\sim$\,6000\,\kms\ in our model he6p0. In these line profiles, each slanted part of the wing corresponds to a range of velocities in the model ejecta that contributes to emission in the \neiifs\ line. In our model s15p2, the O/Ne/Mg shells are located in regions between 700 and 1000\,\kms, between 1500 and 1700\,\kms, and around 2000\,\kms\ (see Fig.~\ref{fig_s15p2_init_comp}). In model he6p0, the contributing shells are between 2000 and 3500\,\kms, between 4000 and 4500\,\kms, and between 5200 and 6000\,\kms\ (see Fig.~\ref{fig_he6p0_init_comp}). Each ledge in the profile wing corresponds to regions that contribute no flux (a gap in velocity space with no material from the original O/Ne/Mg shell). Hence, as for Ni lines before (Figs.~\ref{fig_nick_s15p2} and \ref{fig_nick_he6p0}), the \neiifs\ line can be used  to constrain the distribution in velocity space of the \iso{20}Ne-rich material from the O/Ne/Mg shell of the preSN star.

The bottom panel of Fig.~\ref{fig_ne2_evol} shows the evolution of the fractional integrated flux radiated in \neiifs\ from about 200 to 500\,d and for all models in this study (over that time span, the fractional decay power absorbed by the O-rich material stays roughly constant in all models, except in models s10p0 and s12p0 in which it drops significantly; Fig.~\ref{fig_frac_edep}). In all cases, the fractional flux radiated in \neiifs\ increases with time as the ejecta density drops. However, in red supergiant star explosion models, the total line flux is weak, rising from about 0.01\,\% at 200\,d up to about 1.0\,\% at 500\,d. A trend toward greater \neiifs\ fractional flux with main sequence mass is present, but weak and scattered. In He-star explosion models, the \neiifs\ fractional flux is much greater, rising from about 1\,\% at 200\,d up to 10--30\,\% at 520\,d. Ejecta with a greater expansion rate, and therefore ionization, show stronger \neiifs\ --- of all three He-star explosion models, it is the model with the lowest Ne abundance (i.e., he6p0) that has the strongest \neiifs\ line flux. As indicated in Table~\ref{tab_init}, the Ne yield is comparable between models s15p2 and he6p0, or between models s21p5 and he12p0, and thus the difference in the fractional line flux is dominated by the differences in ejecta structure (i.e., expansion rate, and subsequently ionization) rather than by differences in Ne yield.

These simulations suggest that within an homogeneous set (H-rich progenitors only, or H-deficient progenitors only), the \neiifs\ may be used as a substitute to \oidoub\ (results for this line are shown in Fig.~\ref{fig_opt_lines}) for evaluating the progenitor or the preSN mass (the \oidoub\ diagnostic also suffers from degeneracies in that context). In some models and epochs, the infrared range is essential to constrain ejecta cooling since much power may be channeled into this spectral region, with \neiifs\ acting as a major coolant of the O-rich material (thus weakening dramatically \oidoub\ as well as the optical luminosity).

\begin{figure*}
   \centering
    \begin{subfigure}[b]{0.33\textwidth}
       \centering
       \includegraphics[width=\textwidth]{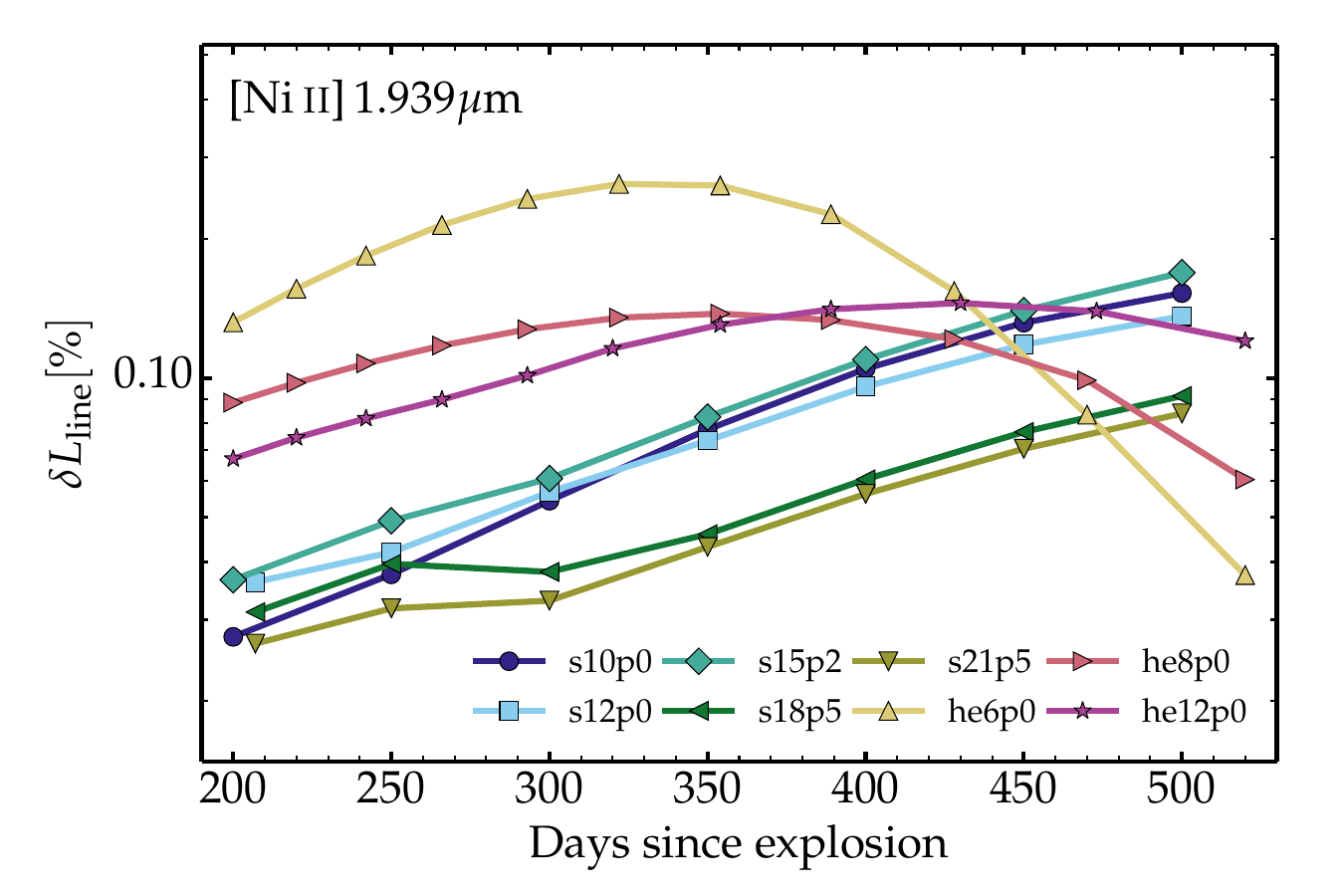}
    \end{subfigure}
    \hfill
    \begin{subfigure}[b]{0.33\textwidth}
       \centering
       \includegraphics[width=\textwidth]{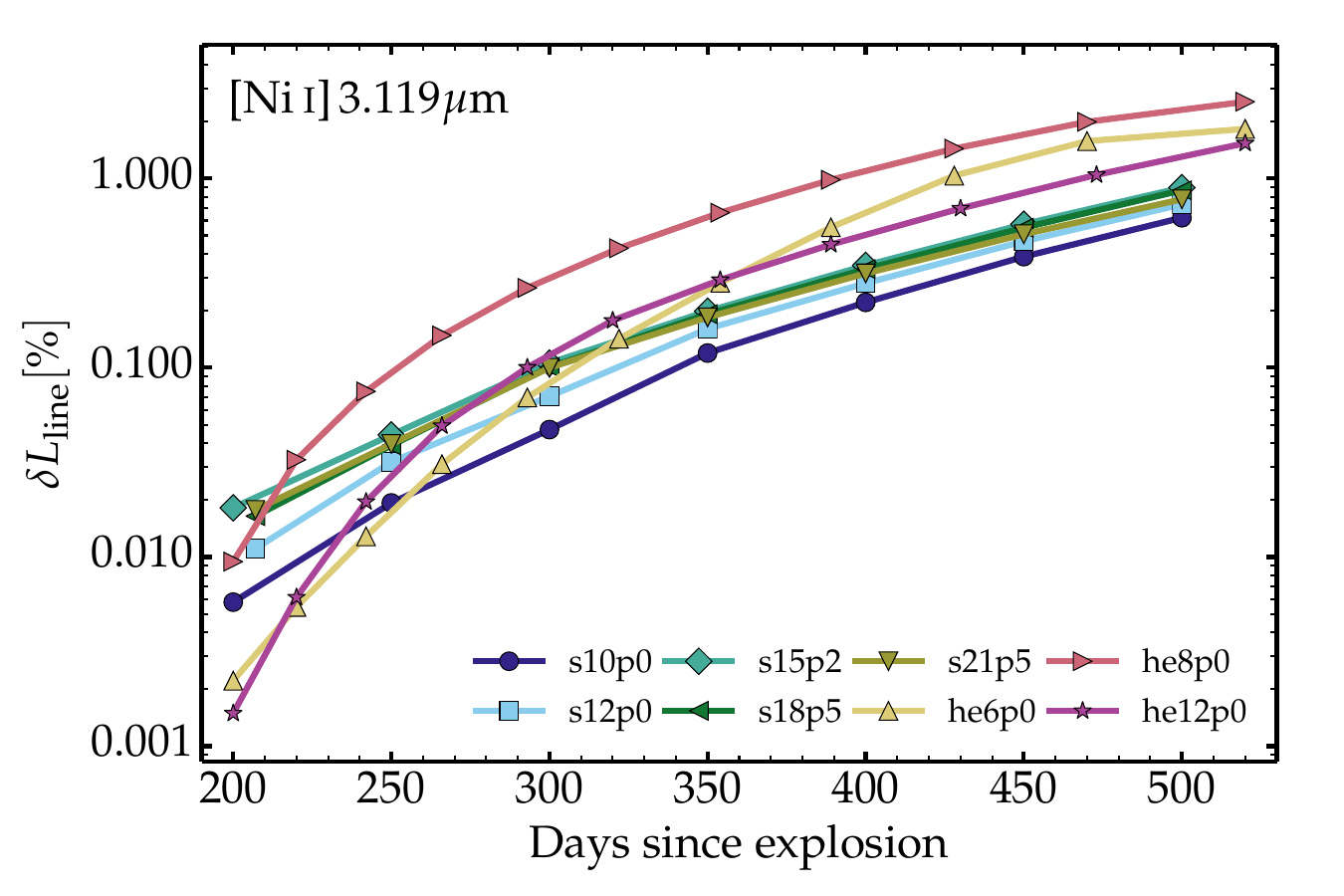}
    \end{subfigure}
     \hfill
    \begin{subfigure}[b]{0.33\textwidth}
       \centering
       \includegraphics[width=\textwidth]{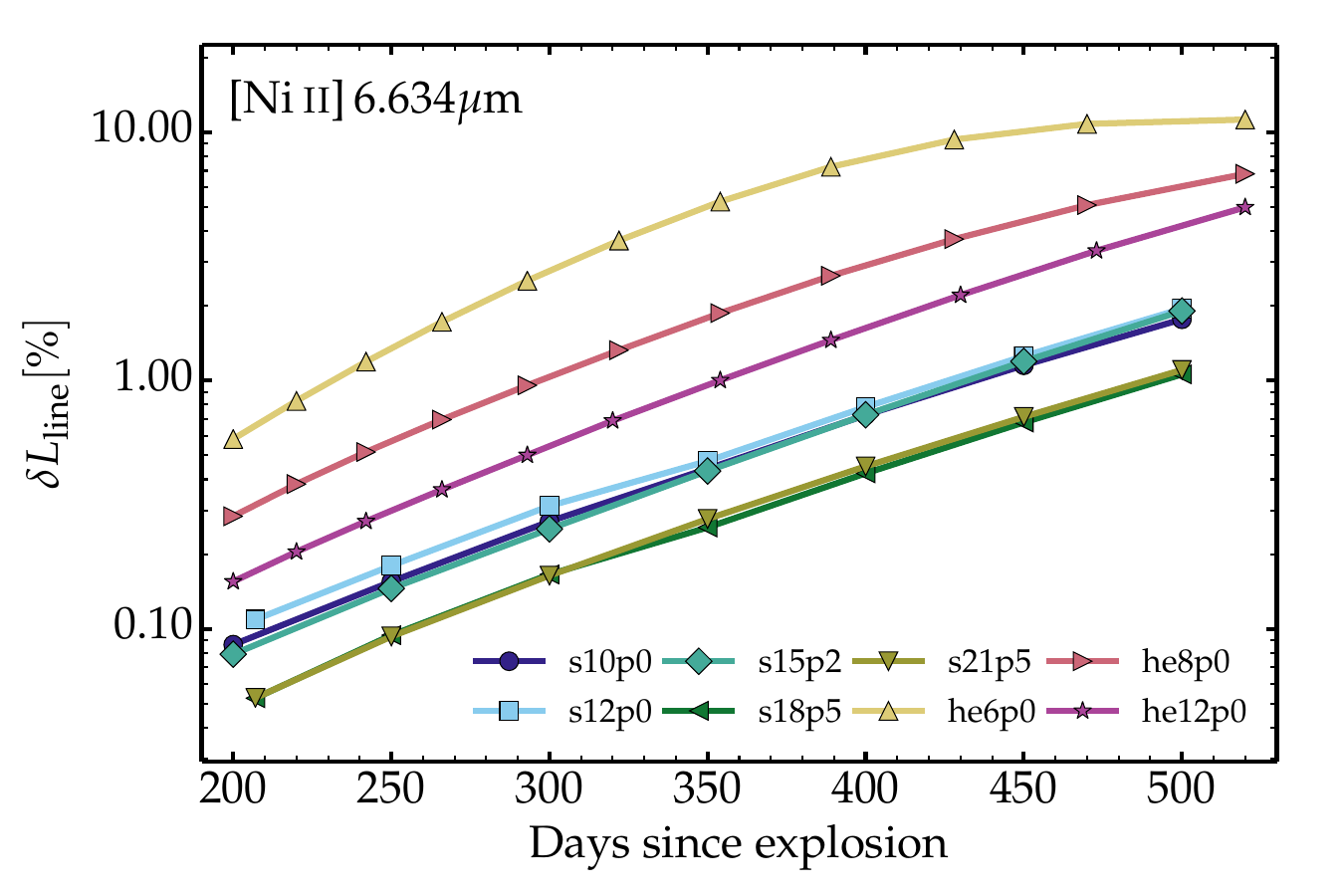}
    \end{subfigure}
     \hfill
    \begin{subfigure}[b]{0.33\textwidth}
       \centering
       \includegraphics[width=\textwidth]{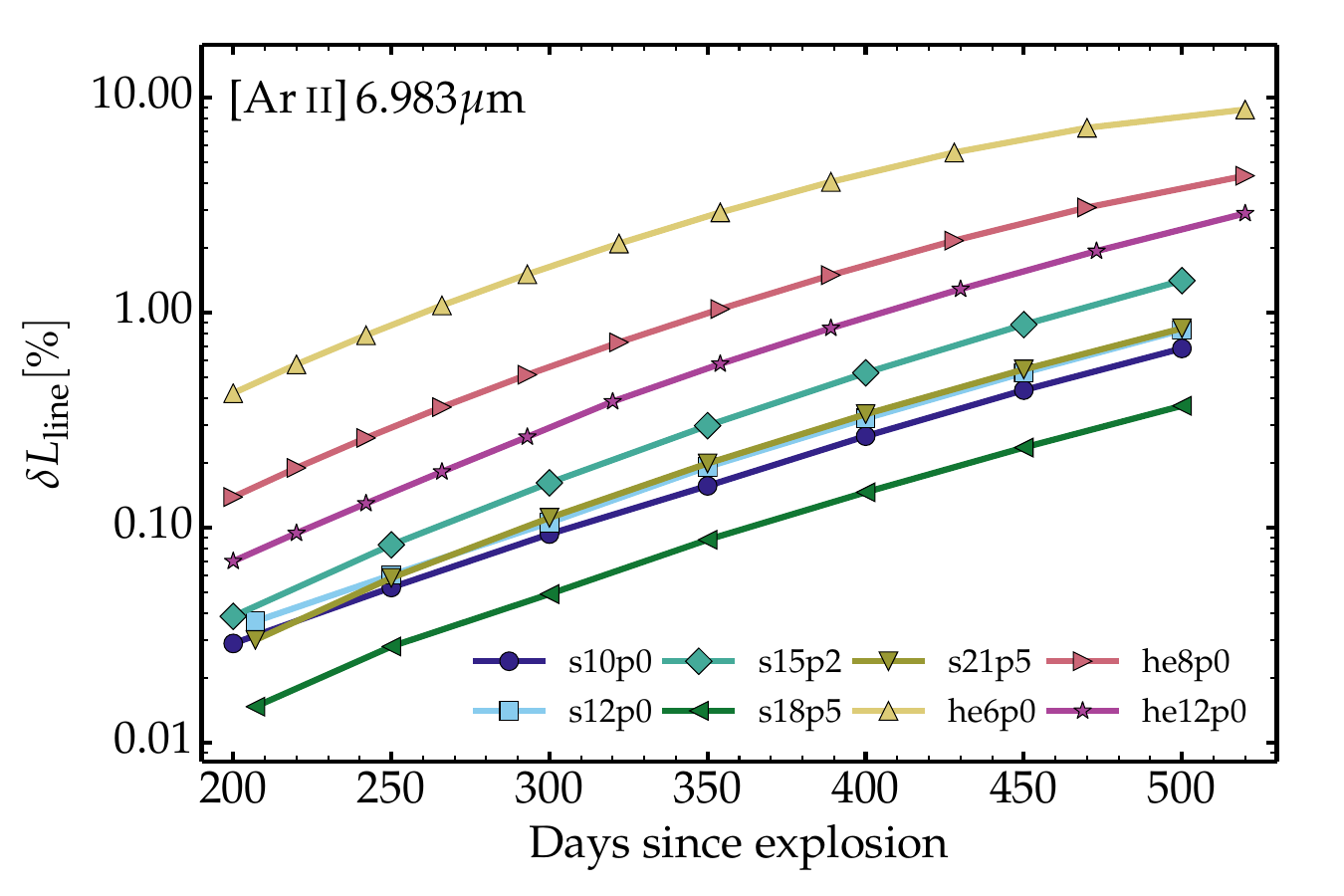}
    \end{subfigure}
    \hfill
    \begin{subfigure}[b]{0.33\textwidth}
       \centering
       \includegraphics[width=\textwidth]{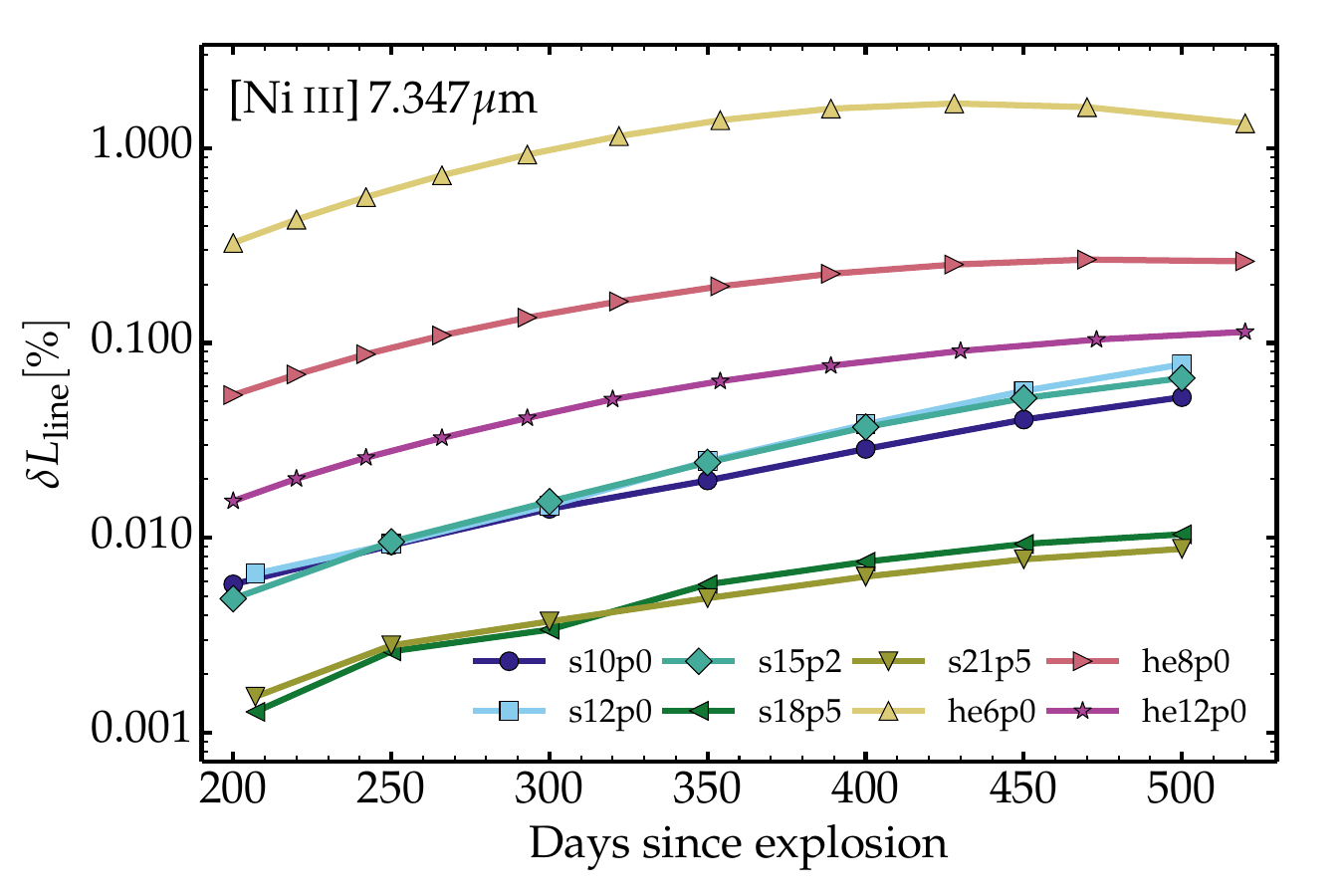}
    \end{subfigure}
     \hfill
    \begin{subfigure}[b]{0.33\textwidth}
       \centering
       \includegraphics[width=\textwidth]{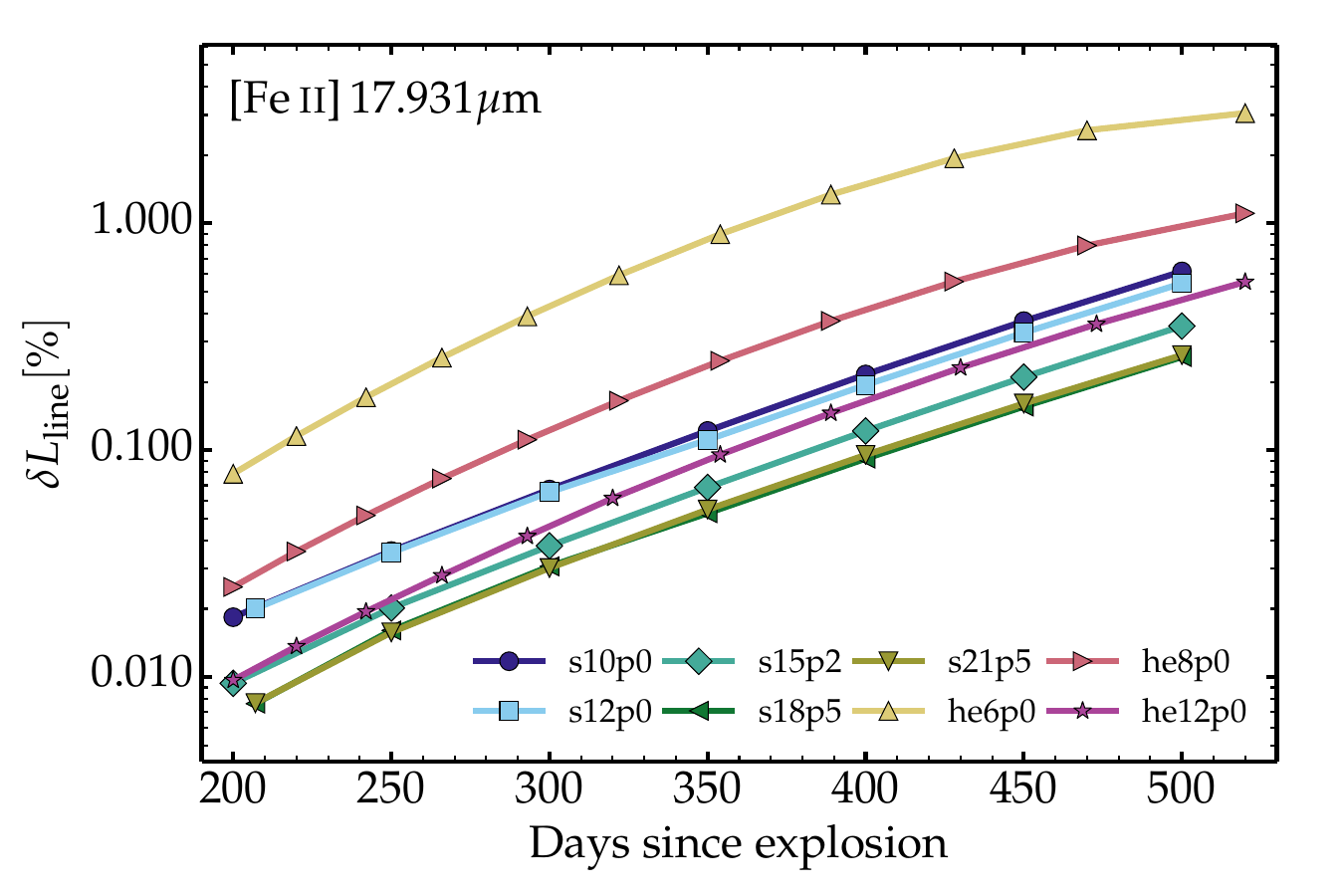}
    \end{subfigure}
\caption{Evolution from $\sim$\,200\,d to $\sim$\,500\,d of the fractional flux associated with a selection of Ni, Ar, and Fe lines in the infrared for our sample of red supergiant star and He-star explosion models. (See Section~\ref{sect_ir_lines} for discussion.)
\label{fig_ir_lines}
}
\end{figure*}

\section{The diversity of infrared spectral lines in core-collapse SNe}
\label{sect_ir_lines}

In this section, we document the evolution of a number of infrared lines from about 200 to 500\,d in our model sample. This extends what was presented earlier in detail for models s15p2 (Section~\ref{sect_s15p2}) and he6p0 (Section~\ref{sect_he6p0}), or for the full model sample in the case of \neiifs\ (Section~\ref{sect_neon}). The evolution of the fractional line flux for the optical lines \oidoub, H$\alpha$, and \caiidoub\ is shown in the appendix (Fig.~\ref{fig_opt_lines}) as well as in previous papers \citep{dessart_sn2p_21,dessart_snibc_21,dessart_snibc_23}. Because of the numerical procedure, the line strength of Fe-group elements are underestimated at the 10\,\% level in models s10p0, s12p0, s18p5, and s21p5 (see Section~\ref{sect_setup} for discussion).

Figure~\ref{fig_ir_lines} shows the evolution of the fractional line flux radiated in [Ni\two]\,1.939\mic, [Ni\one]\,3.119\mic, [Ni\two]\,6.634\mic, [Ar\two]\,6.983\mic, [Ni\three]\,7.347\mic, and [Fe\two]\,17.931\mic. We recover the trends discussed in Sections~\ref{sect_s15p2}--\ref{sect_he6p0}, with a general increase with time in all cases except for [Ni\two]\,1.939\mic\ in the He-star explosion models (probably because the lower level of that transition lies above the ground state and thus weaker than for example [Ni\two]\,6.634\mic, which is a transition between the lowest two levels of Ni\two). The more massive, slower expanding, denser H-rich ejecta exhibit systematically lower fractional fluxes of at most 1\,\%  in those lines compared to H-deficient ejecta. This is not an ionization effect since, for example, the Ni\one, Ni\two, and Ni\three\ line fluxes are greater in model he6p0 than in any of the H-rich ejecta models at say 350\,d. What likely dominates here is the combination of a lower ejecta density and a greater fraction of decay power absorbed by the Fe/Si-rich material in He-star explosion models (see also the bottom-right panel of Fig.~\ref{fig_frac_edep}).

The diversity within each set of H-rich and H-deficient models is more subtle. The decay power absorbed in the Fe/Si-rich material in models s15p2 and s21p5 is up two times greater than that in the lower mass models, which boosts the power to be radiated in Ni or Ar lines. This is, however, mitigated by the difference in ejecta density (or expansion rate $V_{\rm m}$; see Table~\ref{tab_init}) with $V_{\rm m}$ being lower in s15p2 and s21p5. In the H-deficient models, the ejecta density seems to drive the difference between the faster expanding model he6p0 compared to the slower expanding model he12p0, the latter developing strong forbidden lines in the infrared at later times.

Because of these complications, it may be difficult to draw firm conclusions on Ar or Ni yields from an observed infrared spectrum of a core-collapse SN. Indeed, the metal yields from models s15p2 and he6p0 are comparable (see Table~\ref{tab_init}) in spite of the drastic difference between their infrared spectra, largely driven by the dynamical properties of their ejecta. However, the morphology of these metal lines offers an excellent constraint on the spatial or velocity distribution of these metals in the ejecta (as shown for models s15p2 in Fig.~\ref{fig_nick_s15p2} and he6p0 in Fig.~\ref{fig_nick_he6p0}). Stable Ni but also Ar and Fe trace directly the material produced through explosive nucleosynthesis (i.e., \nifs), and could be used to compare to the corresponding distribution of Ne from the O/Ne/Mg material and inferred from \neiifs\ (Section~\ref{sect_neon}).

\section{Evidence of \nifs\ decay inferred from the evolution Fe and Co lines}
\label{sect_fe_co}

The evolution of the strength of Fe and Co lines may provide evidence of \nifs\ decay as the power source in our ejecta models, as an alternative to the evolution of the bolometric luminosity at nebular times. Indeed, the mass of Fe and Co in the ejecta regions originally rich in \nifs\ varies by about a factor of ten over the time span from 200 to 500\,d, and thus for a given power, one would naively expect the Fe lines to strengthen considerably more than the Co lines. The exact mass ratio to consider can be ambiguous since isotopes other than \iso{56}Fe and \iso{56}Co are present in the region of formation of Fe and Co lines (i.e., essentially exclusively those regions where explosive nucleosynthesis took place). The decay of \iso{57}Ni to \iso{57}Co prevents the Co abundance from continuously dropping at 200-500\,d. The challenge is to identify Fe and Co lines for that purpose.

The high ionization of SNe Ia ejecta is conducive to the formation of [Fe\three] and [Co\three] lines, which have strong emission in the optical at 0.469\mic\ and 0.589\mic. The evolution of the line strength of these two emission features at nebular times gives unambiguous evidence of \nifs\ decay as the power source in SNe Ia \citep{kuchner_co3_94}, or at least in the representative sample of SNe Ia used by these authors. Unfortunately, massive star explosions are typically denser and cooler, so that these Fe\three\ and Co\three\ lines are absent from their optical spectra. However, infrared spectra of massive star explosions contain numerous lines from Fe and Co, as discussed in previous sections and observed in the past (see, for example, \citealt{wooden_87A_ir_93} or \citealt{kotak_04et_09}).

Figure~\ref{fig_fe_co_s15p2_he6p0} shows the evolution of the fractional flux in a variety of Fe and Co lines in the infrared for models s15p2 and he6p0. Compared to the optical lines of Fe\three\ and Co\three\ present in SN Ia spectra, we see that these metal lines (from a mixture of once- and twice-ionized Fe and Co) are weak, representing less than 1\,\% of the total power in model s15p2 and at most a few percent in model he6p0. In most cases and for both models, this fractional line flux increases in time for both Fe and Co lines, and in general this increase is greater for the Fe lines relative to Co lines. This is as expected given the ten fold increase in Fe mass relative to Co mass in the Fe/Si-rich shell of these two models. When estimating the Fe and Co masses, we only account for material within the Fe/Si-rich shells (i.e., where these lines form). Furthermore, we account for all Fe and Co isotopes since they all contribute to radiating the decay power absorbed in these regions.

But we also see great disparity between lines. For example, in model s15p2, [Co\two]\,10.520\mic\ and [Fe\two]\,17.931\mic\ follow essentially the same trajectory, and thus would have a constant flux ratio from 200 to 500\,d, as if there was no change in the relative Fe and Co abundances. But other lines exhibit a flux ratio that follows closely the evolution of the Fe and Co masses in the Fe/Si-rich shell, such as the flux ratio of [Fe\two]\,17.931\mic\ and [Co\two]\,14.735\mic\ in model he6p0 (Fig.~\ref{fig_fe17_co14_he6p0}). Unfortunately, the same choice of lines for model s15p2 yields a severe mismatch with the evolution of $M_{\rm Fe}$/$M_{\rm Co}$(Fe-sh). Evidence of \nifs\ decay as the power source of our ejecta models (and by extension of observed core-collapse SNe) is thus not robustly revealed by the evolution of the Fe and Co infrared emission lines in those core-collapse SN ejecta. One potential reason for this complexity, or potential lack of sensitivity, is the efficient cooling of the Fe/Si-rich material by lines other that Fe and Co, and specifically by Ar\two, Ca\two, and Ni\two\ lines (stable Ni is as abundant as Co in the Fe/Si-rich material at about 300\,d, and it dominates over Co both in abundance and as a coolant at later times). Optical-depth effects might also play a role. Most infrared lines of Fe\two\ and Co\two\ are optically thick but Fe\two\ lines are more optically thick and may be affected for longer.

\begin{figure}
\centering
\includegraphics[width=0.9\hsize]{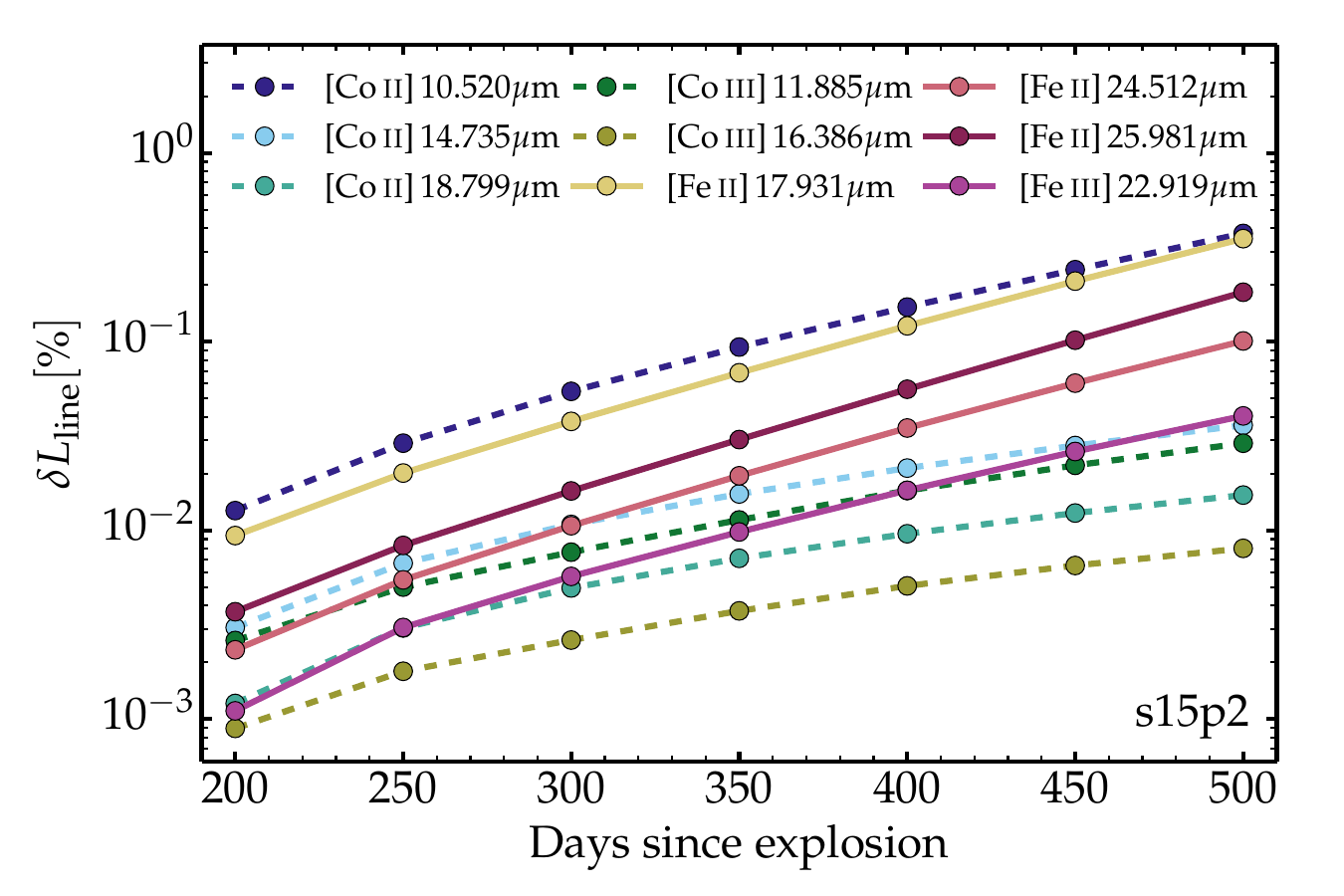}
\includegraphics[width=0.9\hsize]{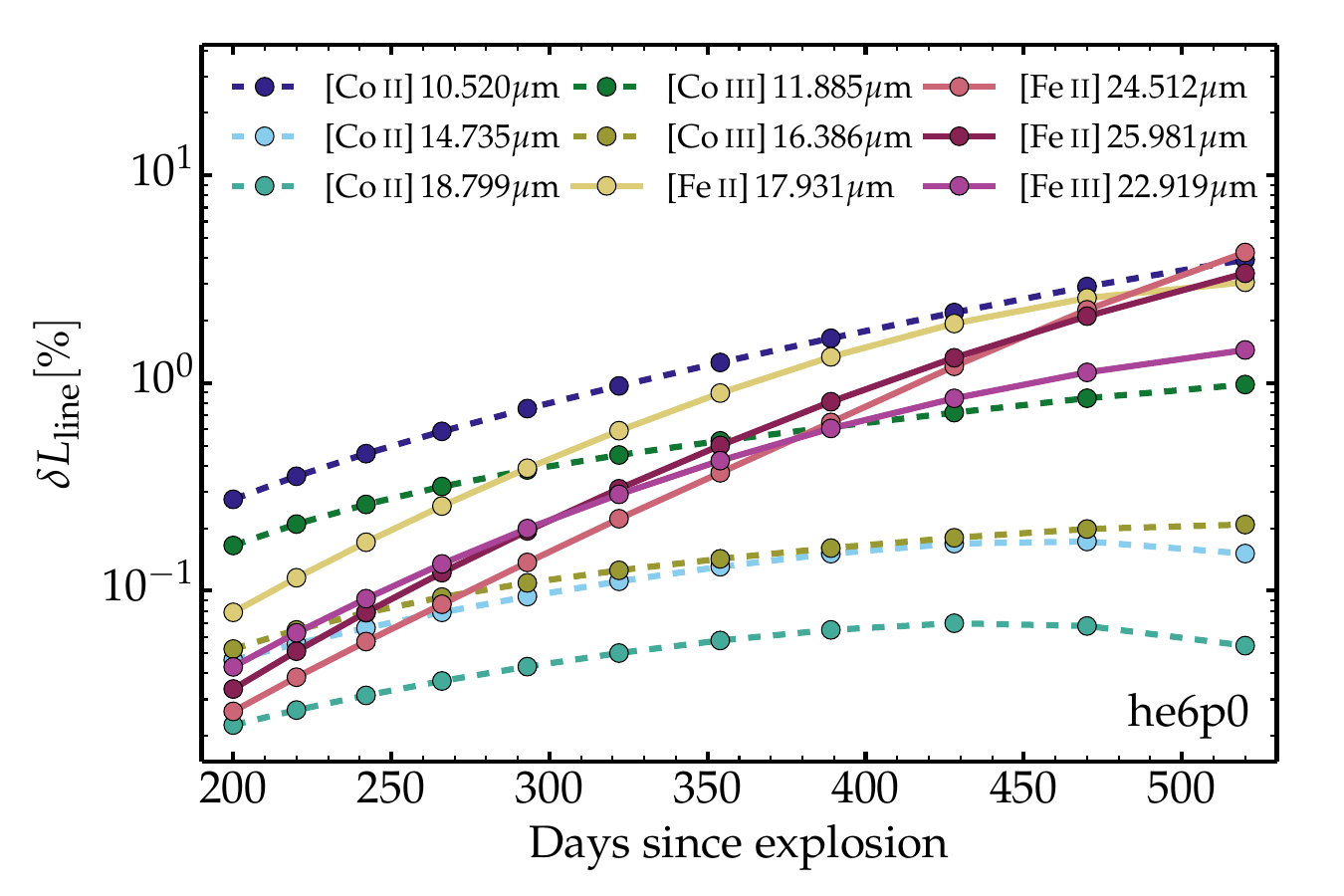}
\caption{Evolution from 200 to about 500\,d of the strength of some Fe (solid) and Co (dashed) lines in the infrared and for models s15p2 (top) and he6p0 (bottom).}
\label{fig_fe_co_s15p2_he6p0}
\end{figure}

\section{Conclusions}
\label{sect_conc}

We have presented the infrared, nebular-phase properties of a sample of massive star explosion models that arise from red supergiant stars (models originally from \citealt{sukhbold_ccsn_16}) or from H-deficient, stripped stars (models originally from \citealt{woosley_he_19} and \citealt{ertl_ibc_20}). This work complements our previous studies which focused on the photometric and spectroscopic properties in the optical primarily \citep{dessart_sn2p_21,dessart_snibc_21,D22_lsst,dessart_snibc_23}. Because many of the models discussed in those works have comparable properties (say model s14p5 and s15p2 or he5p0 and he6p0) or do not seem to have obvious observational counterparts (e.g., he3p30), we selected a subset of models with greater differences (primarily in initial mass), namely model s10p0, s12p0, s15p2, s18p5, and s21p5 for those with hydrogen at core collapse, and he6p0, he8p0, and he12p0 for those without. To improve the physical consistency of our calculations, we rerun most models with a better account of the full isotopic composition, up-to-date atomic data and model atoms, and treated the two-step decay chains associated with the parent isotopes \iso{44}Ti, \iso{48}Cr, \iso{52}Fe, \nifs, and \iso{57}Ni (exceptions are models s10p0, s12p0, s18p5, and s21p5, which accounted only for the \nifs\ decay chain, as in \citealt{dessart_sn2p_21}). The time span covers from $\sim$\,200 to $\sim$\,500\,d. Dust is typically expected to form after about 500\,d in the inner ejecta of Type II SNe \citep{sarangi_dust_22}, as also observed in SN\,1987A \citep{lucy_dust_89}, and thus it is largely absent over the time span covered here. Because of space limitation, a comparison to other work is provided in Appendix~\ref{sect_comp}.

\begin{figure}[t]
\centering
\includegraphics[width=0.9\hsize]{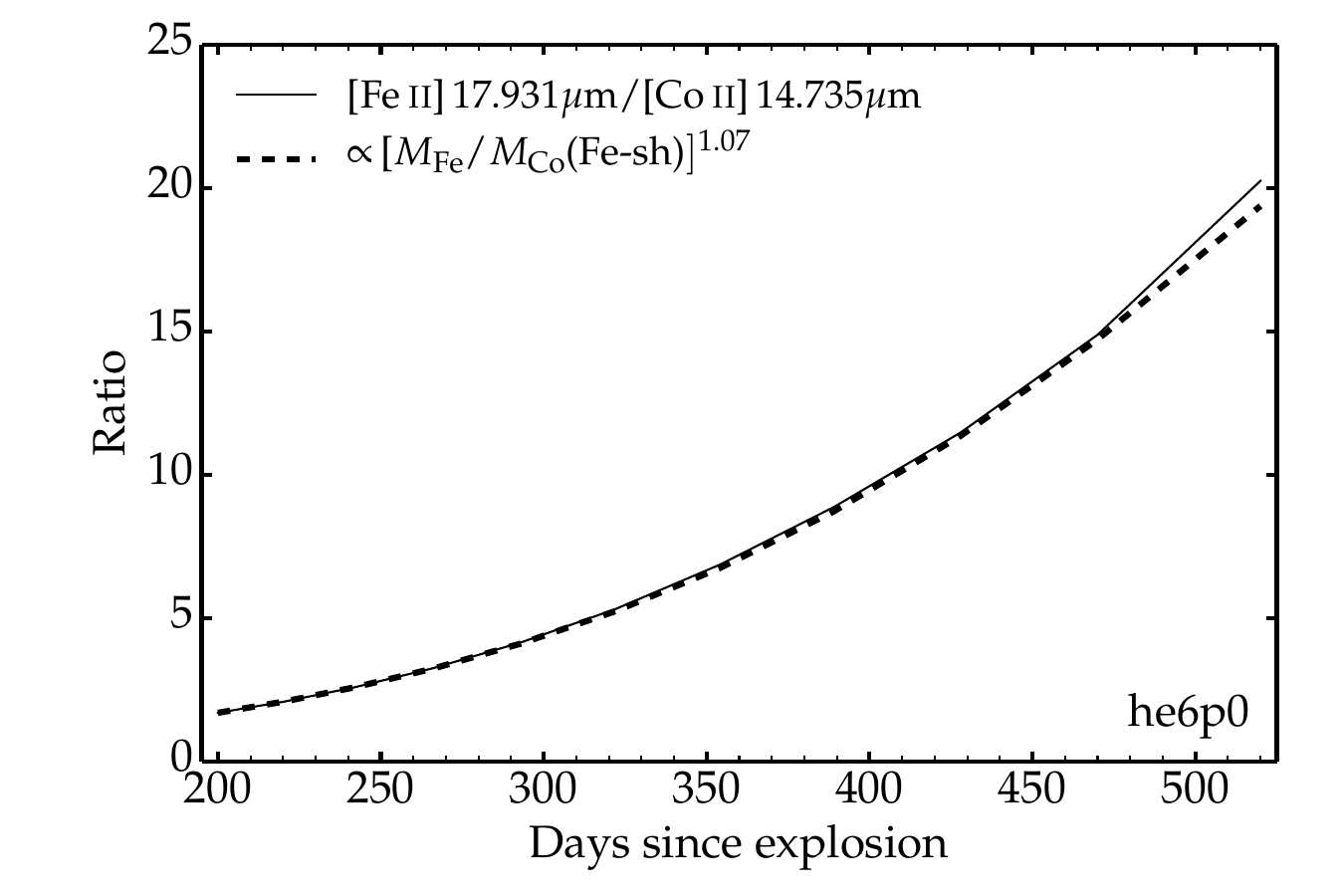}
\caption{Evolution from 200 to 500\,d of the ratio of the total flux emitted in [Fe\two]\,17.931\mic\ and [Co\two]\,14.735\mic\ in model he6p0 (solid). The dashed line shows the corresponding evolution of the total Fe and Co mass within the Fe/Si-rich material, adjusted through an additional power of 1.07.}
\label{fig_fe17_co14_he6p0}
\end{figure}
 
The focus of this work is on the emission from atoms and ions in the ejecta material, which will serve in a subsequent step as the initial conditions for simulations that account for molecule formation and cooling \citep{mcleod_mol_24}. Molecules such as CO, SiO, SiS impact core collapse SN spectra by potentially modifying the temperature and ionization of the emitting gas and radiating a fraction of the decay-power absorbed in selective ejecta regions (see, e.g., \citealt{liljegren_ibc_mol_23}). For example, CO forms in the O/C shell (i.e, or in blobs or clumps of C-rich O-rich material that was originally located in the O/C shell of the preSN star) and contributes to cooling that gas, thereby depleting flux that would otherwise have come out in lines such as \oidoub. This molecular cooling is lacking in the present simulations. We thus overestimate some lines fluxes by some amount and do not predict any infrared emission for example from CO, in contradiction with observations such as those of the Type II-peculiar SN\,1987A \citep{spyromilio_co_87A_88}, the Type II-P SNe 2004et \citep{kotak_04et_09} and SN\,2017eaw \citep{rho_co_17eaw_18} or the Type Ic SN\,2020oi \citep{rho_20oi_21}.

However, whilst molecular emission can be efficient, its impact is limited at nebular times to the regions where molecules form. For example, in our model s15p2, only 5\,\% of the total decay power absorbed by the ejecta is absorbed by the O/C-rich material. And within those regions, CO molecules may account for some but probably not the entire power absorbed, which means that CO emission is fundamentally limited to be at most just a few percent of the luminosity in that model. In our model he6p0, the decay power absorbed in the O/C-rich shell is 10--15\,\% of the total (in both these models, this fraction represents a quarter of the power absorbed in the O-rich shell), so the impact is potentially greater, but it is not clear molecules can form in great abundance in those lower density, fast expanding ejecta (relative to Type II SN ejecta). In other words, there may be more power for the CO molecules to radiate but their formation is perhaps not as efficient. In the simulations of \citet{liljegren_ibc_mol_23}, large CO masses are predicted in the case of a high compression of the O/C shell material, but compression of that material following the \nifs\ bubble effect (see, e.g., \citealt{woosley_87A_late_88}) is likely less efficient in SNe Ibc than in SNe II.\footnote{This less efficient bubble effect in SNe Ibc would arise from a combination of factors. SNe Ibc have a comparable \nifs\ mass to SNe II but have a much larger expansion rate (the \nifs-rich material is present at larger velocities, and thus higher specific kinetic energies than in SNe II). Nonlocal energy deposition starts earlier and leads to a spatial ``spreading'' of the influence of decay heating. Their ejecta turn optically thin at about 20\,d (rather than about 100\,d), which would stop any pressure build-up from decay heating at earlier times than in SNe II.\label{footnote_nibubble}} Furthermore, molecular cooling will alter the total flux emitted in lines but much less (and perhaps not at all) their profile shapes, which represent a key diagnostic for example for chemical mixing. Finally, numerous lines are largely unaffected by molecular formation, such as \neiifs\ since neither CO nor SiO forms in the O/Ne/Mg material where \iso{20}Ne is abundant but where both C and Si are underabundant (see, e.g., \citealt{liljegren_ibc_mol_23}). So, there is much to learn from nebular-phase radiative-transfer models even when molecular formation and cooling are ignored.

In the Type II SN models, about 75\,\% of the flux is emitted below 1\mic\ at all times (this rises to 90\,\% below 2\mic), with little evolution over time, so the infrared contains a small fraction of the total flux (molecular emission would raise this fraction a little -- see above). In the infrared, our model spectra exhibit a myriad of H\one\ lines from the Paschen up to the seventh series, He\one\ lines at 1.083 and 2.058\mic, essentially no lines of C, N, or O, weak lines of Na\one\ and Mg\one, but the strongest line from intermediate mass elements is \neiifs. At higher atomic mass, there are strong emission lines from [S\one] at 1.082\mic, as well as 25.242\mic, and from Ar with \ariimir.

However, many lines in the infrared arise from Fe, Co, and Ni, which are present throughout the spectrum from 1 to 30\mic. Lines from Ni arise from the presence of stable Ni in the regions where explosive nucleosynthesis took place and include [Ni\one] at 3.119 and 7.505\mic, [Ni\two] at 1.939, 6.634, 10.679, and 12.725\mic, and [Ni\three]\,7.347\mic. This wealth of emission lines from iron-group elements, which are predicted to be present in all our models, contrasts with the scarcity of such lines in the optical. One exception is \nkiiopt, which was unambiguously identified in SN\,2012ec \citep{jerkstrand_ni_15}, although this was in a large part due to the unusual weakness of \caiidoub. The associated emission line profiles can directly reveal the magnitude of \nifs-mixing in the explosion. There are also numerous lines of Fe with [Fe\two] at 17.931, 24.512, or 25.981\mic, and Co lines with [Co\two] at 10.520 or 14.735\mic. All these Fe, Co, and Ni transitions in the infrared are forbidden and essentially all strengthen with time as the ejecta density drops from 200 to 500\,d. Combined with the luminosity drop by a factor of about 20 over that time span, many of these emission lines retain a near-constant flux (or luminosity), which should make them as detectable at 500\,d as at 200\,d. Results for other Type II SN models in our sample are presented in Figs.~\ref{fig_opt_lines}--\ref{fig_frac_edep}--\ref{fig_line_evol_all}.

In the He-star explosion models (which would correspond to Type Ib or Ic SNe), the infrared properties are qualitatively analogous but with marked quantitative differences arising from the lower ejecta density (the models have essentially the same explosion energy but a lower ejecta mass relative to the Type II SN models). This leads to a greater escape of $\gamma$-rays and a greater fraction of the power arising from the local deposition of positrons within the Fe/Si-rich regions. Consequently, the continuum flux level is weaker, forbidden lines are stronger, and the ionization is higher, leading to strong emission from \neiifs, \nkiimir, \ariimir, \nkiiimir, or [Co\two]\,10.520\mic. The spectral energy distribution exhibits a drastic shift to the infrared with time. In model he6p0, the fraction of the flux emitted beyond 1\mic\ is 15\,\% at 200\,d but nearly 80\,\% at 520\,d. This increase in the strength of infrared lines in He-star explosion models is even greater than in model s15p2, such that numerous lines in the infrared evolve essentially at constant luminosity. Results for other He-star explosion models in our sample are presented in Figs.~\ref{fig_opt_lines}--\ref{fig_frac_edep}--\ref{fig_line_evol_all}.

We investigated whether Ne, whose abundance increases with He-core mass or preSN mass (in a similar fashion to O), could be used as a progenitor mass indicator. There is a tendency for models with a greater Ne mass to exhibit a greater flux in \neiifs. However, there is a scatter within each model type (i.e., Type II or Type Ibc SN model), as well as between ejecta models of different initial mass. For example, models s15p2 and he6p0 have a similar Ne yield (and O yield), but the fractional line flux radiated in \neiifs\ is 10 to 100 times greater in the He-star explosion model at all times (this contrast is reduced when considering the actual line flux since the luminosity is greater in model s15p2 because of the greater $\gamma$-ray trapping efficiency; this contrast is also smaller when considering \oidoub\ -- see Fig.~\ref{fig_opt_lines}). The profile of the \neiifs\ emission can nonetheless serve to map the distribution in velocity space of the Ne-rich material. In He-star explosion models, \neiifs\ and the infrared as a whole is needed to capture the full SN luminosity since the optical represents a decreasing fraction of the total flux as time passes.

We find that Fe and Co lines generally strengthen in time in all models but each line tends to follow its own trajectory. This likely arises from the fact that each line has a different sensitivity to the density, temperature, and ionization evolution of the region where they form -- many of these lines correspond to transitions between excited states, which would thus tend to weaken in time for the benefit of transitions tied to the ground state. The other reason is that the Fe/Si-rich material cools through numerous lines of species other than Fe and Co, and in particular Ar, Ca, and Ni, and the role of these other coolants also varies in time. We thus find that the evolution of the Fe and Co line fluxes does not provide a robust evidence of \nifs\ decay as the ejecta power source of our core-collapse SN models, even though it is the only power source in all the simulations of this work.

The present simulations will now be compared to full optical to mid-infrared observations of nearby core-collapse SNe (see, for example, \citealt{dessart_24ggi_25}). They will also be used as initial conditions for radiative transfer calculations that include molecular formation and cooling (McLeod et al., in prep.).

\begin{acknowledgements}

LD acknowledges fruitful discussion with Wynn Jacobson-Gal\'an and Rubina Kotak. This work was granted access to the HPC resources of TGCC under the allocations 2023 -- A0150410554 and 2024 -- A0170410554 on Irene-Rome made by GENCI, France.

\end{acknowledgements}

\onecolumn

\appendix

\section{Additional information and results for our sample of core-collapse SN models}
\label{sect_add}

Figures~\ref{fig_s15p2_init_comp} and \ref{fig_he6p0_init_comp} illustrate the ejecta properties (i.e., composition and velocity versus Lagrangian mass) for the H-rich SN model s15p2 and the H-deficient he6p0 model. The complicated composition results from the inherent onion-shell structure of the original, unmixed 1D ejecta models from \citet{sukhbold_ccsn_16} and \citet{ertl_ibc_20}, but also from the shuffling in mass space that we applied. As discussed in \citet{dessart_shuffle_20}, the approach in all cases was to split all shells (but only a small fraction of the outermost shell) into three subshells and redistribute them in the same order, starting from the innermost ejecta layer. We thus obtain a pattern of alternating shells of different composition repeated three times, connecting in the outer ejecta with the outermost shell that was largely excluded from the shuffling. The present H-rich (H-poor) ejecta models are the same as those presented in \citet{dessart_sn2p_21} \citep{dessart_snibc_21}.

To complement the results presented in the main text for infrared lines, we illustrate the evolution of the fractional luminosity emerging in the strong optical lines of \oidoub, H$\alpha$, and \caiidoub\ in Fig.~\ref{fig_opt_lines}. Figure~\ref{fig_frac_edep} also shows the evolution of the fractional powers absorbed in the ejecta, its regions of distinct composition, as well as the relative contribution of positrons, which are absorbed locally in the \nifs-rich regions, as a function of time. Finally, we show in Fig.~\ref{fig_line_evol_all} the evolution of the fractional luminosity emerging in a sample of infrared lines in models other than s15p2 and he6p0, for which a similar information is provided in the right panel of Figs~\ref{fig_cum_lum_s15p2} and \ref{fig_cum_lum_he6p0}.

The explosion energy of our H-rich ejecta models (all taken from \citealt{sukhbold_ccsn_16}) is typically below 10$^{51}$\,erg and thus probably on the low side (e.g., compared to SN\,1987A). So, we tested the impact of increasing the kinetic energy of the ejecta, which for simplicity was achieved by scaling both the radius and the velocity by some factor $\alpha$ (this leaves the SN age unaffected but raises the kinetic energy by a factor $\alpha^2$) and scaling the densities correspondingly by $1/\alpha^3$ in order the maintain the same ejecta mass and yields. When raising in that manner the ejecta kinetic energy by a factor of two in model s15p2 at 350\,d, the decay power absorbed dropped from 87.9\,\% to 68.1\,\%, which led to an overall reduction of the emergent flux throughout the optical and infrared ranges -- the decay-power absorbed was reduced following the enhanced escape of $\gamma$-rays. A second effect is the different spatial distribution of that absorbed decay power. Lines forming in the H-rich ejecta weakened (e.g., H\one\ lines, including those from the Balmer series in the optical and from the higher series in the infrared), whereas the \oidoub\ line flux increased relative to H$\alpha$. The same effect would have been produced by reducing the ejecta mass. This trend suggests that some diversity in Type II nebular-phase spectra may be driven by variations in the trapping efficiency of their ejecta (i.e., driven by variations in $M_{\rm ej}$ or $E_{\rm kin}$) rather than variation in oxygen yield.  However, in this comparison, strong metal lines in the infrared (e.g., \nkiimir, \ariimir, or \neiifs) retained a comparable strength or even strengthened, likely because the reduced ejecta density favored the emission from these forbidden lines relative to recombination lines such as those of H\one. This complex sensitivity of nebular-phase spectral properties to variations in ejecta properties (not limited to composition variations) is a challenge for radiative-transfer modeling.

\begin{figure*}[h]
\centering
\includegraphics[width=0.8\hsize]{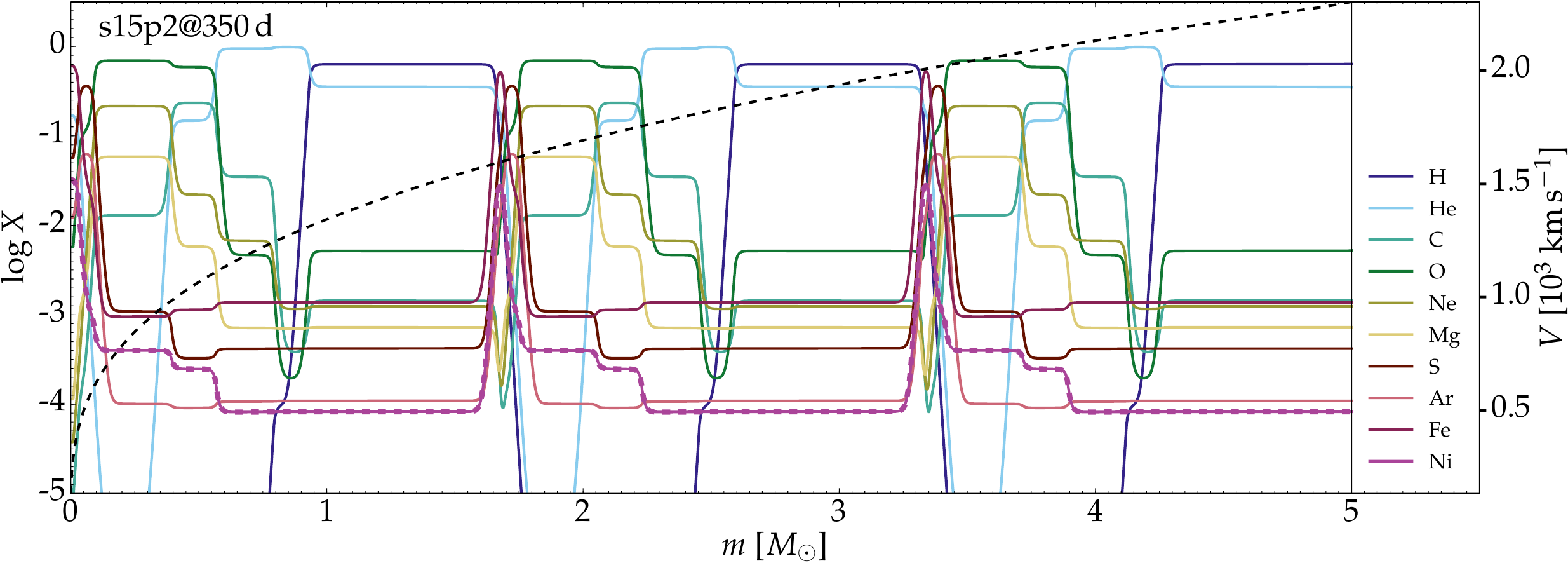}
\caption{Composition profile for the H-rich, red supergiant star explosion model s15p2 at 350\,d. We show the variation of the mass fraction versus Lagrangian mass in the inner 5\,\msun\ for species H, He, C, O, Ne, Mg, S, Ar, Fe, and Ni (with \iso{58}Ni shown as a thick dashed line) -- these abundances account for radioactive decay where appropriate. The thin dashed line corresponds to the velocity profile (see $y$-axis at right).
\label{fig_s15p2_init_comp}
}
\end{figure*}

\begin{figure*}[h]
\centering
\includegraphics[width=0.8\hsize]{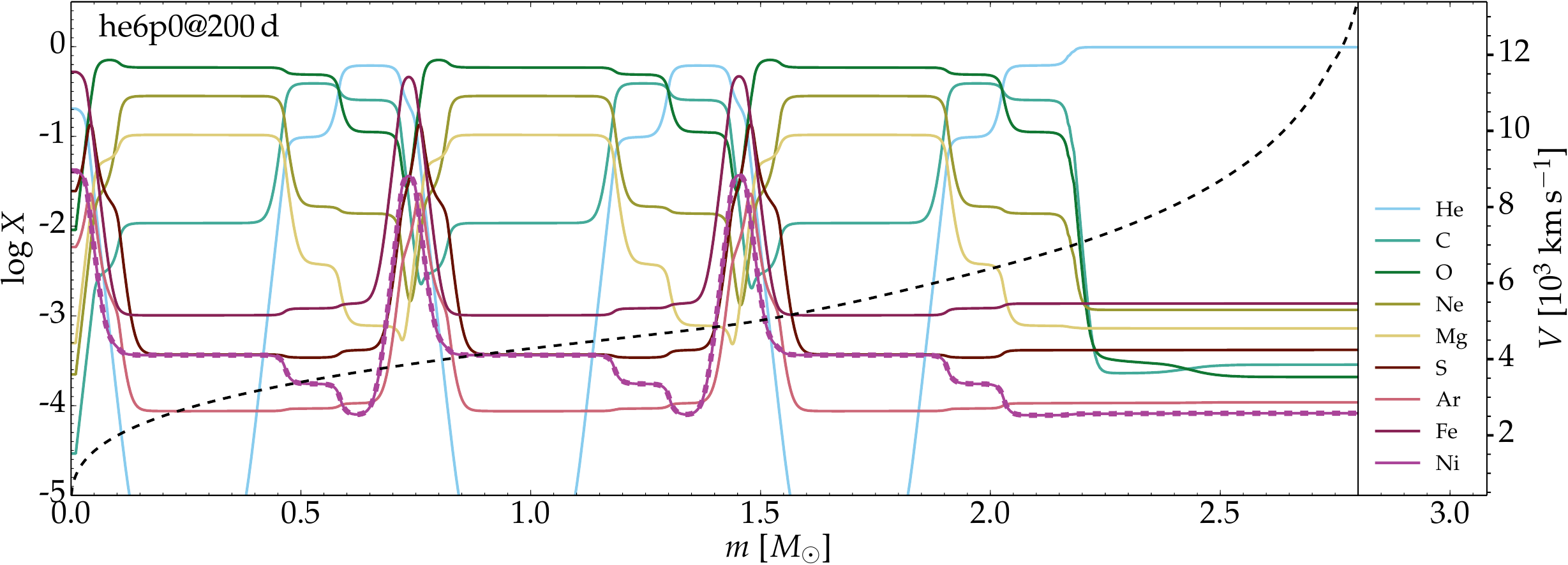}
\caption{Same as Fig.~\ref{fig_s15p2_init_comp} but for the H-deficient, He-star explosion model he6p0 at 200\,d.
\label{fig_he6p0_init_comp}
}
\end{figure*}

\begin{figure*}[h]
\centering
\includegraphics[width=0.33\hsize]{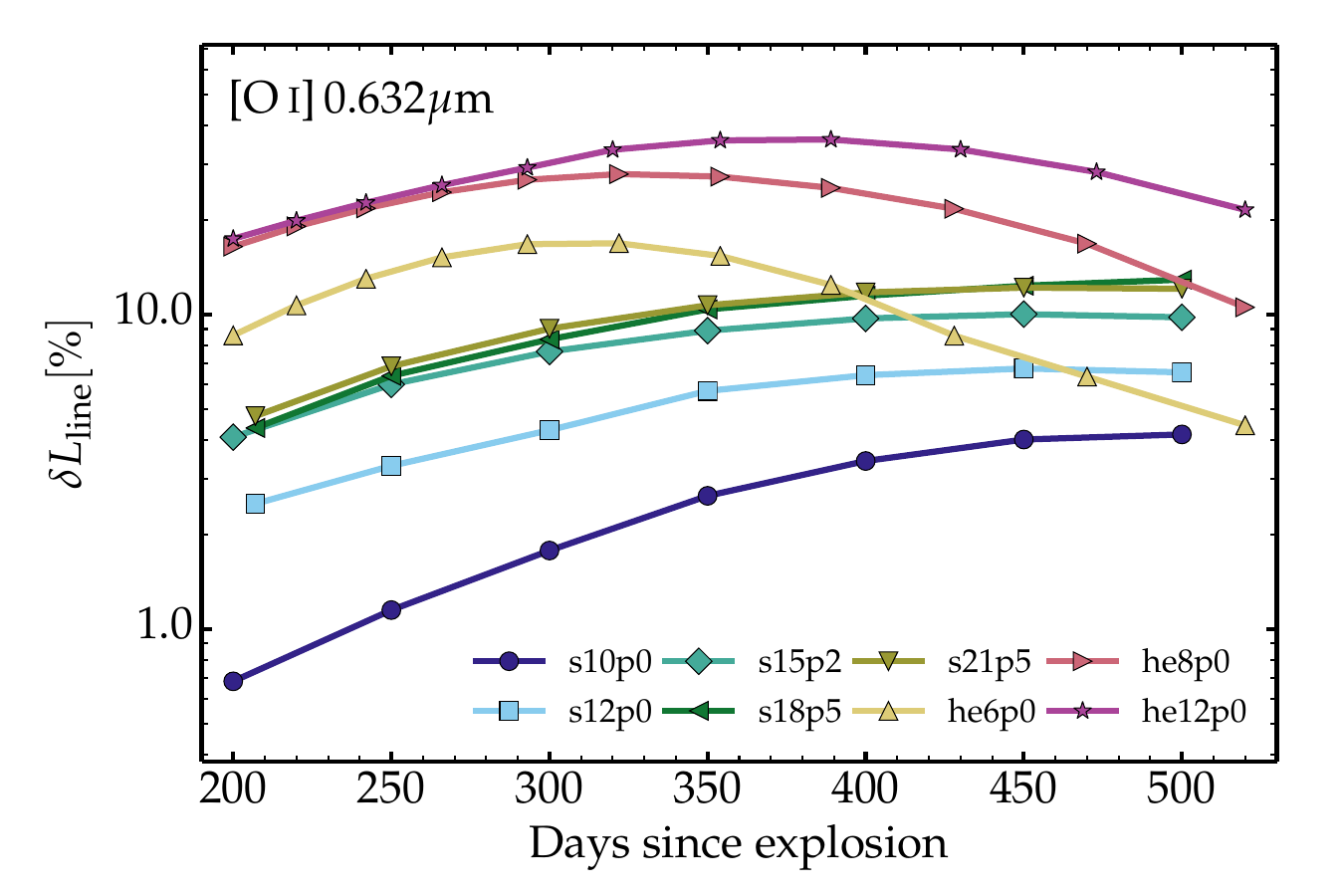}
\includegraphics[width=0.33\hsize]{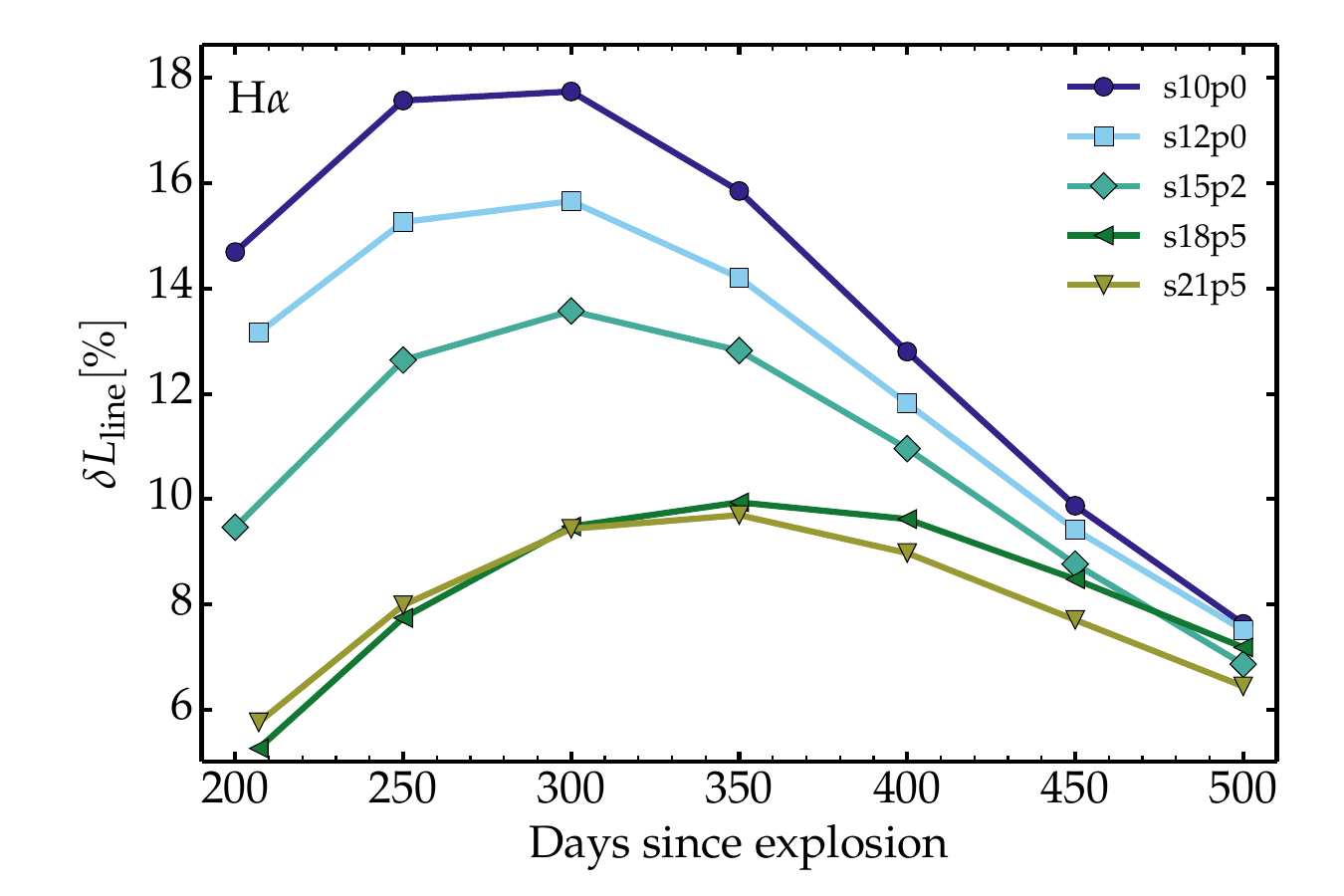}
\includegraphics[width=0.33\hsize]{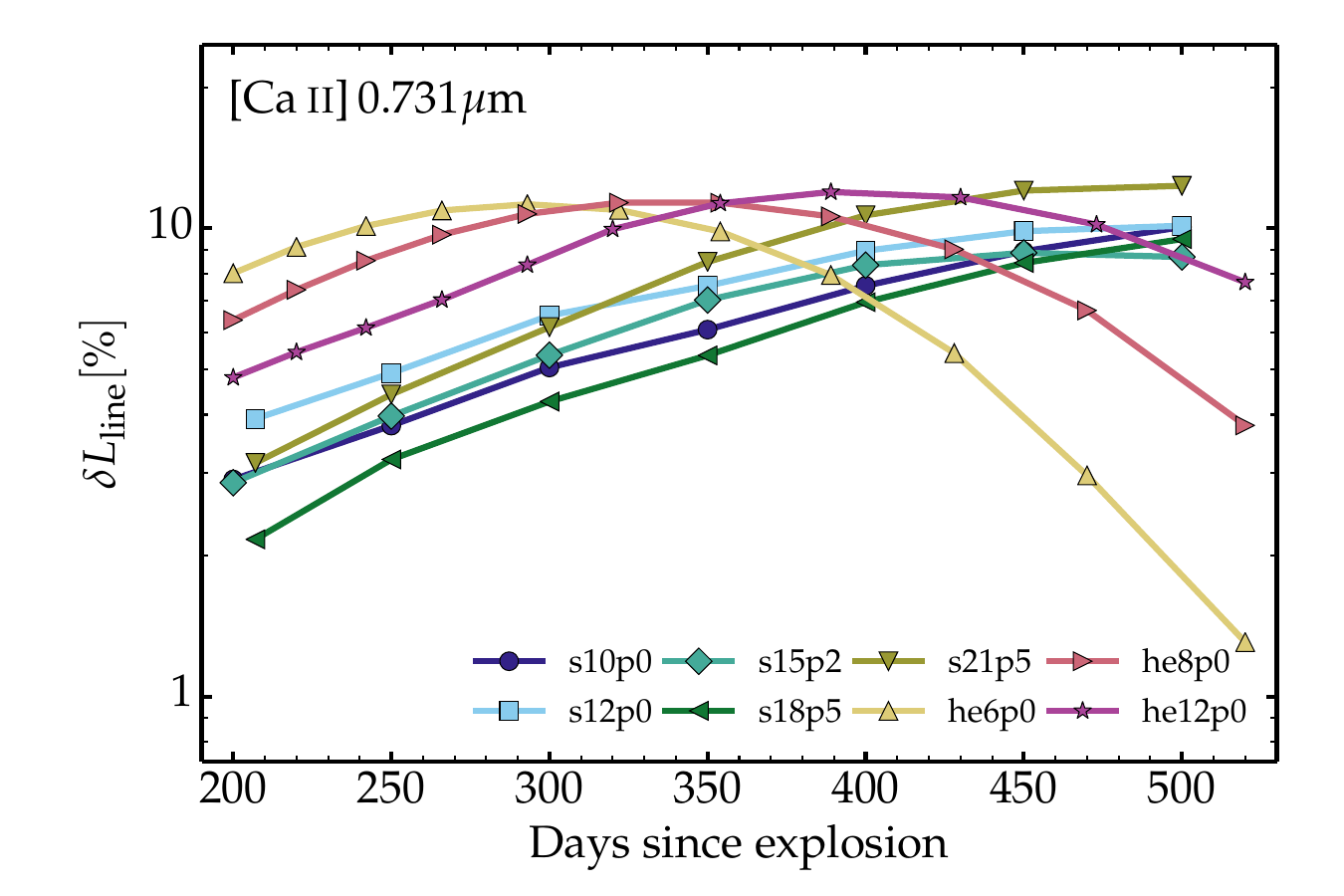}
\caption{Same as Fig.~\ref{fig_ir_lines} but for important line coolants present in the optical range. From left to right, we show the model results for \oidoub, H$\alpha$, and \caiidoub.}
\label{fig_opt_lines}
\end{figure*}

\begin{figure*}
   \centering
    \begin{subfigure}[b]{0.33\textwidth}
       \centering
       \includegraphics[width=\textwidth]{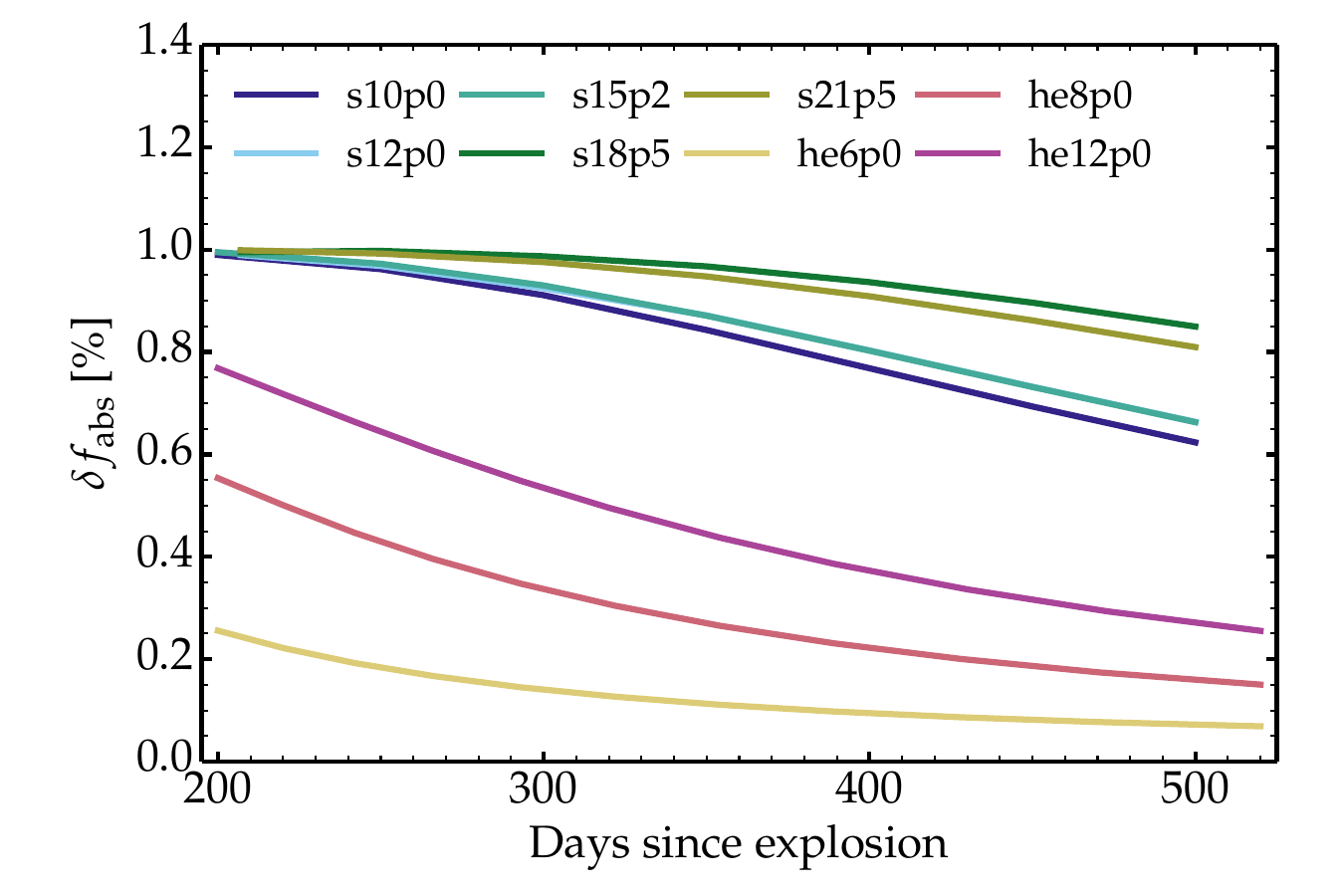}
    \end{subfigure}
    \hfill
    \begin{subfigure}[b]{0.33\textwidth}
       \centering
       \includegraphics[width=\textwidth]{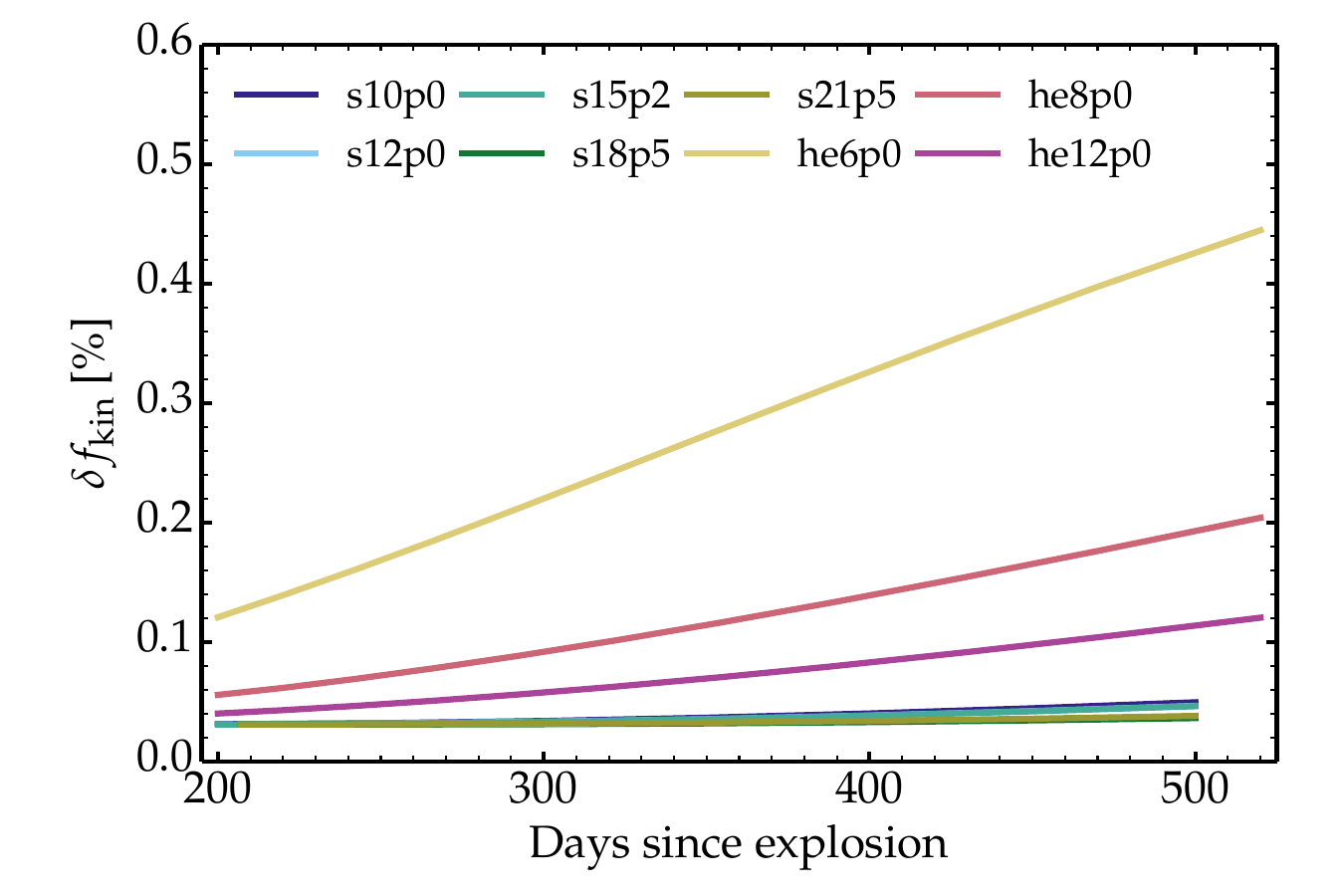}
    \end{subfigure}
     \hfill
    \begin{subfigure}[b]{0.33\textwidth}
       \centering
       \includegraphics[width=\textwidth]{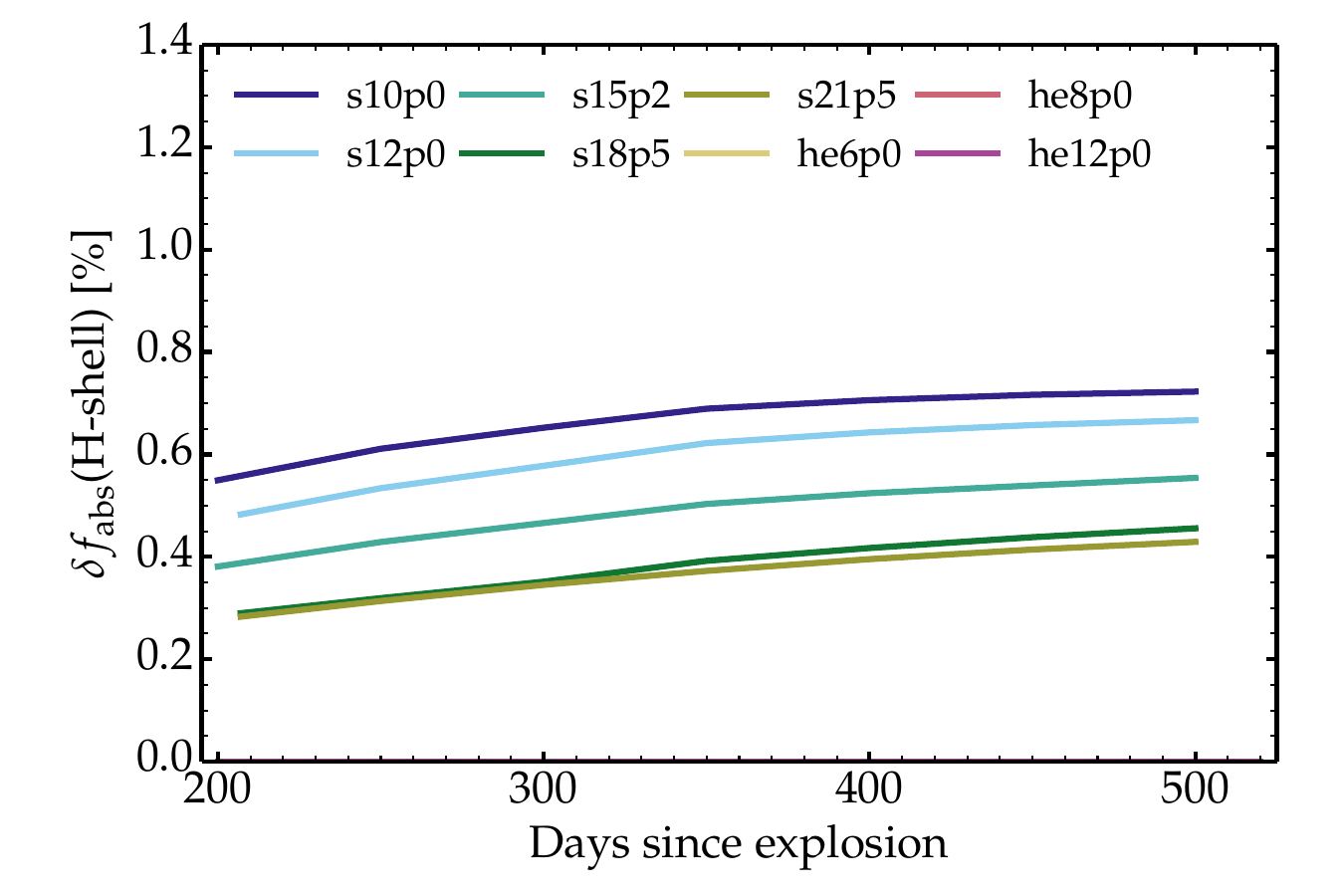}
    \end{subfigure}
     \hfill
    \begin{subfigure}[b]{0.33\textwidth}
       \centering
       \includegraphics[width=\textwidth]{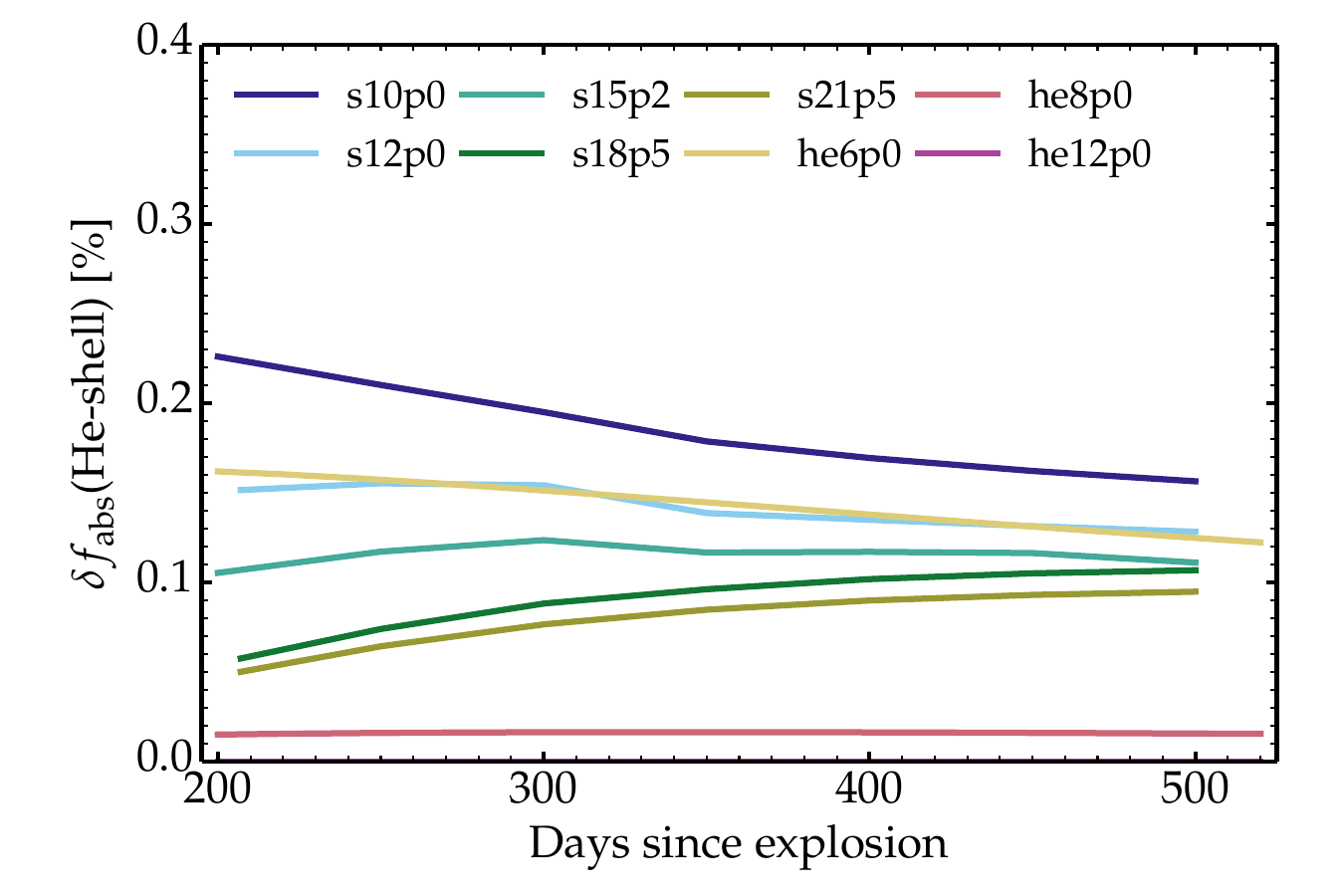}
    \end{subfigure}
    \hfill
    \begin{subfigure}[b]{0.33\textwidth}
       \centering
       \includegraphics[width=\textwidth]{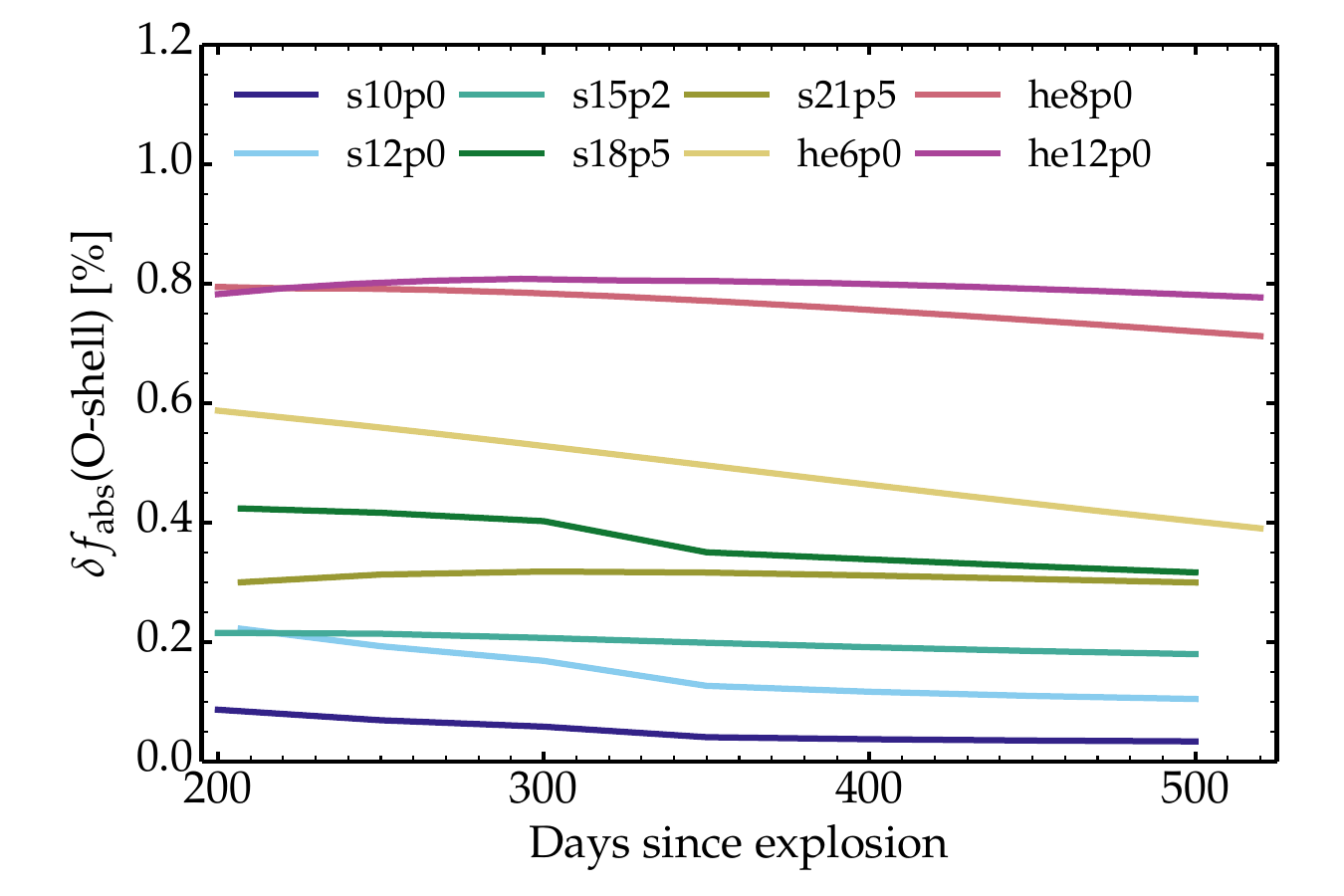}
    \end{subfigure}
     \hfill
    \begin{subfigure}[b]{0.33\textwidth}
       \centering
       \includegraphics[width=\textwidth]{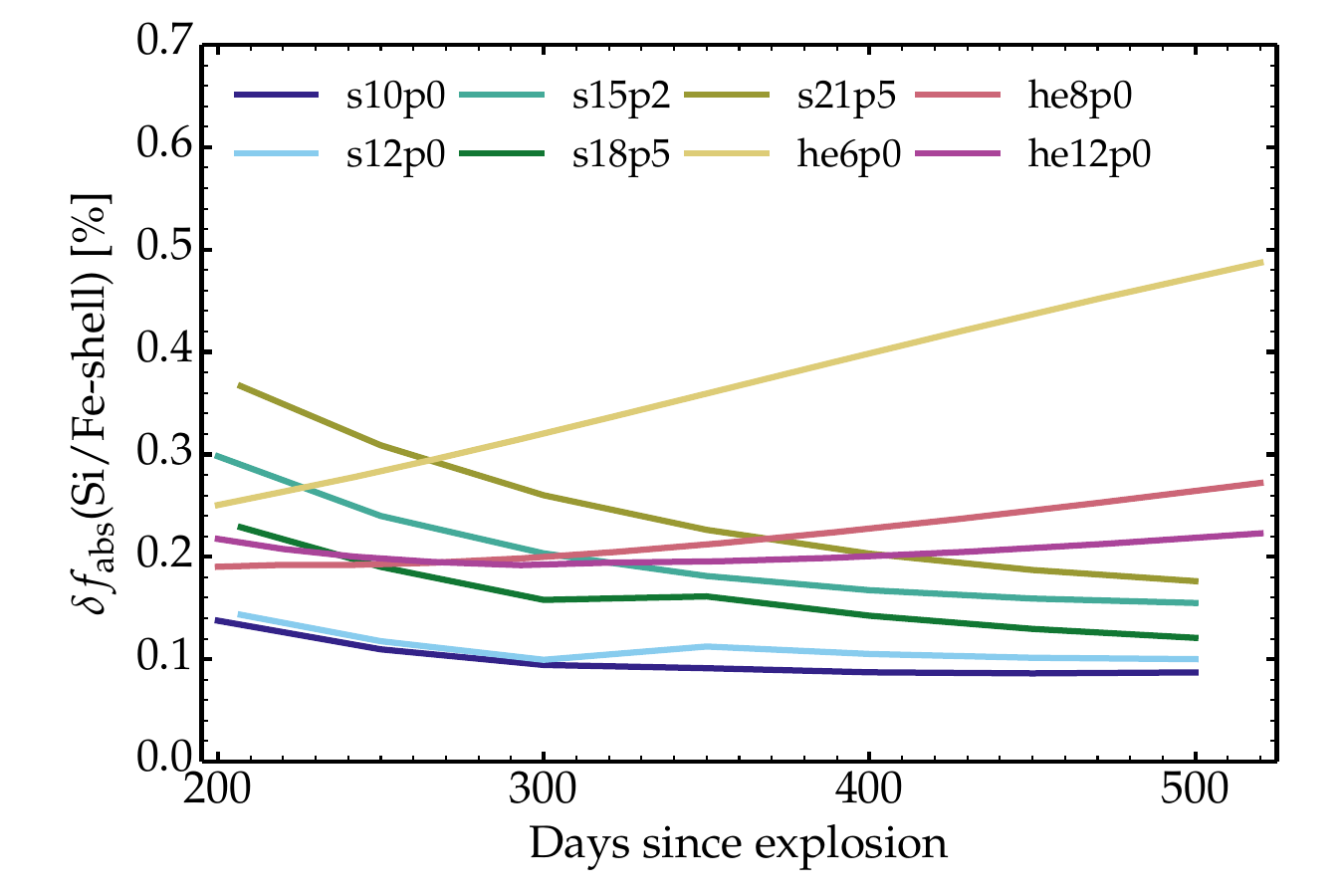}
    \end{subfigure}
\caption{Evolution of various powers as a function of the total decay power emitted (top-left panel) or absorbed (all other panels) by the ejecta for our set of simulations and over the time span from about 200 to about 500\,d after explosion. From left to right and top to bottom, we show the fractional power absorbed in the ejecta, that fraction coming from positrons, and the fractions absorbed in the H-rich (He-star explosion models are excluded since H deficient), He-rich, O-rich ,and Fe/Si-rich material.
\label{fig_frac_edep}
}
\end{figure*}

\begin{figure*}
   \centering
    \begin{subfigure}[b]{0.33\textwidth}
       \centering
       \includegraphics[width=\textwidth]{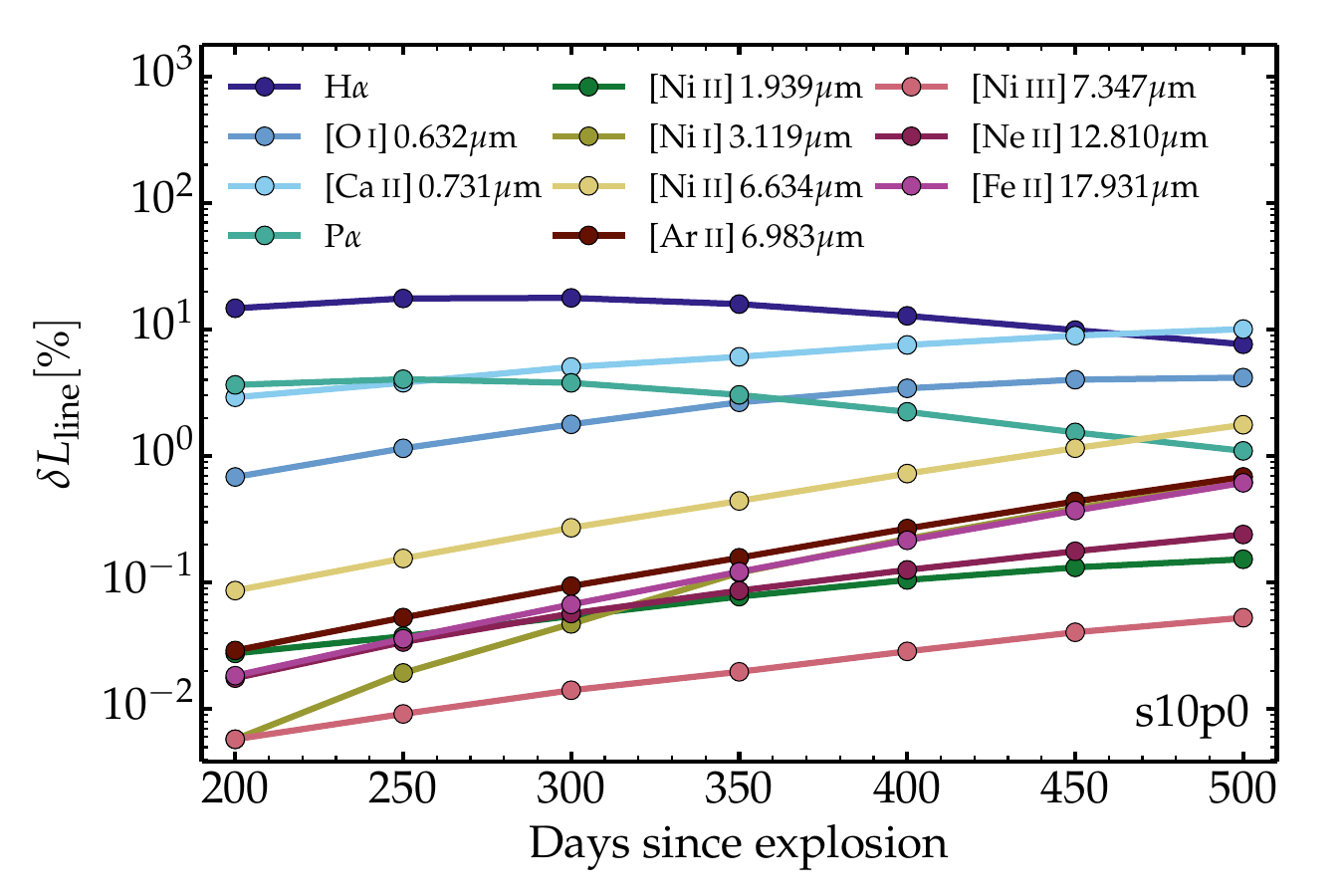}
    \end{subfigure}
     \hfill
    \begin{subfigure}[b]{0.33\textwidth}
       \centering
       \includegraphics[width=\textwidth]{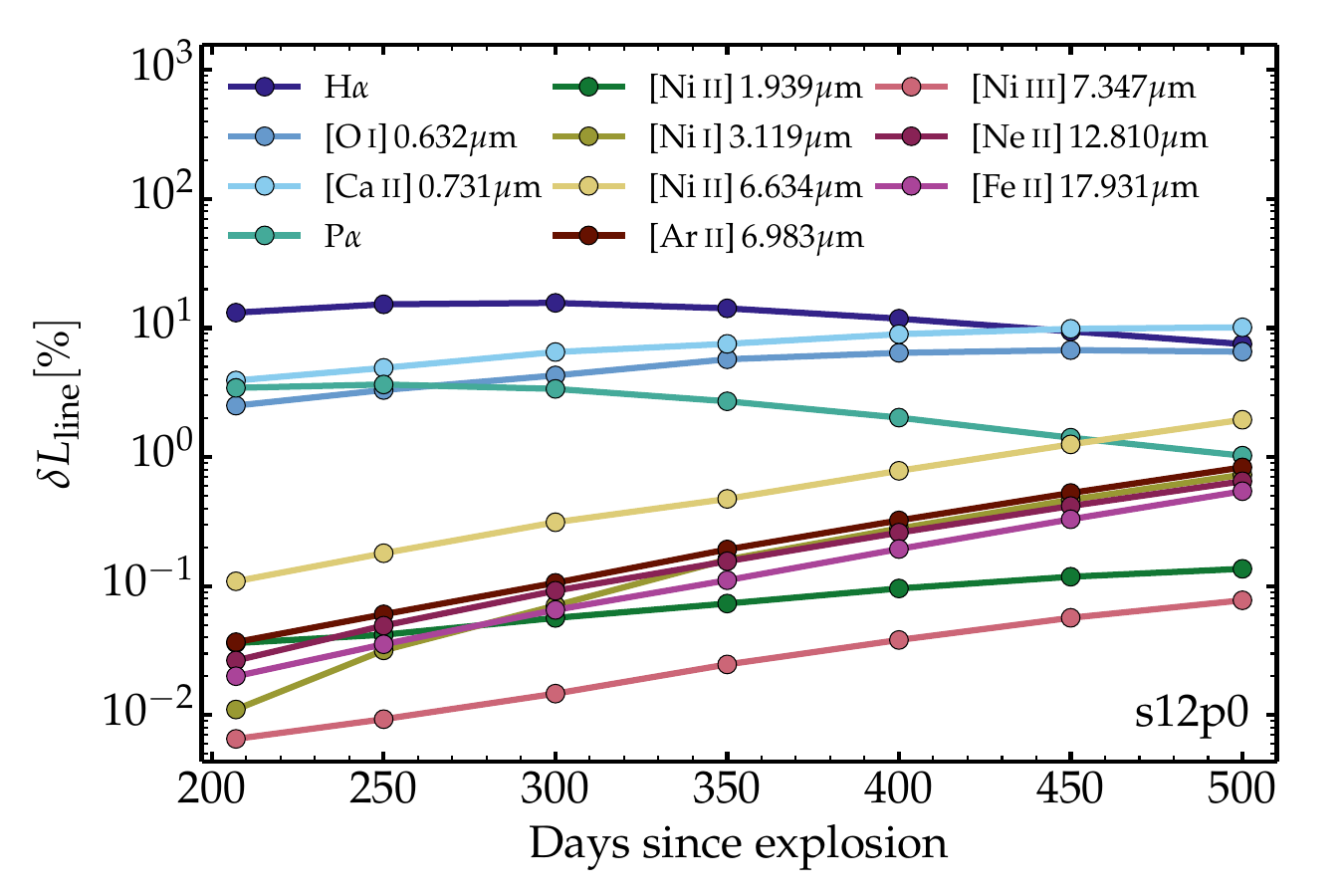}
    \end{subfigure}
    \hfill
    \begin{subfigure}[b]{0.33\textwidth}
       \centering
       \includegraphics[width=\textwidth]{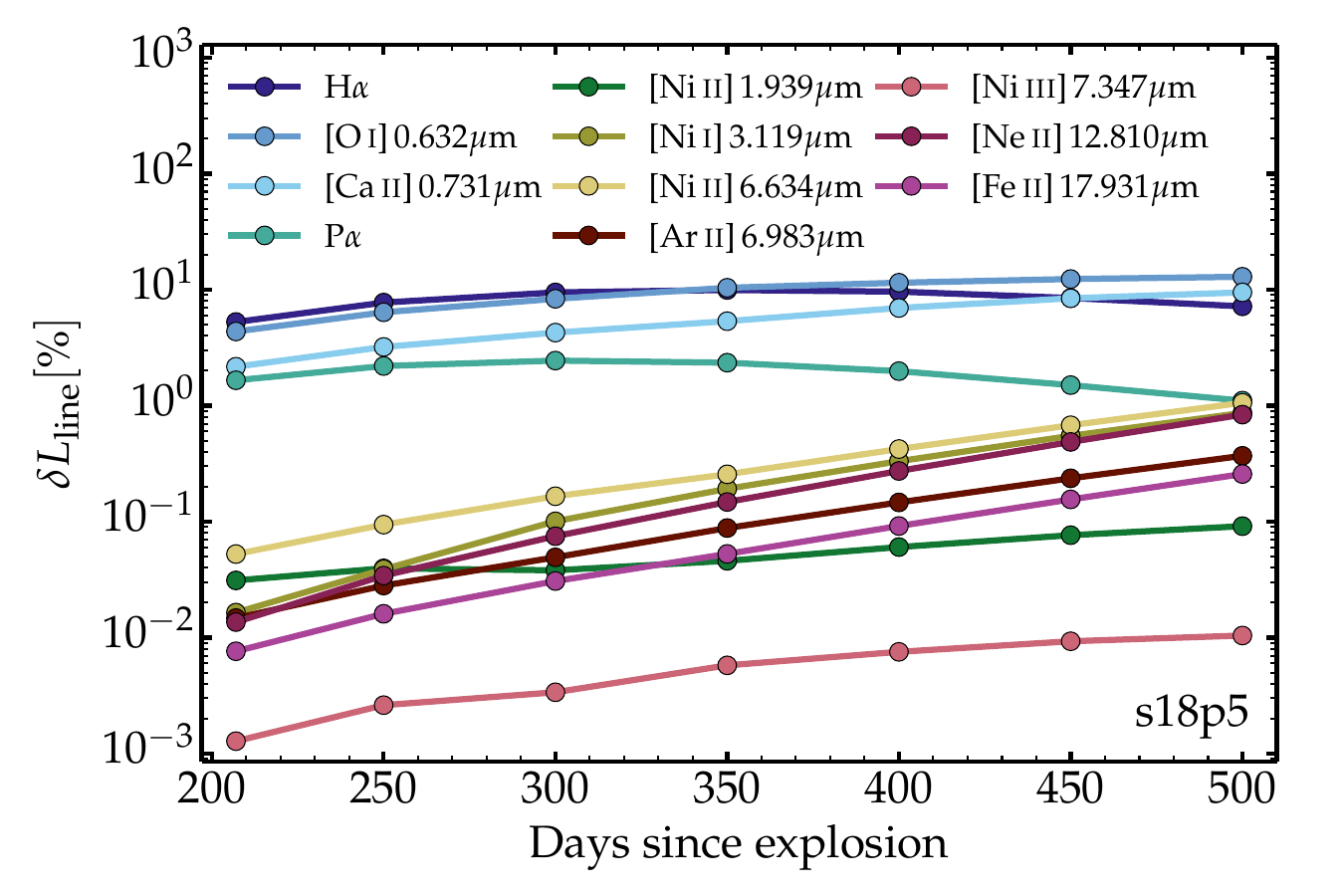}
    \end{subfigure}
     \hfill
    \begin{subfigure}[b]{0.33\textwidth}
       \centering
       \includegraphics[width=\textwidth]{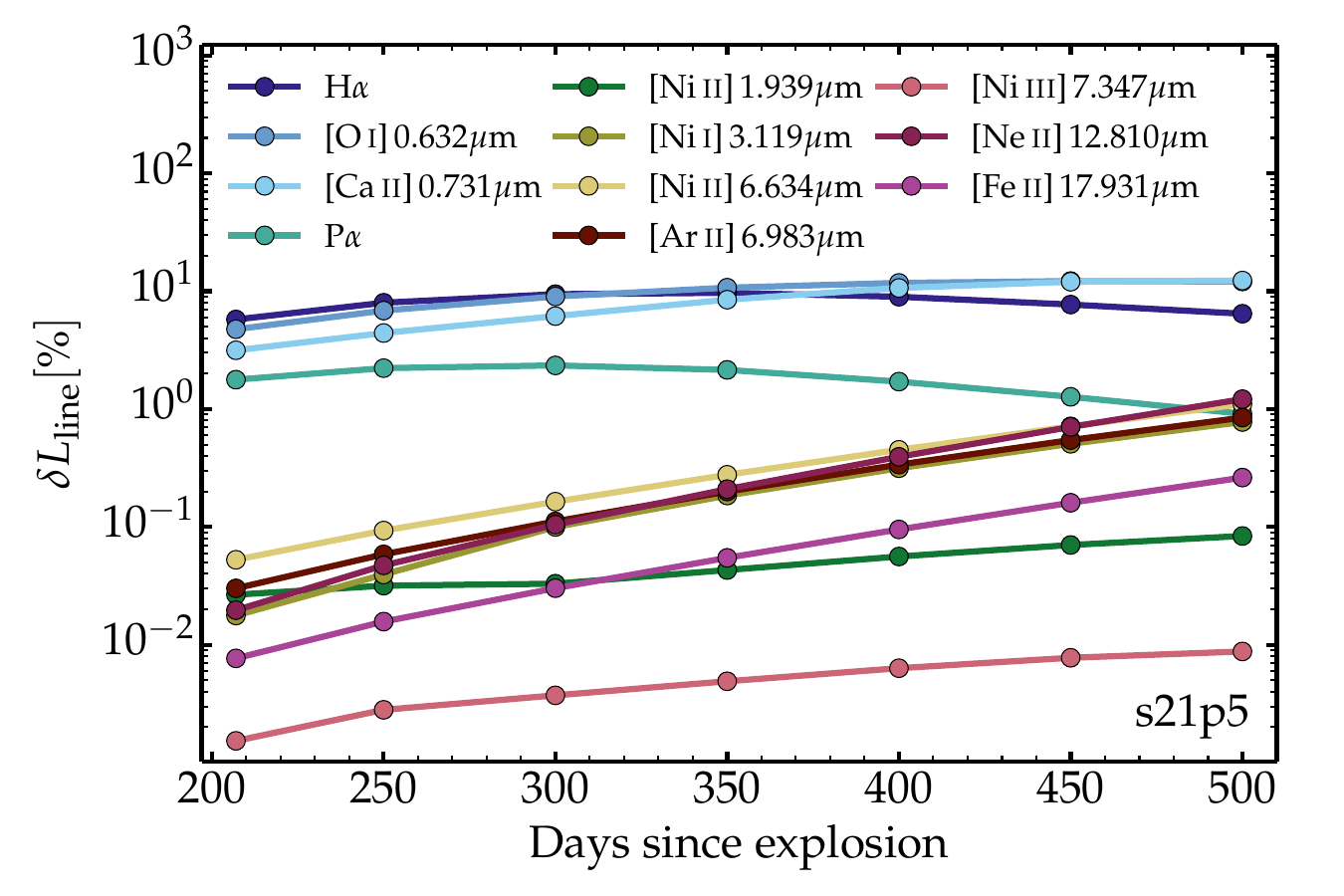}
    \end{subfigure}
     \hfill
    \begin{subfigure}[b]{0.33\textwidth}
       \centering
       \includegraphics[width=\textwidth]{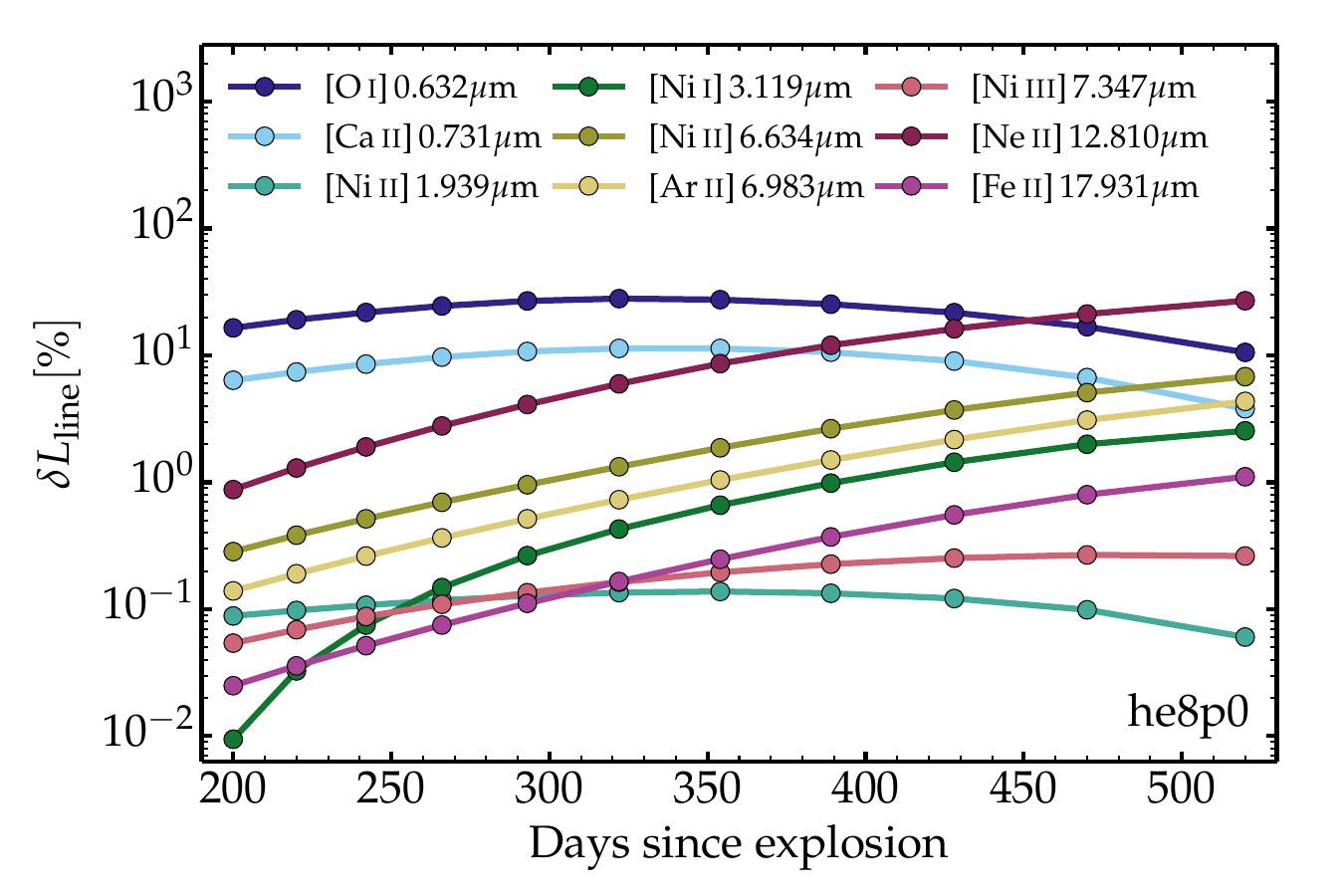}
    \end{subfigure}
     \hfill
    \begin{subfigure}[b]{0.33\textwidth}
       \centering
       \includegraphics[width=\textwidth]{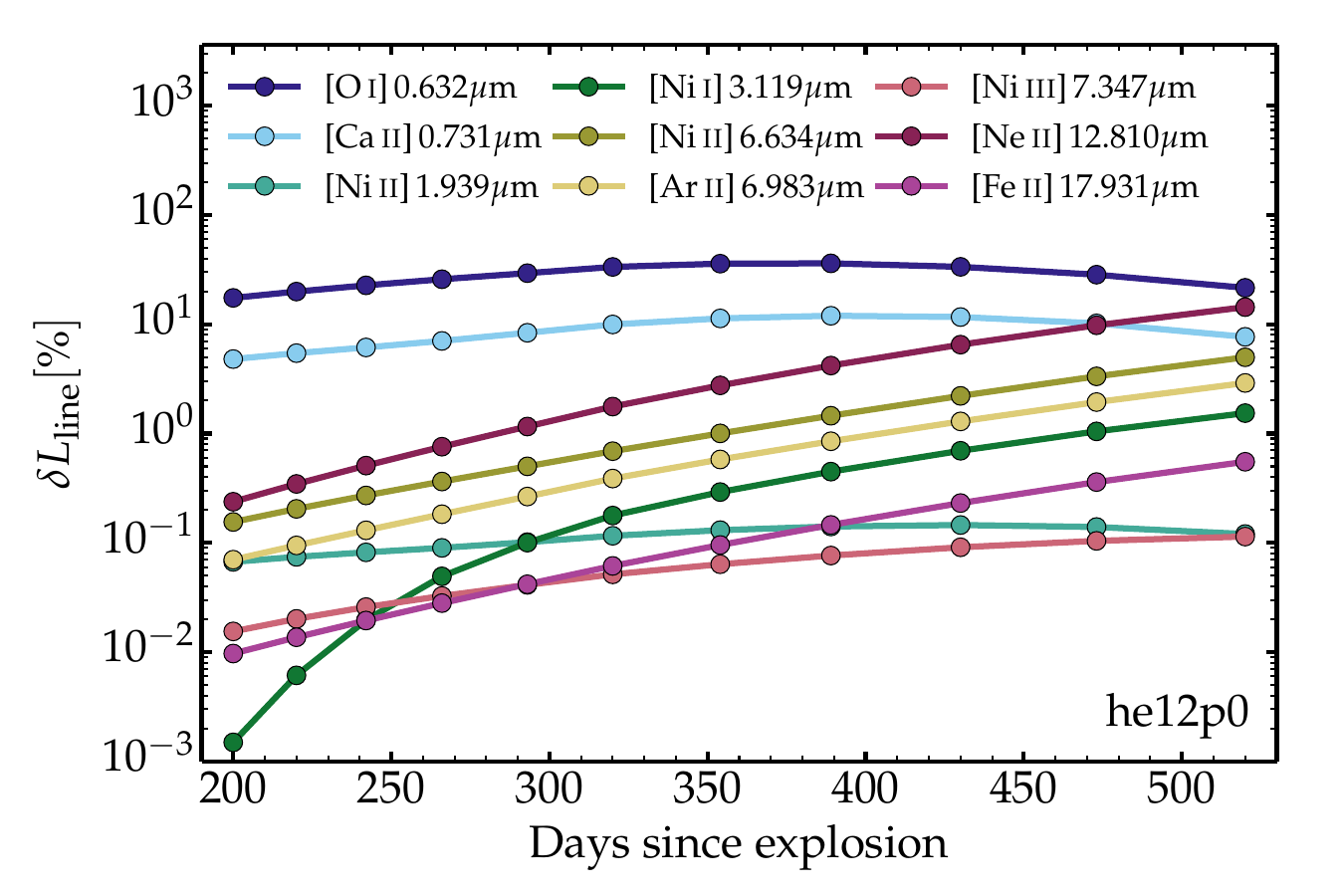}
    \end{subfigure}
\caption{Evolution of a few strong coolants across the optical and infrared spectral ranges for models other than s15p2 (shown in the right panel of Fig.~\ref{fig_cum_lum_s15p2}) and he6p0 (shown in the right panel of Fig.~\ref{fig_cum_lum_he6p0}).
\label{fig_line_evol_all}
}
\end{figure*}

\section{Comparison to earlier work}
\label{sect_comp}

We can compare the results for our type II SN models to those of \citet{jerkstrand_04et_12} who employed similar explosion models (12--19\,\msun\ stars from \citealt{WH07} instead of the models from \citealt{sukhbold_ccsn_16} used here) -- similar comparisons have been presented in \citet{dessart_sn2p_21} but mostly for the optical diagnostics. We neglected molecules and dust, whereas \citet{jerkstrand_04et_12} treated them, although in an approximate manner: molecular cooling was introduced as a prescribed ``thermostat'' for the material from the O/Si and O/C shells where SiO and CO are expected to form, and molecular and dust emission were both added as a separate infrared emission component to the emergent spectrum (this was later improved by modeling molecules with a molecular network and then treating molecules with the gas in the NLTE solution \citep{liljegren_co_20} or in the full radiative transfer solution \citep{liljegren_ibc_mol_23}). Chemical mixing was treated with a comparable aim (i.e., by allowing for macroscopic mixing without microscopic mixing) but this is achieved in the Monte-Carlo code \sumo\ through a virtual grid in which the inner, mixed ejecta is composed of spherical clumps of distinct composition and density and randomly distributed within some prescribed velocity \citep{jerkstrand_87a_11}, whereas in the grid-based code \cmfgen, this is achieved by means of a shuffling in mass space of spherical shells from the 1D, unmixed ejecta \citep{dessart_shuffle_20}. Line overlap and blanketing is taken into account in \cmfgen, whereas the Sobolev approximation is used in \sumo. The temperature solution or the sources of atomic data may also deviate between the two codes. The shuffled-shell structure in the \cmfgen\ input models forces photons emitted within the inner ejecta layers to cross all overlying shells before escaping and may thus overestimate optical-depth effects relative to \sumo\ models.

One difference is the prediction by \sumo\ of sizable variations in density amongst clumps of distinct composition, for example when modeling nebular-phase spectra of Type II SNe such as 2004et \citep{jerkstrand_04et_12}. Similar density variations have been inferred by \citet{li_87A_93} or \citet{spyromilio_pinto_91} in the context of SN\,1987A, although \citet{KF98b} find little sensitivity of their results to the filling factors given to material of distinct composition. The impact of clumping was studied by \citet{dessart_sn2p_21} but was found to be limited unless it induced an ionization shift. In the largely recombined ejecta of their Type II SN models, \citet{dessart_sn2p_21} found that even for a factor of ten in clumping (i.e., a uniform 10\,\% filling factor), the \oidoub\ line flux decreased by order 10\,\%, causing a small strengthening of Na\one\,D and Mg\one]\,4571\,\AA. \citet{dessart_sn2p_21} also implemented a radial stretching of the Fe-rich shells to mimic the \nifs-bubble effect but this was found to have negligible effect on the resulting Type II SN spectra, in contrast to the findings of \citet{li_87A_93} or \citet{jerkstrand_04et_12}. Using smooth ejecta models, \citet{dessart_snibc_21} reproduced satisfactorily the optical observations of a number of well observed Type II SNe one year after explosion, including for example SN\,1987A, SN\,2012aw, or SN\,2013ej. One might argue that infrared metal lines are better diagnostics of ejecta clumping, but the unclumped s15p2 model presented here was found to yield a satisfactory match to infrared metal lines observed in Type II SN\,2024ggi \citep{dessart_24ggi_25}.

\citet{dessart_snibc_21} found a greater impact of clumping on Type Ibc model spectra at 200\,d because these ejecta were partially ionized when smooth, such that clumping enhanced recombination, for example quenching Fe\two\ emission for the benefit of \oidoub\ emission from the O-rich material. However, when clumping was raised beyond a factor of a few, such models overpredicted the strength of Na\one\,D and Mg\one]\,4571\,\AA. As we surmise in Footnote~\ref{footnote_nibubble}, the density variations arising from the \nifs-bubble effect in SNe Ibc should be weaker than in SNe II. Recently, \citet{ergon_20acat_24} advocated a strong \nifs-bubble effect and clumping in the Type IIb SN\,2020acat to explain its nonstandard, overly luminous peak and short rise time, which is perplexing. One alternative to extreme clumping in SN\,2020acat may be large scale ejecta asymmetry and high-velocity \nifs\ along our line of sight.

Different studies also predict distinct emitting regions for a variety of lines. \citet{dessart_sn2p_21} find that \caiidoub\ emission arises from the Fe/Si-rich material, whereas \citet{jerkstrand_04et_12} find that it arises from primordial Ca in the H-rich gas, as found earlier by \citet{li_mccray_ca2_93} and \citet{KF98a}. When modeling SN Ibc ejecta, which are H deficient, \citet{dessart_snibc_21} found the same, Fe/Si-rich material at the origin of the \caiidoub\ emission. Another difference is the \ariimir\ emission, which was found here to arise from the Si-rich material, whereas \citet{jerkstrand_04et_12} found that it arises from the He/C shell. \neiifs\ is predicted somewhat stronger in our model s15p2 than in the 15\,\msun\ model of \citet{jerkstrand_04et_12}, with an increasing offset of a factor of 4--5 at 500\,d, although this may in part be due to the use of a gray dust in \citet{jerkstrand_04et_12}, something that may overestimate the dust optical-depth in the mid-infrared. 

Understanding the origin of these differences is beyond the scope of this paper but the smooth-ejecta model s15p2 presented here yields a satisfactory match to the observed optical-to-infrared nebular-phase spectra of SN\,2024ggi \citep{dessart_24ggi_25}, which is also a close analog in the optical range and at nebular times of SN\,2004et (Dessart et al., in prep). What the present modeling indicates is not that such ejecta are not clumped, but rather than the impact of clumping on nebular-phase spectra is typically small (and function of ejecta ionization) in the simulations with \cmfgen\ based on shuffled-shell ejecta and thus justifies conducting a study with smooth ejecta.  However, it does raise a concern about the robustness of the inferred ejecta clumping from nebular-phase spectral modeling, since if it holds, it should invalidate models that ignore clumping. This requires further study.


\begin{thebibliography}{59}
\expandafter\ifx\csname natexlab\endcsname\relax\def\natexlab#1{#1}\fi

\bibitem[{{Ashall} {et~al.}(2024){Ashall}, {Hoeflich}, {Baron}, {Shahbandeh},
  {DerKacy}, {Medler}, {Shappee}, {Tucker}, {Fereidouni}, {Mera}, {Andrews},
  {Baade}, {Bostroem}, {Brown}, {Burns}, {Burrow}, {Cikota}, {de Jaeger}, {Do},
  {Dong}, {Dominguez}, {Fox}, {Galbany}, {Hsiao}, {Krisciunas}, {Khaghani},
  {Kumar}, {Lu}, {Maund}, {Mazzali}, {Morrell}, {Patat}, {Pfeffer}, {Phillips},
  {Schmidt}, {Stangl}, {Stevens}, {Stritzinger}, {Suntzeff}, {Telesco}, {Wang},
  \& {Yang}}]{ashall_21aefx_24}
{Ashall}, C., {Hoeflich}, P., {Baron}, E., {et~al.} 2024, \apj, 975, 203

\bibitem[{{Basko}(1994)}]{basko_56ni_94}
{Basko}, M. 1994, \apj, 425, 264

\bibitem[{{Blondin} {et~al.}(2023){Blondin}, {Dessart}, {Hillier},
  {Ramsbottom}, \& {Storey}}]{blondin_21aefx_23}
{Blondin}, S., {Dessart}, L., {Hillier}, D.~J., et al. 2023, \aap, 678, A170

\bibitem[{{DerKacy} {et~al.}(2023){DerKacy}, {Ashall}, {Hoeflich}, {Baron},
  {Shappee}, {Baade}, {Andrews}, {Bostroem}, {Brown}, {Burns}, {Burrow},
  {Cikota}, {de Jaeger}, {Do}, {Dong}, {Dominguez}, {Galbany}, {Hsiao},
  {Karamehmetoglu}, {Krisciunas}, {Kumar}, {Lu}, {Evans}, {Maund}, {Mazzali},
  {Medler}, {Morrell}, {Patat}, {Phillips}, {Shahbandeh}, {Stangl}, {Stevens},
  {Stritzinger}, {Suntzeff}, {Telesco}, {Tucker}, {Valenti}, {Wang}, {Yang},
  {Jha}, \& {Kwok}}]{derkacy_21aefx_23}
{DerKacy}, J.~M., {Ashall}, C., {Hoeflich}, P., {et~al.} 2023, \apjl, 945, L2

\bibitem[{{Dessart}(2024)}]{dessart_pm_24}
{Dessart}, L. 2024, \aap, 692, A204

\bibitem[{{Dessart} {et~al.}(2023{\natexlab{a}}){Dessart}, {Guti{\'e}rrez},
  {Kuncarayakti}, {Fox}, \& {Filippenko}}]{dessart_late_23}
{Dessart}, L., {Guti{\'e}rrez}, C.~P., {Kuncarayakti}, H., et al. 2023{\natexlab{a}}, \aap, 675, A33

\bibitem[{{Dessart} \& {Hillier}(2020)}]{dessart_shuffle_20}
{Dessart}, L. \& {Hillier}, D.~J. 2020, \aap, 643, L13

\bibitem[{{Dessart} {et~al.}(2021{\natexlab{a}}){Dessart}, {Hillier},
  {Sukhbold}, {Woosley}, \& {Janka}}]{dessart_snibc_21}
{Dessart}, L., {Hillier}, D.~J., {Sukhbold}, T., et al. 2021{\natexlab{a}}, \aap, 656, A61

\bibitem[{{Dessart} {et~al.}(2021{\natexlab{b}}){Dessart}, {Hillier},
  {Sukhbold}, {Woosley}, \& {Janka}}]{dessart_sn2p_21}
{Dessart}, L., {Hillier}, D.~J., {Sukhbold}, T., et al. 2021{\natexlab{b}}, \aap, 652, A64

\bibitem[{{Dessart} {et~al.}(2023{\natexlab{b}}){Dessart}, {Hillier},
  {Woosley}, \& {Kuncarayakti}}]{dessart_snibc_23}
{Dessart}, L., {Hillier}, D.~J., {Woosley}, S.~E., et al.
  2023{\natexlab{b}}, \aap, 677, A7

\bibitem[{{Dessart} {et~al.}(2025){Dessart}, {Kotak}, {Jacobson-Galan}, {Das},
  {Fremling}, {Kasliwal}, {Qin}, \& {Rose}}]{dessart_24ggi_25}
{Dessart}, L., {Kotak}, R., {Jacobson-Galan}, W., {et~al.} 2025,
  arXiv:2507.05803

\bibitem[{{Dessart} {et~al.}(2022){Dessart}, {Prieto}, {Hillier},
  {Kuncarayakti}, \& {Hueichapan}}]{D22_lsst}
{Dessart}, L., {Prieto}, J.~L., {Hillier}, D.~J., et al. 2022, \aap, 666, L14

\bibitem[{{Ergon} {et~al.}(2024){Ergon}, {Lundqvist}, {Fransson},
  {Kuncarayakti}, {Das}, {De}, {Ferrari}, {Fremling}, {Medler}, {Maeda},
  {Pastorello}, {Sollerman}, \& {Stritzinger}}]{ergon_20acat_24}
{Ergon}, M., {Lundqvist}, P., {Fransson}, C., {et~al.} 2024, \aap, 683, A241

\bibitem[{{Ertl} {et~al.}(2020){Ertl}, {Woosley}, {Sukhbold}, \&
  {Janka}}]{ertl_ibc_20}
{Ertl}, T., {Woosley}, S.~E., {Sukhbold}, T., \& {Janka}, H.~T. 2020, \apj,
  890, 51

\bibitem[{{Fransson} \& {Chevalier}(1989)}]{fransson_chevalier_89}
{Fransson}, C. \& {Chevalier}, R.~A. 1989, \apj, 343, 323

\bibitem[{{Gerardy} {et~al.}(2002){Gerardy}, {Fesen}, {Nomoto}, {Maeda},
  {Hoflich}, \& {Wheeler}}]{gerardy_co_00ew_02}
{Gerardy}, C.~L., {Fesen}, R.~A., {Nomoto}, K., {et~al.} 2002, \pasj, 54, 905

\bibitem[{{Hillier} \& {Dessart}(2012)}]{HD12}
{Hillier}, D.~J. \& {Dessart}, L. 2012, \mnras, 424, 252

\bibitem[{{Hillier} \& {Miller}(1998)}]{hm98}
{Hillier}, D.~J. \& {Miller}, D.~L. 1998, \apj, 496, 407

\bibitem[{{Hsiao} {et~al.}(2019){Hsiao}, {Phillips}, {Marion}, {Kirshner},
  {Morrell}, {Sand}, {Burns}, {Contreras}, {Hoeflich}, {Stritzinger},
  {Valenti}, {Anderson}, {Ashall}, {Baltay}, {Baron}, {Banerjee}, {Davis},
  {Diamond}, {Folatelli}, {Freedman}, {F{\"o}rster}, {Galbany}, {Gall},
  {Gonz{\'a}lez-Gait{\'a}n}, {Goobar}, {Hamuy}, {Holmbo}, {Kasliwal},
  {Krisciunas}, {Kumar}, {Lidman}, {Lu}, {Nugent}, {Perlmutter}, {Persson},
  {Piro}, {Rabinowitz}, {Roth}, {Ryder}, {Schmidt}, {Shahbandeh}, {Suntzeff},
  {Taddia}, {Uddin}, \& {Wang}}]{hsiao_nir_19}
{Hsiao}, E.~Y., {Phillips}, M.~M., {Marion}, G.~H., {et~al.} 2019, \pasp, 131,
  014002

\bibitem[{{Hunter} {et~al.}(2009){Hunter}, {Valenti}, {Kotak}, {Meikle},
  {Taubenberger}, {Pastorello}, {Benetti}, {Stanishev}, {Smartt}, {Trundle},
  {Arkharov}, {Bufano}, {Cappellaro}, {Di Carlo}, {Dolci}, {Elias-Rosa},
  {Frandsen}, {Fynbo}, {Hopp}, {Larionov}, {Laursen}, {Mazzali}, {Navasardyan},
  {Ries}, {Riffeser}, {Rizzi}, {Tsvetkov}, {Turatto}, \&
  {Wilke}}]{hunter_etal_07gr_07bi}
{Hunter}, D.~J., {Valenti}, S., {Kotak}, R., {et~al.} 2009, \aap, 508, 371

\bibitem[{{Jerkstrand} {et~al.}(2011){Jerkstrand}, {Fransson}, \&
  {Kozma}}]{jerkstrand_87a_11}
{Jerkstrand}, A., {Fransson}, C., \& {Kozma}, C. 2011, \aap, 530, A45

\bibitem[{{Jerkstrand} {et~al.}(2012){Jerkstrand}, {Fransson}, {Maguire},
  {Smartt}, {Ergon}, \& {Spyromilio}}]{jerkstrand_04et_12}
{Jerkstrand}, A., {Fransson}, C., {Maguire}, K., {et~al.} 2012, \aap, 546, A28

\bibitem[{{Jerkstrand} {et~al.}(2015){Jerkstrand}, {Smartt}, {Sollerman},
  {Inserra}, {Fraser}, {Spyromilio}, {Fransson}, {Chen}, {Barbarino},
  {Dall'Ora}, {Botticella}, {Della Valle}, {Gal-Yam}, {Valenti}, {Maguire},
  {Mazzali}, \& {Tomasella}}]{jerkstrand_ni_15}
{Jerkstrand}, A., {Smartt}, S.~J., {Sollerman}, J., {et~al.} 2015, \mnras, 448,
  2482

\bibitem[{{Kotak} {et~al.}(2006){Kotak}, {Meikle}, {Pozzo}, {van Dyk},
  {Farrah}, {Fesen}, {Filippenko}, {Foley}, {Fransson}, {Gerardy},
  {H{\"o}flich}, {Lundqvist}, {Mattila}, {Sollerman}, \&
  {Wheeler}}]{kotak_05af_06}
{Kotak}, R., {Meikle}, P., {Pozzo}, M., {et~al.} 2006, \apjl, 651, L117

\bibitem[{{Kotak} {et~al.}(2005){Kotak}, {Meikle}, {van Dyk}, {H{\"o}flich}, \&
  {Mattila}}]{kotak_04dj_05}
{Kotak}, R., {Meikle}, P., {van Dyk}, S.~D., et al. 2005, \apjl, 628, L123

\bibitem[{{Kotak} {et~al.}(2009){Kotak}, {Meikle}, {Farrah}, {Gerardy},
  {Foley}, {Van Dyk}, {Fransson}, {Lundqvist}, {Sollerman}, {Fesen},
  {Filippenko}, {Mattila}, {Silverman}, {Andersen}, {H{\"o}flich}, {Pozzo}, \&
  {Wheeler}}]{kotak_04et_09}
{Kotak}, R., {Meikle}, W.~P.~S., {Farrah}, D., {et~al.} 2009, \apj, 704, 306

\bibitem[{{Kozma} \& {Fransson}(1998{\natexlab{a}})}]{KF98a}
{Kozma}, C. \& {Fransson}, C. 1998{\natexlab{a}}, \apj, 496, 946

\bibitem[{{Kozma} \& {Fransson}(1998{\natexlab{b}})}]{KF98b}
{Kozma}, C. \& {Fransson}, C. 1998{\natexlab{b}}, \apj, 497, 431

\bibitem[{{Kuchner} {et~al.}(1994){Kuchner}, {Kirshner}, {Pinto}, \&
  {Leibundgut}}]{kuchner_co3_94}
{Kuchner}, M.~J., {Kirshner}, R.~P., {Pinto}, P.~A., et al. 1994,
  \apjl, 426, L89

\bibitem[{{Kwok} {et~al.}(2023){Kwok}, {Jha}, {Temim}, {Fox}, {Larison},
  {Camacho-Neves}, {Brenner Newman}, {Pierel}, {Foley}, {Andrews}, {Badenes},
  {Barna}, {Bostroem}, {Deckers}, {Fl{\"o}rs}, {Garnavich}, {Graham}, {Graur},
  {Hosseinzadeh}, {Howell}, {Hughes}, {Johansson}, {Kendrew}, {Kerzendorf},
  {Maeda}, {Maguire}, {McCully}, {O'Brien}, {Rest}, {Sand}, {Shahbandeh},
  {Strolger}, {Szalai}, {Ashall}, {Baron}, {Burns}, {DerKacy}, {Evans},
  {Fisher}, {Galbany}, {Hoeflich}, {Hsiao}, {de Jaeger}, {Karamehmetoglu},
  {Krisciunas}, {Kumar}, {Lu}, {Maund}, {Mazzali}, {Medler}, {Morrell},
  {Phillips}, {Shappee}, {Stritzinger}, {Suntzeff}, {Telesco}, {Tucker}, \&
  {Wang}}]{kwok_21aefx_23}
{Kwok}, L.~A., {Jha}, S.~W., {Temim}, T., {et~al.} 2023, \apjl, 944, L3

\bibitem[{{Kwok} {et~al.}(2024){Kwok}, {Siebert}, {Johansson}, {Jha},
  {Blondin}, {Dessart}, {Foley}, {Hillier}, {Larison}, {Pakmor}, {Temim},
  {Andrews}, {Auchettl}, {Badenes}, {Barnabas}, {Bostroem}, {Brenner Newman},
  {Brink}, {Bustamante-Rosell}, {Camacho-Neves}, {Clocchiatti}, {Coulter},
  {Davis}, {Deckers}, {Dimitriadis}, {Dong}, {Farah}, {Filippenko},
  {Fl{\"o}rs}, {Fox}, {Garnavich}, {Padilla Gonzalez}, {Graur}, {Hambsch},
  {Hosseinzadeh}, {Howell}, {Hughes}, {Kerzendorf}, {Saux}, {Maeda}, {Maguire},
  {McCully}, {Mihalenko}, {Newsome}, {O'Brien}, {Pearson}, {Pellegrino},
  {Pierel}, {Polin}, {Rest}, {Rojas-Bravo}, {Sand}, {Schwab}, {Shahbandeh},
  {Shrestha}, {Smith}, {Strolger}, {Szalai}, {Taggart}, {Terreran}, {Terwel},
  {Tinyanont}, {Valenti}, {Vink{\'o}}, {Wheeler}, {Yang}, {Zheng}, {Ashall},
  {DerKacy}, {Galbany}, {Hoeflich}, {de Jaeger}, {Lu}, {Maund}, {Medler},
  {Morell}, {Shappee}, {Stritzinger}, {Suntzeff}, {Tucker}, \&
  {Wang}}]{kwok_22pul_24}
{Kwok}, L.~A., {Siebert}, M.~R., {Johansson}, J., {et~al.} 2024, \apj, 966, 135

\bibitem[{{Li} \& {McCray}(1993)}]{li_mccray_ca2_93}
{Li}, H. \& {McCray}, R. 1993, \apj, 405, 730

\bibitem[{{Li} {et~al.}(1993){Li}, {McCray}, \& {Sunyaev}}]{li_87A_93}
{Li}, H., {McCray}, R., \& {Sunyaev}, R.~A. 1993, \apj, 419, 824

\bibitem[{{Liljegren} {et~al.}(2023){Liljegren}, {Jerkstrand}, {Barklem},
  {Nyman}, {Brady}, \& {Yurchenko}}]{liljegren_ibc_mol_23}
{Liljegren}, S., {Jerkstrand}, A., {Barklem}, P.~S., {et~al.} 2023, \aap, 674,
  A184

\bibitem[{{Liljegren} {et~al.}(2020){Liljegren}, {Jerkstrand}, \&
  {Grumer}}]{liljegren_co_20}
{Liljegren}, S., {Jerkstrand}, A., \& {Grumer}, J. 2020, \aap, 642, A135

\bibitem[{{Liu} \& {Dalgarno}(1994)}]{liu_dalgarno_94}
{Liu}, W. \& {Dalgarno}, A. 1994, \apj, 428, 769

\bibitem[{{Liu} \& {Dalgarno}(1995)}]{liu_dalgarno_95}
{Liu}, W. \& {Dalgarno}, A. 1995, \apj, 454, 472

\bibitem[{{Liu} {et~al.}(1992){Liu}, {Dalgarno}, \& {Lepp}}]{liu_dalgarno_92}
{Liu}, W., {Dalgarno}, A., \& {Lepp}, S. 1992, \apj, 396, 679

\bibitem[{{Lucy} {et~al.}(1989){Lucy}, {Danziger}, {Gouiffes}, \&
  {Bouchet}}]{lucy_dust_89}
{Lucy}, L.~B., {Danziger}, I.~J., {Gouiffes}, C., \& {Bouchet}, P. 1989, in IAU
  Colloq. 120: Structure and Dynamics of the Interstellar Medium, Vol. 350, 164

\bibitem[{{Maguire} {et~al.}(2012){Maguire}, {Jerkstrand}, {Smartt},
  {Fransson}, {Pastorello}, {Benetti}, {Valenti}, {Bufano}, \&
  {Leloudas}}]{maguire_2p_12}
{Maguire}, K., {Jerkstrand}, A., {Smartt}, S.~J., {et~al.} 2012, \mnras, 420,
  3451

\bibitem[{{McLeod} {et~al.}(2024){McLeod}, {Hillier}, \&
  {Dessart}}]{mcleod_mol_24}
{McLeod}, C., {Hillier}, D.~J., \& {Dessart}, L. 2024, \mnras, 532, 549

\bibitem[{{Meikle} {et~al.}(2011){Meikle}, {Kotak}, {Farrah}, {Mattila}, {Van
  Dyk}, {Andersen}, {Fesen}, {Filippenko}, {Foley}, {Fransson}, {Gerardy},
  {H{\"o}flich}, {Lundqvist}, {Pozzo}, {Sollerman}, \&
  {Wheeler}}]{meikle_04dj_11}
{Meikle}, W.~P.~S., {Kotak}, R., {Farrah}, D., {et~al.} 2011, \apj, 732, 109

\bibitem[{{Rho} {et~al.}(2021){Rho}, {Evans}, {Geballe}, {Banerjee},
  {Hoeflich}, {Shahbandeh}, {Valenti}, {Yoon}, {Jin}, {Williamson}, {Modjaz},
  {Hiramatsu}, {Howell}, {Pellegrino}, {Vink{\'o}}, {Cartier}, {Burke},
  {McCully}, {An}, {Cha}, {Pritchard}, {Wang}, {Andrews}, {Galbany}, {Van Dyk},
  {Graham}, {Blinnikov}, {Joshi}, {P{\'a}l}, {Kriskovics}, {Ordasi}, {Szakats},
  {Vida}, {Chen}, {Li}, {Zhang}, \& {Yan}}]{rho_20oi_21}
{Rho}, J., {Evans}, A., {Geballe}, T.~R., {et~al.} 2021, \apj, 908, 232

\bibitem[{{Rho} {et~al.}(2018){Rho}, {Geballe}, {Banerjee}, {Dessart}, {Evans},
  \& {Joshi}}]{rho_co_17eaw_18}
{Rho}, J., {Geballe}, T.~R., {Banerjee}, D.~P.~K., {et~al.} 2018, \apjl, 864,
  L20

\bibitem[{{Sarangi}(2022)}]{sarangi_dust_22}
{Sarangi}, A. 2022, \aap, 668, A57

\bibitem[{{Sarangi} {et~al.}(2025){Sarangi}, {Zsiros}, {Szalai}, {Martinez},
  {Shahbandeh}, {Fox}, {Van Dyk}, {Filippenko}, {Bersten}, {De Looze},
  {Ashall}, {Temim}, {Jencson}, {Rest}, {Milisavljevic}, {Dessart}, {Dwek},
  {Smith}, {Tinyanont}, {Brink}, {Zheng}, {Clayton}, \&
  {Andrews}}]{sarangi_05af_25}
{Sarangi}, A., {Zsiros}, S., {Szalai}, T., {et~al.} 2025, arXiv e-prints,
  arXiv:2504.20574

\bibitem[{{Shahbandeh} {et~al.}(2022){Shahbandeh}, {Hsiao}, {Ashall}, {Teffs},
  {Hoeflich}, {Morrell}, {Phillips}, {Anderson}, {Baron}, {Burns}, {Contreras},
  {Davis}, {Diamond}, {Folatelli}, {Galbany}, {Gall}, {Hachinger}, {Holmbo},
  {Karamehmetoglu}, {Kasliwal}, {Kirshner}, {Krisciunas}, {Kumar}, {Lu},
  {Marion}, {Mazzali}, {Piro}, {Sand}, {Stritzinger}, {Suntzeff}, {Taddia}, \&
  {Uddin}}]{shahbandeh_nir_sesn_22}
{Shahbandeh}, M., {Hsiao}, E.~Y., {Ashall}, C., {et~al.} 2022, \apj, 925, 175

\bibitem[{{Shahbandeh} {et~al.}(2023){Shahbandeh}, {Sarangi}, {Temim},
  {Szalai}, {Fox}, {Tinyanont}, {Dwek}, {Dessart}, {Filippenko}, {Brink},
  {Foley}, {Jencson}, {Pierel}, {Zs{\'\i}ros}, {Rest}, {Zheng}, {Andrews},
  {Clayton}, {De}, {Engesser}, {Gezari}, {Gomez}, {Gonzaga}, {Johansson},
  {Kasliwal}, {Lau}, {De Looze}, {Marston}, {Milisavljevic}, {O'Steen},
  {Siebert}, {Skrutskie}, {Smith}, {Strolger}, {Van Dyk}, {Wang}, {Williams},
  {Williams}, {Xiao}, \& {Yang}}]{shahbandeh_jwst_23}
{Shahbandeh}, M., {Sarangi}, A., {Temim}, T., {et~al.} 2023, \mnras, 523, 6048

\bibitem[{{Spyromilio} {et~al.}(1988){Spyromilio}, {Meikle}, {Learner}, \&
  {Allen}}]{spyromilio_co_87A_88}
{Spyromilio}, J., {Meikle}, W.~P.~S., {Learner}, R.~C.~M., et al.
  1988, \nat, 334, 327

\bibitem[{{Spyromilio} \& {Pinto}(1991)}]{spyromilio_pinto_91}
{Spyromilio}, J. \& {Pinto}, P.~A. 1991, in ESO Conf. and Workshop Proc., 37, 423

\bibitem[{{Sukhbold} {et~al.}(2016){Sukhbold}, {Ertl}, {Woosley}, {Brown}, \&
  {Janka}}]{sukhbold_ccsn_16}
{Sukhbold}, T., {Ertl}, T., {Woosley}, S.~E., et al. 2016, \apj, 821, 38

\bibitem[{{Szalai} {et~al.}(2025){Szalai}, {Zs{\'\i}ros}, {Jencson}, {Fox},
  {Shahbandeh}, {Sarangi}, {Temim}, {De Looze}, {Smith}, {Filippenko}, {Van
  Dyk}, {Andrews}, {Ashall}, {Clayton}, {Dessart}, {Dulude}, {Dwek}, {Gomez},
  {Johansson}, {Milisavljevic}, {Pierel}, {Rest}, {Tinyanont}, {Brink}, {De},
  {Engesser}, {Foley}, {Gezari}, {Kasliwal}, {Lau}, {Marston}, {O'Steen},
  {Siebert}, {Skrutskie}, {Strolger}, {Wang}, {Williams}, {Williams}, {Xiao},
  \& {Zheng}}]{szalai_93j_25}
{Szalai}, T., {Zs{\'\i}ros}, S., {Jencson}, J., {et~al.} 2025, \aap, 697, A132

\bibitem[{{Weil} {et~al.}(2020){Weil}, {Fesen}, {Patnaude}, \&
  {Milisavljevic}}]{weil_17eaw_20}
{Weil}, K.~E., {Fesen}, R.~A., {Patnaude}, D.~J, et al.  2020,
  \apj, 900, 11

\bibitem[{{Wooden} {et~al.}(1993){Wooden}, {Rank}, {Bregman}, {Witteborn},
  {Tielens}, {Cohen}, {Pinto}, \& {Axelrod}}]{wooden_87A_ir_93}
{Wooden}, D.~H., {Rank}, D.~M., {Bregman}, J.~D., {et~al.} 1993, \apjs, 88, 477

\bibitem[{{Woosley}(1988)}]{woosley_87A_late_88}
{Woosley}, S.~E. 1988, \apj, 330, 218

\bibitem[{{Woosley}(2019)}]{woosley_he_19}
{Woosley}, S.~E. 2019, \apj, 878, 49

\bibitem[{{Woosley} \& {Heger}(2007)}]{WH07}
{Woosley}, S.~E. \& {Heger}, A. 2007, \physrep, 442, 269

\bibitem[{{Woosley} {et~al.}(2002){Woosley}, {Heger}, \& {Weaver}}]{whw02}
{Woosley}, S.~E., {Heger}, A., \& {Weaver}, T.~A. 2002, Rev. of Mod. Phys., 74, 1015

\bibitem[{{Zs{\'\i}ros} {et~al.}(2024){Zs{\'\i}ros}, {Szalai}, {De Looze},
  {Sarangi}, {Shahbandeh}, {Fox}, {Temim}, {Milisavljevic}, {Van Dyk}, {Smith},
  {Filippenko}, {Brink}, {Zheng}, {Dessart}, {Jencson}, {Johansson}, {Pierel},
  {Rest}, {Tinyanont}, {Niculescu-Duvaz}, {Barlow}, {Wesson}, {Andrews},
  {Clayton}, {De}, {Dwek}, {Engesser}, {Foley}, {Gezari}, {Gomez}, {Gonzaga},
  {Kasliwal}, {Lau}, {Marston}, {O'Steen}, {Siebert}, {Skrutskie}, {Strolger},
  {Wang}, {Williams}, {Williams}, \& {Xiao}}]{zsiros_80K_24}
{Zs{\'\i}ros}, S., {Szalai}, T., {De Looze}, I., {et~al.} 2024, \mnras, 529,
  155

\end{thebibliography}
\end{document}